\documentclass[11pt, usenames,dvipsnames]{article}
\pdfoutput=1

\usepackage{bm,wrapfig,float,array,mathtools}
\usepackage{amsfonts}
\usepackage{amssymb}
\usepackage{cite}
\usepackage{amsmath}
\usepackage{color}
\usepackage{graphicx}
\usepackage{hyperref}
\usepackage{float}
\usepackage[T1]{fontenc}
\usepackage[utf8]{inputenc}
\usepackage{alphabeta}
\usepackage{fancyvrb}
\usepackage[dvipsnames]{xcolor}
\usepackage{bold-extra}
\usepackage{lmodern}
\usepackage{cleveref}
\usepackage[normalem]{ulem}
\usepackage{tikz-cd}
\newcommand{\RNum}[1]{\uppercase\expandafter{\romannumeral #1\relax}}
\usepackage[dvipsnames]{xcolor}
\usepackage[framemethod=TikZ]{mdframed}
\usepackage{amsthm}
\usetikzlibrary{patterns}

\newcounter{state}[section]
\renewcommand{\thestate}{\arabic{section}.\arabic{state}}
\newenvironment{state}[2][]{%
\refstepcounter{state}%
\ifstrempty{#1}%
{\mdfsetup{%
frametitle={%
\tikz[baseline=(current bounding box.east),outer sep=0pt]
\node[anchor=east,rectangle,fill=NavyBlue!50]
{\strut Statement~\thestate};}}
}%
{\mdfsetup{%
frametitle={%
\tikz[baseline=(current bounding box.east),outer sep=0pt]
\node[anchor=east,rectangle,fill=NavyBlue!20]
{\strut Statement~\thestate:~#1};}}%
}%
\mdfsetup{innertopmargin=10pt,linecolor=NavyBlue!20,%
linewidth=2pt,topline=true,%
frametitleaboveskip=\dimexpr-\ht\strutbox\relax
}
\begin{mdframed}[]\relax%
\label{#2}}{\end{mdframed}}

\newcounter{mainstate}[section]

\newcounter{note}[section]
\renewcommand{\thenote}{\arabic{section}.\arabic{note}}
\newenvironment{note}[2][]
{%
\refstepcounter{note}%
\ifstrempty{#1}%
{\mdfsetup{%
frametitle={%
\tikz[baseline=(current bounding box.east),outer sep=0pt]
\node[anchor=east,rectangle,fill=PineGreen!30]
{\strut };}}
}%
{\mdfsetup{%
frametitle={%
\tikz[baseline=(current bounding box.east),outer sep=0pt]
\node[anchor=east,rectangle,fill=PineGreen!30]
{\strut Definition~\thenote:~#1};}}%
}%
\mdfsetup{innertopmargin=10pt,linecolor=PineGreen!30,%
linewidth=2pt,topline=true,%
frametitleaboveskip=\dimexpr-\ht\strutbox\relax
}
\begin{mdframed}[]\relax%
\label{#2}}{\end{mdframed}}

\newenvironment{tech}[2][]
{%
\ifstrempty{#1}%
{\mdfsetup{%
frametitle={%
\tikz[baseline=(current bounding box.east),outer sep=0pt]
\node[anchor=east,rectangle,fill=red!20]
{\strut };}}
}%
{\mdfsetup{%
frametitle={%
\tikz[baseline=(current bounding box.east),outer sep=0pt]
\node[anchor=east,rectangle,fill=red!20]
{\strut #1};}}%
}%
\mdfsetup{innertopmargin=10pt,linecolor=red!20,%
linewidth=2pt,topline=true,%
frametitleaboveskip=\dimexpr-\ht\strutbox\relax
}
\begin{mdframed}[]\relax%
\label{#2}}{\end{mdframed}}

\newcounter{example}[section]
\renewcommand{\theexample}{\arabic{section}.\arabic{example}}
\newenvironment{example}[2][]{%
\refstepcounter{example}%
\ifstrempty{#1}%
{\mdfsetup{%
frametitle={%
\tikz[baseline=(current bounding box.east),outer sep=0pt]
\node[anchor=east,rectangle,fill=orange!50]
{\strut Example~\theexample};}}
}%
{\mdfsetup{%
frametitle={%
\tikz[baseline=(current bounding box.east),outer sep=0pt]
\node[anchor=east,rectangle,fill=orange!20]
{\strut Example~\theexample:~#1};}}%
}%
\mdfsetup{innertopmargin=10pt,linecolor=orange!20,%
linewidth=2pt,topline=true,%
frametitleaboveskip=\dimexpr-\ht\strutbox\relax
}
\begin{mdframed}[]\relax%
\label{#2}}{\end{mdframed}}

\graphicspath{ {figures/} }

\usepackage{tikz}
\usepackage{diagbox}
\usetikzlibrary{positioning}
\usetikzlibrary{calc}
\usetikzlibrary{decorations.pathreplacing,calligraphy}

\usepackage{xstring}
\usetikzlibrary{decorations.pathmorphing} 
\usetikzlibrary{decorations.markings} 
\usetikzlibrary{arrows} 
\usetikzlibrary{shapes} 
\usetikzlibrary{matrix} 
\usetikzlibrary{positioning} 
\usepackage[english]{babel} 
\usepackage[autostyle]{csquotes}
\usepackage{pifont}
\usetikzlibrary{shapes.multipart}

\newcommand{\bea}{\begin{eqnarray}}
\newcommand{\eea}{\end{eqnarray}}
\newcommand{\be}{\begin{equation}}
\newcommand{\ee}{\end{equation}}
\newcommand{\ba}{\begin{aligned}}
\newcommand{\ea}{\end{aligned}}
\newcommand{\bit}{\begin{itemize}}
\newcommand{\eit}{\end{itemize}}
\newcommand{\ben}{\begin{enumerate}}
\newcommand{\een}{\end{enumerate}}

\renewcommand{\ni}{\noindent}
\newcommand{\id}{\text{id}}

\newcommand{\Tr}{\text{Tr}}

\newcommand{\wt}{\widetilde}
\newcommand{\wh}{\widehat}
\newcommand{\ot}{\otimes}
\newcommand{\half}{\frac{1}{2}}

\newcommand{\Z}{{\mathbb Z}}
\newcommand{\R}{{\mathbb R}}
\newcommand{\bC}{{\mathbb C}}

\renewcommand{\P}{{\mathbb P}}

\newcommand{\bG}{{\mathbb G}}

\newcommand{\sym}{\text{sym}}
\newcommand{\phys}{\text{phys}}
\newcommand{\Bsym}{\mathfrak{B}^{\sym}}

\newcommand{\Bphys}{\mathfrak{B}^{\phys}}

\newcommand{\Spin}{\text{Spin}}

\newcommand{\cA}{\mathcal{A}}

\newcommand{\cD}{\mathcal{D}}
\newcommand{\cE}{\mathcal{E}}

\newcommand{\cG}{\mathcal{G}}
\newcommand{\cH}{\mathcal{H}}
\newcommand{\cI}{\mathcal{I}}

\newcommand{\cL}{\mathcal{L}}
\newcommand{\cM}{\mathcal{M}}
\newcommand{\cN}{\mathcal{N}}
\newcommand{\cO}{\mathcal{O}}
\newcommand{\cP}{\mathcal{P}}

\newcommand{\cS}{\mathcal{S}}
\newcommand{\cT}{\mathcal{T}}

\newcommand{\cW}{\mathcal{W}}
\newcommand{\cX}{\mathcal{X}}
\newcommand{\cY}{\mathcal{Y}}
\newcommand{\cZ}{\mathcal{Z}}
\newcommand{\Bock}{\text{Bock}}

\renewcommand{\L}{\mathsf{\Lambda}}

\newcommand{\fT}{\mathfrak{T}}
\newcommand{\fB}{\mathfrak{B}}
\newcommand{\fZ}{\mathfrak{Z}}
\newcommand{\fe}{\mathfrak{e}}
\newcommand{\ff}{\mathfrak{f}}
\newcommand{\fg}{\mathfrak{g}}

\newcommand{\su}{\mathfrak{su}}
\renewcommand{\sp}{\mathfrak{sp}}
\newcommand{\so}{\mathfrak{so}}
\renewcommand{\u}{\mathfrak{u}}

\newcommand{\Rep}{\mathsf{Rep}}

\newcommand{\bD}{\mathsf{D}}

\newcommand{\G}[1]{G^{(#1)}}
\newcommand{\whG}[1]{\wh G^{(#1)}}
\renewcommand{\H}[1]{H^{(#1)}}
\newcommand{\whH}[1]{\wh H^{(#1)}}
\newcommand{\K}[1]{K^{(#1)}}

\begin{document}

\baselineskip=18pt  
\numberwithin{equation}{section}  
\allowdisplaybreaks  

\thispagestyle{empty}

\vspace*{0.8cm} 
\begin{center}
{{\huge Lectures on Generalized Symmetries}}

 \vspace*{1.5cm}
{\large Lakshya Bhardwaj$^1$, Lea E.\ Bottini$^1$, Ludovic Fraser-Taliente$^2$, Liam Gladden$^2$,\\ Dewi S.W.\ Gould$^1$, Arthur Platschorre$^2$ and Hannah Tillim$^{2,3}$}\\

 \vspace*{.2cm} 
{\small \textit{(Based on a course given by Lakshya Bhardwaj)}}
\end{center}

  \vspace*{.5cm}
$^1$ {\it  Mathematical Institute, University of Oxford, Andrew-Wiles Building,  Woodstock Road, Oxford OX2 6GG, UK}

$^2$ {\it  Rudolf Peierls Centre for Theoretical Physics, University of Oxford, Parks Road, Oxford OX1 3PU, UK}

$^3$ {\it  William H. Miller III Department of Physics and Astronomy, Johns Hopkins University, Baltimore, MD 21218, USA}

\smallskip

\vspace*{2cm}

\noindent
These are a set of lecture notes on generalized global symmetries in quantum field theory. The focus is on invertible symmetries with a few comments regarding non-invertible symmetries. The main topics covered are the basics of higher-form symmetries and their properties including 't Hooft anomalies, gauging and spontaneous symmetry breaking. We also introduce the useful notion of symmetry topological field theories (SymTFTs). Furthermore, an introduction to higher-group symmetries describing mixings of higher-form symmetries is provided. Some advanced topics covered include the encoding of higher-form symmetries in holography and geometric engineering constructions in string theory. Throughout the text, all concepts are consistently illustrated using gauge theories as examples.

\newpage

\tableofcontents

\section{Introduction}
The aim of these lecture notes is to introduce the subject of generalized global symmetries in quantum field theory (QFT). This is a subject that has witnessed intense research activity over the last decade, with the research in this field connecting to broad research areas of theoretical and phenomenological high-energy physics including quantum gravity, and theoretical condensed matter physics. The main source of this excitement is the possibility of exploiting these generalized symmetries in ways similar to how global symmetries have been utilized for probing a variety of physical phenomena in QFT and related topics. Global symmetries and their 't Hooft anomalies are invariant under renormalization group (RG) flows and thus can be used to probe strongly coupled physics beyond the reach of the standard tools of perturbation theory. Related to this is the fact that spontaneous breaking of global symmetries provides a powerful paradigm \`a la Landau and Ginzburg for understanding the phase structure and organizing the phase diagram in the infrared (IR). As we will discuss, all of these concepts familiar from the study of standard global symmetries, are also applicable in the study of generalized global symmetries.

\paragraph{Related Literature.}
Let us note at the outset that these notes only introduce what are known as \textbf{invertible} generalized symmetries. These include higher-form and higher-group symmetries. Much of recent effort in this field over the last couple of years has been focused on the study of \textbf{non-invertible symmetries}. For a set of excellent introductory notes on the topic of non-invertible symmetries, we refer the reader to the recently appeared \cite{Schafer-Nameki:2023jdn}. To a reader without background in this field and wishing to obtain an understanding up to the cutting edge of this field, we would recommend a paired reading of this set of notes and of \cite{Schafer-Nameki:2023jdn}. We also note the recent manuscript \cite{Brennan:2023mmt} that provides an introduction to generalized global symmetries aimed at high energy phenomenologists. Our approach will be complementary to theirs and aimed at high energy theorists. Another set of notes were provided in \cite{Gomes:2023ahz}, which are at a more introductory level.

\paragraph{Prerequisites.}
These notes are aimed at the level of an intermediate or advanced graduate student. The physics prerequisites include an advanced understanding of quantum field theory, including its non-perturbative aspects. The reader is assumed to be familiar with non-abelian gauge theories, magnetic monopoles, background field formalism, anomalies, spontaneous symmetry breaking, and Higgs mechanism. Some knowledge of supersymmetry is also assumed in some examples, but it is not necessary to understand the fundamental concepts. The math prerequisites include a good understanding of differential geometry and algebraic topology, at the level used in advanced quantum field theory and string theory courses. A good understanding of group theory would be extremely beneficial as well.

\paragraph{Organization of the Notes.}
These notes are organised as follows:
\bit
\item In section \ref{hfs}, we introduce the concept of higher-form symmetries as a generalization of the usual notion of global symmetries. This is done by rephrasing global symmetries in terms of topological operators. Many examples of quantum field theories admitting higher-form symmetries are discussed. This includes the Maxwell theory and discrete gauge theories.
\item In section \ref{hfg}, we discuss higher-form symmetries of general abelian and non-abelian gauge theories. From a more theoretical viewpoint, this section can be seen as a collection of many examples. Alternatively, from a more practical viewpoint, this section can be seen as describing many fine and subtle properties of gauge theories that are often not emphasized in a first introduction to gauge theories.
\item In section \ref{hfp}, we discuss various properties associated to higher-form symmetries, which provide the essential toolkit used when one employs these symmetries for physical applications. This includes a discussion of 't Hooft anomalies, symmetry protected topological (SPT) phases, gauging and spontaneous symmetry breaking of these symmetries. We also introduce the concept of symmetry topological field theory (SymTFT).
\item In section \ref{hgs}, we introduce higher-group symmetries. These are new symmetry structures that can be interpreted as arising from ``mixings'' of higher-form symmetries. We discuss various types of higher-group symmetries, involving continuous and discrete higher-form symmetries, and describe how they arise in various gauge theories. However, we do not delve into their properties like 't Hooft anomalies, gauging and spontaneous breaking. By this point in the text, we trust that the reader would be well-equipped to easily gather these concepts from the listed references.
\item In section \ref{adv}, we provide brief introductory expositions to the study of higher-form symmetries in geometric engineering constructions of quantum field theories via string theory, and the study of higher-form symmetries using holographic duality.
\eit

\section{Introduction to Higher-Form Symmetries}\label{hfs}
The aim of this section is to introduce $p$-form symmetries. These symmetries generalize the usual global symmetries, which in this language are referred to as 0-form symmetries.

We will follow the seminal work \cite{Gaiotto:2014kfa}, though this is not to say that this was the first work discussing such ideas. In fact, many of the ideas involved can be traced to works by various communities in late 1980s and early 1990s in the study of non-perturbative aspects of gauge theories, discrete gauge theories, rational CFTs and Chern-Simons theories. See \cite{Alford:1992yx,alford1990discrete,coleman1992quantum,coleman1991growing,alford1990interactions,alford1992new,Bais:1993ax,Bais:1991pe,alford1991zero,alvarez1989hidden,alvarez1990duality,dijkgraaf1990topological,dijkgraaf1991quasi,dijkgraaf1989operator,Moore:1988qv,Elitzur:1989nr,Moore:1989vd,Bucher:1991bc,vafa1986modular,witten1989quantum} (in no particular order) for a small sample of these works. Additionally, the works \cite{Frohlich:2004ef,Pantev:2005rh,Pantev:2005wj,Pantev:2005zs,Freed:2006ya,Freed:2006yc,Frohlich:2006ch,Fuchs:2007tx,Witten:2009at,Frohlich:2009gb,Nussinov:2009zz,Seiberg:2010qd,Gaiotto:2010be,Davydov:2010rm,Seiberg:2011dr,Chen:2011pg,Gu:2012ib,Aharony:2013hda,Gukov:2013zka,Kapustin:2013qsa,Kapustin:2013uxa,Razamat:2013opa,Freed:2014eja,Kapustin:2014gua,Kapustin:2014tfa,Kapustin:2014dxa,Kapustin:2014zva,Dierigl:2014xta} were precursors of \cite{Gaiotto:2014kfa}.

\subsection{0-Form Symmetries}
Let us begin by reviewing the well-known case of 0-form symmetries. A standard way to describe such a symmetry is in terms of a unitary operator
\be
U(t)\,,
\ee
acting on the Hilbert space at time $t$ which commutes with the Hamiltonian generating time evolution, i.e.
\be\label{t1}
U(t)=U(t')\,, \qquad \forall~t,t'\,.
\ee
Moreover, $U(t)$ acts on all of space, because it acts on local operators located at any point in space. The action is implemented by conjugation
\be\label{a1}
U(t)\cO(\vec{x},t)U^{-1}(t)=\cO'(\vec{x},t)\,,
\ee
which transforms a local operator $\cO(\vec{x},t)$ into another local operator $\cO'(\vec{x},t)$.

The crucial insight of \cite{Gaiotto:2014kfa} was that the above properties of the operator $U(t)$ can be rephrased as follows:

\begin{state}[0-Form Symmetries are Topological Codimension-1 Operators]{0fs}
A 0-form symmetry is a codimension-1 operator $U$ which is topological and invertible.
\end{state}

\ni Let us unpack this statement further: 
\paragraph{Insertion Along General Codimension-1 Submanifolds.}
The codimension-1 operator $U$ can be inserted along any codimension-1 manifold $\Sigma_{d-1}$ (which may or may not be compact) inside the $d$-dimensional spacetime. After insertion, the operator is denoted as
\be
U(\Sigma_{d-1})\,.
\ee
In particular, we can choose 
\be
\Sigma_{d-1} =\{\text{Spatial slice at a particular time $t$}\}\equiv\Sigma_t\,,
\ee
in which case the inserted operator can be identified with the unitary operator $U(t)$ acting on the Hilbert space that we discussed above
\be\label{u}
U(\Sigma_t)\equiv U(t)\,.
\ee

\begin{tech}[Operators in Correlation Functions]{}
Note that $U(\Sigma_{d-1})$ is best interpreted as a codimension-1 operator inside a correlation function rather than an operator acting on a Hilbert space. This is because if a particular time direction has been chosen, then the submanifold $\Sigma_{d-1}$ need not be a spatial slice.
\end{tech}

\begin{figure}
\centering
\scalebox{1.1}{
\begin{tikzpicture}[scale=1.3]
\draw[fill=red!20, thick] (0,0) ellipse (0.5 and 1);
\draw[fill=red!20, pattern color=red, thick] (4,0) ellipse (0.7 and 1.2);
\draw[thick,blue] (0,1) to [out=-5, in=125] (4.3,1.1);
\draw[thick,blue] (0,-1) to [out=5, in=235] (4.3,-1.1);
\node[black] at (0,0) {$\Sigma_{d-1}$}; 
\node[black] at (4,0) {$\Sigma'_{d-1}$}; 
\node[blue] at (2,0) {$\Sigma_d$}; 
\end{tikzpicture}
}
\caption{Two $(d-1)$-dimensional manifolds $\Sigma_{d-1}, \Sigma'_{d-1}$ which are related by a topological deformation. $\Sigma_d$ is a $d$-dimensional manifold whose boundary is formed by $\Sigma'_{d-1}$ and an orientation-reversed copy of $\Sigma_{d-1}$. Placing a topological operator $U$ along $\Sigma_{d-1}$ is equivalent to placing it along $\Sigma'_{d-1}$.}
\label{fig:stokesontopops}
\end{figure}
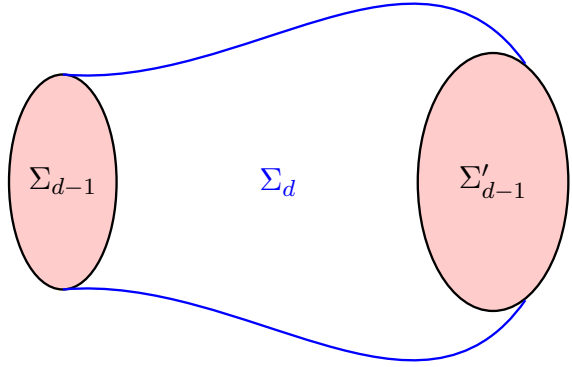

\paragraph{Topological Nature.}
The fact that $U$ is a topological operator means that we have
\be\label{t2}
U(\Sigma_{d-1})=U(\Sigma'_{d-1})\,,
\ee
if $\Sigma'_{d-1}$ is a codimension-1 submanifold of spacetime that can be obtained by topologically deforming $\Sigma_{d-1}$, i.e.\ if we have
\be\label{d}
\Sigma'_{d-1}-\Sigma_{d-1}=\partial\Sigma_d\,,
\ee
where $\partial\Sigma_d$ is the boundary of a $d$-dimensional submanifold $\Sigma_d$ of spacetime. See figure \ref{fig:stokesontopops}. This topological nature (\ref{t2}) of the codimension-1 operator $U(\Sigma_{d-1})$ is a generalization of the fact (\ref{t1}) that the unitary operator $U(t)$ is time independent, as substituting
\be\label{s}
\Sigma_{d-1}=\Sigma_t,\qquad \Sigma'_{d-1}=\Sigma_{t'}\,,
\ee
in (\ref{t2}) recovers the familiar statement (\ref{t1}). 

\paragraph{Invertibility Property.}
The fact that $U$ is invertible means that there exists another topological codimension-1 operator $U'$ such that we have the relationship
\be\label{i}
U(\Sigma_{d-1})U'(\Sigma_{d-1})=1\,.
\ee
That is, inserting both $U$ and $U'$ along the same codimension-1 submanifold $\Sigma_{d-1}$ is equivalent to inserting no operator along $\Sigma_{d-1}$. It is convenient to use the notation
\be
U'\equiv U^{-1}\,,
\ee
in terms of which the equation (\ref{i}) becomes
\be
U(\Sigma_{d-1})U^{-1}(\Sigma_{d-1})=1\,.
\ee
For the operators (\ref{u}) acting on the Hilbert space, the invertibility follows from the fact that they are unitary.

\paragraph{Action on Local Operators.}
The action (\ref{a1}) on local operators can be phrased using the general operators $U(\Sigma_{d-1})$ as the following equation (valid inside a correlation function)
\be\label{a2}
U(\Sigma_{d-1})\cO(x)=\cO'(x)U(\Sigma'_{d-1})\,,
\ee
where
\be
x\equiv(\vec{x},t)\,,
\ee
is a spacetime point lying inside the submanifold $\Sigma_d$ of spacetime relating $\Sigma_{d-1}$ and $\Sigma'_{d-1}$ as in (\ref{d}). Given the equation (\ref{a2}), we say that the 0-form symmetry $U$ acts on the local operator $\cO(x)$ by converting it into the local operator $\cO'(x)$. The equation (\ref{a1}) is recovered by substituting (\ref{s}) into (\ref{a2}) and taking the limit
\be
\Delta t=t'-t\to 0\,.
\ee

\begin{figure}
\centering
\scalebox{1.1}{
  \begin{tikzpicture}
  \draw[red] (0,0) circle (2cm);
  \draw[red] (-2,0) arc (180:360:2 and 0.6);
  \draw[dashed,red] (2,0) arc (0:180:2 and 0.6);
  \fill[fill=black] (0,0) circle (1pt);
\node[red] (n4) at (0,1) {$U(S^{d-1})$};
\node[] (n4) at (-0.5,0) {$\cO(x)~$};
\begin{scope}[shift={(5,0)}]
\draw[red] (1,0) circle (0.5cm);
  \draw[red] (0.5,0) arc (180:360:0.5 and 0.2);
  \draw[dashed,red] (1.5,0) arc (0:180:0.5 and 0.2);
  \fill[fill=black] (0,0) circle (1pt);
\node[] (n4) at (-0.5,0) {$\cO'(x)~$};
\node[red] (n4) at (1,-1) {$U(S^{d-1})$};
\end{scope}
\begin{scope}[shift={(9.5,0)}]
\fill[fill=black] (0,0) circle (1pt);
\node[] (n4) at (-0.5,0) {$\cO'(x)~$};
\end{scope}
\node at (3,0) {=};
\node at (7.5,0) {=};
\end{tikzpicture}
}
\caption{Left: A topological operator $U$ is inserted on a sphere $S^{d-1}$ which links with the point $x$ where local operator $\cO$ is placed. Middle: $U(S^{d-1})$ has been topologically deformed such that it now does not link with $x$. Due to (\ref{a2}), this process modifies the local operator placed at $x$ to be $\cO'$. Right: As $S^{d-1}$ is now the boundary of a $d$-dimensional ball, we can contract away $U(S^{d-1})$.}
\label{fig:spheretopopcontraction}
\end{figure}
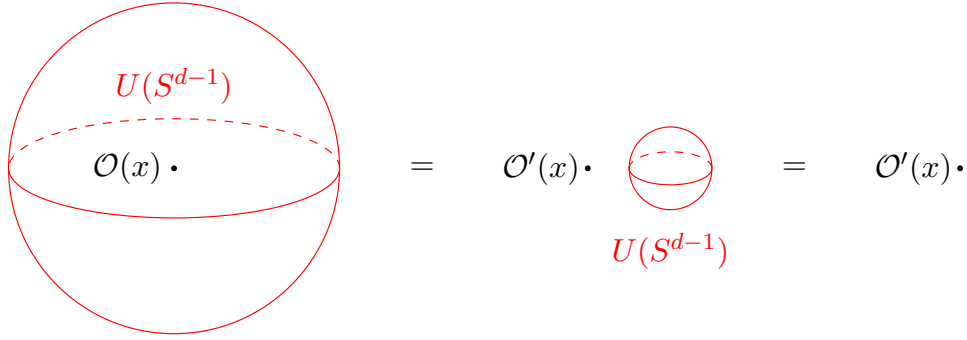

\paragraph{Action Via Linking.}
An equivalent way of describing the action (\ref{a2}) is
\be\label{a3}
U(S^{d-1})\cO(x)=\cO'(x)\,.
\ee
where on the LHS we insert the operator $U$ along a $(d-1)$-dimensional sphere $S^{d-1}$ linking the spacetime point $x$ as shown in the figure \ref{fig:spheretopopcontraction}. This representation (\ref{a3}) of the action can be obtained from (\ref{a2}) as explained in figure \ref{fig:spheretopopcontraction}.

\begin{figure}
\centering
\scalebox{1.1}{
\begin{tikzpicture}[scale=1]
  \draw[red] (0,0) circle (2cm);
  \draw[red] (-2,0) arc (180:360:2 and 0.6);
  \draw[dashed,red] (2,0) arc (0:180:2 and 0.6);
\begin{scope}[scale=1.5]
    \draw[blue] (0,0) circle (2cm);
  \draw[blue] (-2,0) arc (180:360:2 and 0.6);
  \draw[dashed,blue] (2,0) arc (0:180:2 and 0.6);
  \node[blue] (n4) at (0,1.75) {$U_g(S^{d-1})$};
\end{scope}
  \fill[fill=black] (0,0) circle (1pt);
\node[red] (n4) at (0,1.5) {$U_{g'}(S^{d-1})$};
\node[] (n4) at (-0.5,0) {$\cO(x)~$};
\node[rotate=-45] at (3.5,-2.5) {=};
\begin{scope}[shift={(4.5,-5.5)}]
\begin{scope}[scale=1]
    \draw[blue] (0,0) circle (2cm);
  \draw[blue] (-2,0) arc (180:360:2 and 0.6);
  \draw[dashed,blue] (2,0) arc (0:180:2 and 0.6);
  \node[blue] (n4) at (0,1.5) {$U_g(S^{d-1})$};
\end{scope}
  \fill[fill=black] (0,0) circle (1pt);
\node[] (n4) at (-1,0) {$g'\cdot\cO(x)$};
\end{scope}
\node[rotate=90] at (4.5,-8) {=};
\begin{scope}[shift={(4.5,-9)}]
  \fill[fill=black] (0,0) circle (1pt);
\node[] (n4) at (-1,0) {$g\cdot\left(g'\cdot\cO(x)\right)\qquad$};
\end{scope}
\node[rotate=45] at (-3.5,-2.5) {=};
\begin{scope}[shift={(-4.5,-5.5)}]
\begin{scope}[scale=1]
    \draw[purple] (0,0) circle (2cm);
  \draw[purple] (-2,0) arc (180:360:2 and 0.6);
  \draw[dashed,purple] (2,0) arc (0:180:2 and 0.6);
  \node[purple] (n4) at (0,1.5) {$U_{gg'}(S^{d-1})$};
\end{scope}
  \fill[fill=black] (0,0) circle (1pt);
\node[] (n4) at (-0.5,0) {$\cO(x)~$};
\end{scope}
\node[rotate=90] at (-4.5,-8) {=};
\begin{scope}[shift={(-4.5,-9)}]
  \fill[fill=black] (0,0) circle (1pt);
\node[] (n4) at (-1,0) {$(gg')\cdot\cO(x)\quad$};
\end{scope}
\end{tikzpicture}
}
\caption{Beginning with the top configuration, we can either take the left route in which we first fuse the topological operators and then act on the local operator, or we can take the right route in which we sequentially act on the local operator $\cO$. Both routes must yield the same result, and so we obtain equation (\ref{actfus}).}
\label{fusact}
\end{figure}
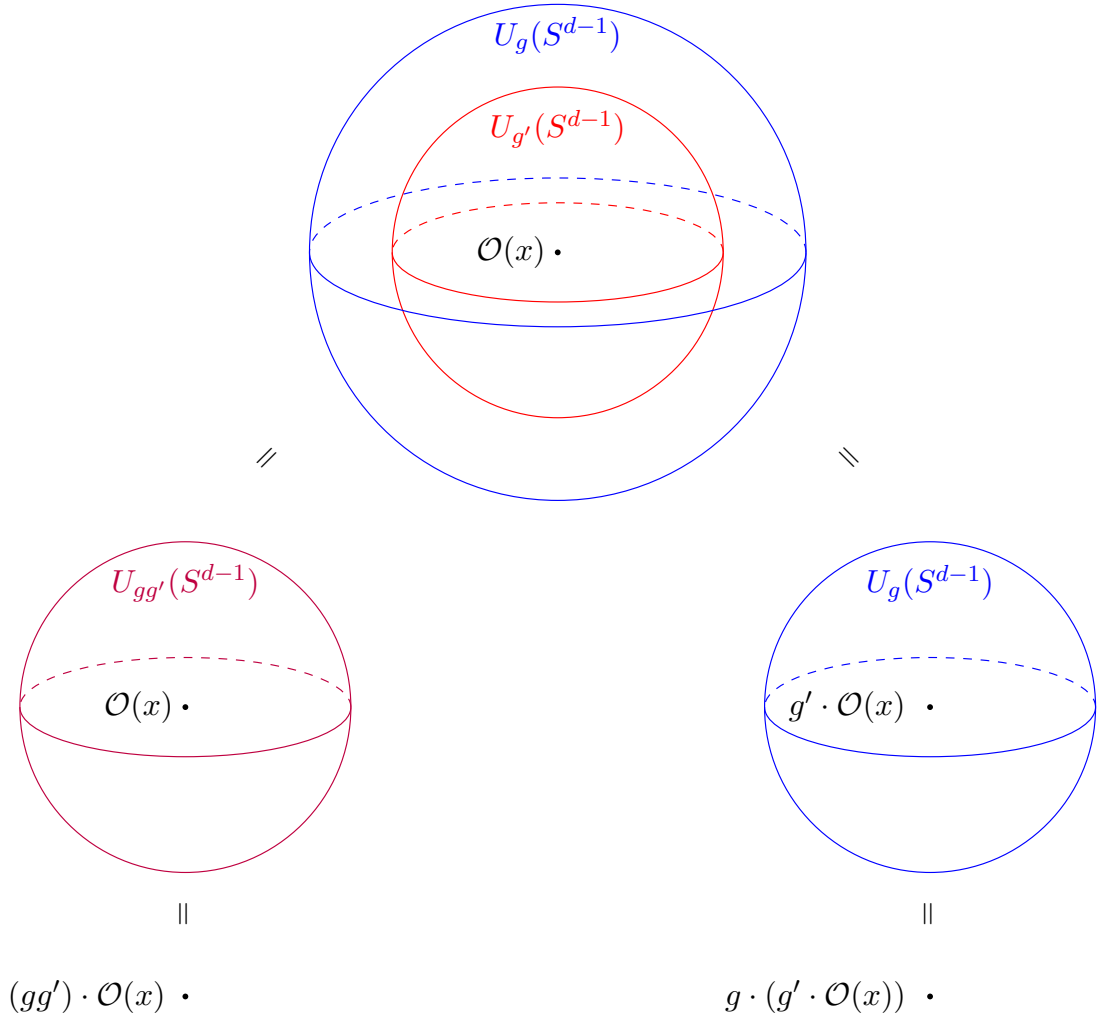

\paragraph{0-Form Symmetry Group.}
In general, 0-form symmetries combine to form a (possibly non-abelian) 0-form symmetry group
\be
G^{(0)}\,.
\ee
This means that we have topological codimension-1 operators
\be
U_g,\qquad \forall~g\in G^{(0)}\,,
\ee
parametrized by the elements of the group $G^{(0)}$. The group multiplication describes the composition of these operators
\be\label{c}
U_g(\Sigma_{d-1})U_{g'}(\Sigma_{d-1})=U_{gg'}(\Sigma_{d-1})\,,
\ee
which is often denoted as a \textbf{fusion rule}
\be
U_g\ot U_{g'}=U_{gg'}\,.
\ee
If we restrict to $\Sigma_{d-1}$'s that are time slices, and so $U$'s that are unitary operators acting on the Hilbert space, (\ref{c}) becomes
\be
U_g(t)U_{g'}(t)=U_{gg'}(t)\,,
\ee
which implies that the Hilbert space forms a representation of the 0-form symmetry group $G^{(0)}$. Similarly, the vector space of local operators (at a spacetime point $x$) also forms a representation of $G^{(0)}$. To see this, use the presentation (\ref{a3}) of the action of operators $U_g$ on local operators
\be
U_g\left(S^{d-1}\right)\cO(x)=g\cdot\cO(x)\,.
\ee
Using (\ref{c}) for $\Sigma_{d-1}=S^{d-1}$ we obtain
\be\label{actfus}
g\cdot\left(g'\cdot\cO(x)\right)=(gg')\cdot\cO(x)\,.
\ee
See figure \ref{fusact}.

\begin{example}[$U(1)$ 0-Form Symmetry]{U10}
The example of 
\be
G^{(0)}=U(1)\,,
\ee
neatly illustrates all of the above concepts. Such a symmetry is associated to a Noether current\footnote{Recently obstructions to the validity of Noether's theorem have been pointed out in \cite{Benedetti:2022zbb}.}, which is a local operator $j^\mu(x)$ satisfying the continuity equation
\be\label{ce}
\partial_\mu j^\mu=0\,.
\ee
From this, one usually defines the charge operator
\be
Q(t)=\int d^{d-1}\vec{x}~~j^0(\vec{x},t)\,,
\ee
which satisfies
\be\label{ce2}
\frac{dQ}{dt}=0\,,
\ee
as a consequence of (\ref{ce}). The unitary operators $U_g(t),~g\in U(1)$ implementing the $U(1)$ global symmetry on the Hilbert space are constructed as
\be
U_g(t)=\exp\left(i\alpha Q(t)\right),\qquad g=e^{i\alpha}\in U(1)\,,
\ee
which satisfy
\be
U_g(t')=U_g(t)\,,
\ee
as a consequence of (\ref{ce2}). 

Let us now describe the associated topological codimension-1 operators $U_g$. The insertion on an arbitrary submanifold $\Sigma_{d-1}$ of spacetime can be described as
\be
U_g(\Sigma_{d-1})=\exp\left(i\alpha\int_{\Sigma_{d-1}} j^{d-1} \right)\,,
\ee
where $j^{d-1}$ is a $(d-1)$-form on spacetime, which in index notation takes the form
\be
j^{d-1}_{\mu_1\mu_2\cdots\mu_{d-1}}=\epsilon_{\mu_1\mu_2\cdots\mu_{d-1}\mu_d}j^{\mu_d}\,,
\ee
where $\epsilon_{\mu_1\mu_2\cdots\mu_{d-1}\mu_d}$ is the completely antisymmetric Levi-Civita tensor. This $(d-1)$-form is closed
\be
dj^{d-1}=0\,,
\ee
as a consequence of (\ref{ce}), which implies the topological property
\be\label{t0}
\ba
U_g(\Sigma'_{d-1})&=U_g(\Sigma_{d-1})\times\exp\left(i\alpha\int_{\Sigma'_{d-1}-\Sigma_{d-1}=\partial\Sigma_d} j^{d-1} \right)\,,\\
&=U_g(\Sigma_{d-1})\times\exp\left(i\alpha\int_{\Sigma_d} dj^{d-1} \right)\,,\\
&=U_g(\Sigma_{d-1})\,.
\ea
\ee
In the presence of a local operator $\cO(x)$ of charge
\be
q\in\Z\,,
\ee
under $U(1)$, the continuity equation is modified to
\be
\cO(x)\partial_\mu j^\mu(x')=q\delta(x-x')\cO(x)\,,
\ee
or in terms of the $(d-1)$-form
\be
\cO(x)dj^{d-1}=q\delta^d(x)\cO(x)\,,
\ee
where $\delta^d(x)$ is the $d$-form Poincaré dual to the delta function $\delta(x-x')$. Using this we explicitly compute the action (\ref{a3}) to be
\be\label{da}
\ba
U_g\left(S^{d-1}\right)\cO(x)&=\exp\left(i\alpha\int_{S^{d-1}} j^{d-1} \right)\cO(x)\,,\\
&=\exp\left(i\alpha\int_{D^d} dj^{d-1} \right)\cO(x)\,,\\
&=\exp\left(i\alpha\int_{D^d} q\delta^d(x) \right)\cO(x)\,,\\
&=\exp\left(iq\alpha\right)\cO(x)\,.
\ea
\ee
where $D^d$ is the $d$-dimensional disk containing the point $x$ whose boundary is $S^{d-1}$.
\end{example}

\subsection{Higher-Form Symmetries}
\subsubsection{General Properties}
The generalization to higher-form symmetries occurs by generalizing the codimension in statement \ref{0fs}.

\begin{note}[$p$-Form Symmetries]{}
A $p$-form symmetry is a codimension-$(p+1)$ operator $U$ which is topological and invertible.
\end{note}
We can unpack the above definition further in a close analogy with the discussion following statement \ref{0fs}.

\paragraph{Insertion Along General Codimension-$(p+1)$ Submanifolds.}
The operator $U$ can be inserted along any codimension-$(p+1)$ submanifold $\Sigma_{d-p-1}$ of spacetime. After insertion, the operator is denoted
\be
U(\Sigma_{d-p-1})\,.
\ee
In particular, we can choose $\Sigma_{d-p-1}$ to lie in a spatial slice at a particular time $t$, in which case the inserted operator $U(\Sigma_{d-p-1})$ can be identified with a unitary operator acting on the Hilbert space at time $t$.

\paragraph{Topological Nature.}
The fact that $U$ is a topological operator means that we have
\be
U(\Sigma_{d-p-1})=U(\Sigma'_{d-p-1})\,,
\ee
if $\Sigma'_{d-p-1}$ is a codimension-$(p+1)$ submanifold of spacetime that can be obtained by topologically deforming $\Sigma_{d-p-1}$, i.e.\ if we have
\be
\Sigma'_{d-p-1}-\Sigma_{d-p-1}=\partial\Sigma_{d-p}\,,
\ee
where $\partial\Sigma_{d-p}$ is the boundary of a $(d-p)$-dimensional submanifold $\Sigma_{d-p}$ of spacetime. This can be depicted analogously to figure \ref{fig:stokesontopops}.

\paragraph{Invertibility Property.}
The fact that $U$ is invertible means that there exists another topological codimension-$(p+1)$ operator $U^{-1}$ such that we have the relationship
\be
U(\Sigma_{d-p-1})U^{-1}(\Sigma_{d-p-1})=1\,.
\ee
That is, inserting both $U$ and $U^{-1}$ along the same codimension-$(p+1)$ submanifold $\Sigma_{d-p-1}$ is equivalent to inserting no operator along $\Sigma_{d-p-1}$. 

\paragraph{$p$-Form Symmetry Group.}
In general, $p$-form symmetries combine to form a $p$-form symmetry group
\be
G^{(p)}.
\ee
This means that we have topological codimension-$(p+1)$ operators
\be
U_g,\qquad \forall~g\in G^{(p)},
\ee
parametrized by the elements of the group $G^{(p)}$. The group multiplication describes the composition of these operators
\be\label{fr}
U_g(\Sigma_{d-p-1})U_{g'}(\Sigma_{d-p-1})=U_{gg'}(\Sigma_{d-p-1})\,,
\ee
which is often denoted as a \textbf{fusion rule}
\be
U_g\ot U_{g'}=U_{gg'}\,.
\ee

\begin{figure}
\centering
\scalebox{1.2}{
\begin{tikzpicture}
\draw [thick,red](-1,1) -- (-1,-2);
\draw [thick,blue](1,1) -- (1,-2);
\node at (3,-0.5) {=};
\draw [thick,blue](5,1) -- (5,-2);
\draw [thick,red](7,1) -- (7,-2);
\node[red] at (-1,-2.5) {$U_g$};
\node[blue] at (1,-2.5) {$U_{g'}$};
\node[red] at (7,-2.5) {$U_g$};
\node[blue] at (5,-2.5) {$U_{g'}$};
\node[rotate=90] at (0,-3.5) {=};
\node[rotate=90] at (6,-3.5) {=};
\draw [thick,purple](0,-4.5) -- (0,-7.5);
\node[purple] at (0,-8) {$U_{gg'}$};
\begin{scope}[shift={(6,0)}]
\draw [thick,brown](0,-4.5) -- (0,-7.5);
\node[brown] at (0,-8) {$U_{g'g}$};
\end{scope}
\draw[-stealth] (1,1.5) .. controls (1,3) and (-2.5,3) .. (-2.5,1.5);
\end{tikzpicture}
}
\caption{As shown in the figure, for $p\ge1$, one can move a topological operator associated to a $p$-form symmetry around a topological operator associated to another $p$-form symmetry. This changes the ordering of the two topological operators, implying that upon fusion the resulting operators are same $U_{gg'}=U_{g'g}$, and hence $p\ge1$-form symmetry groups have to be abelian.}
\label{abhg}
\end{figure}
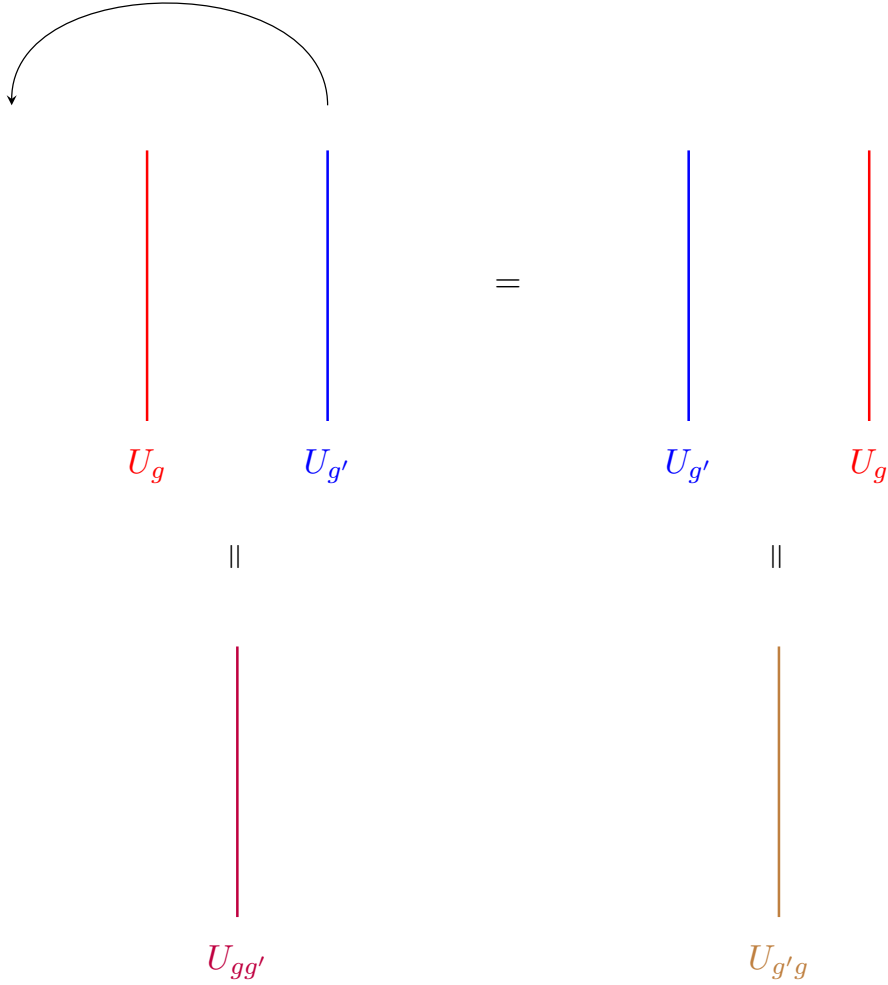

\paragraph{Higher-Form Symmetry Groups are Abelian.}
An important property is that a $p$-form symmetry group $G^{(p)}$ is always abelian for $p\ge1$. This is because we can use topological deformations to change the ordering of two topological operators of codimension greater than one. See figure \ref{abhg}.

\paragraph{Special Case: $U(1)$ $p$-Form Symmetry.}
Generalizing example \ref{U10} discussed above, the special case of 
\be\label{pU1}
G^{(p)}=U(1)
\ee
neatly illustrates all of the above concepts. Such a symmetry is associated to a higher-form generalization of a Noether current, which is a local operator $j^{d-p-1}(x)$ that is a $(d-p-1)$-form on spacetime. The analog of the continuity equation is that this $(d-p-1)$-form is closed
\be\label{cep}
dj^{d-p-1}=0\,.
\ee
Using this we can describe the topological codimension-$(p+1)$ operators constituting the $p$-form symmetry (\ref{pU1}) as
\be
U_g(\Sigma_{d-p-1})=\exp\left(i\alpha\int_{\Sigma_{d-p-1}} j^{d-p-1} \right),\qquad g=e^{i\alpha}\in U(1)\,.
\ee
These operators can be easily shown to be topological by using (\ref{cep}) and following the same argument as presented in (\ref{t0}).

A concrete example realizing $U(1)$ higher-form symmetries is the following.

\begin{example}[Higher-Form Symmetries in the Maxwell Theory]{hfsM}
Consider a $d$-dimensional gauge theory with gauge group $U(1)$ and no matter fields. This is a $d$-dimensional generalization of the usual Maxwell theory, which is obtained by substituting $d=4$.

Let $A$ be the $U(1)$ gauge field, which is a 1-form on spacetime. Its field strength is
\be
F=dA\,,
\ee
which is a 2-form on spacetime. Since the square of the exterior derivative vanishes, this 2-form is closed
\be
dF=0\,,
\ee
which is also known as the Bianchi identity. Using $F$ as a Noether current, we deduce the existence of a $(d-3)$-form symmetry
\be
G^{(d-3)}=U(1)\,,
\ee
for which the topological surface operators are
\be
U^{(m)}_g(\Sigma_{2})=\exp\left(i\alpha\int_{\Sigma_{2}} F \right),\qquad g=e^{i\alpha}\in U(1)\,.
\ee
The superscript in $U^{(m)}_g$ stands for the fact that this $U(1)$ $(d-3)$-form symmetry is known as the \textbf{magnetic higher-form symmetry} of the pure $U(1)$ gauge theory under study.

Moreover, by the equations of motion, we have
\be
d\star F=0\,,
\ee
where $\star F$ is a $(d-2)$-form, the Hodge dual of the field strength $F$. Using $\star F$ as a Noether current, we deduce the existence of a 1-form symmetry
\be
G^{(1)}=U(1)\,,
\ee
for which the topological codimension-2 operators are
\be\label{Ue}
U^{(e)}_g(\Sigma_{d-2})=\exp\left(i\alpha\int_{\Sigma_{d-2}} \star F \right),\qquad g=e^{i\alpha}\in U(1)\,.
\ee
We write the superscript in $U^{(e)}_g$ because this $U(1)$ 1-form symmetry is known as the \textbf{electric 1-form symmetry} of the pure $U(1)$ gauge theory under study. As we will see in example \ref{hfcMw}, this is essentially Gauss's law.
\end{example}

\subsubsection{Action of $p$-Form Symmetries}
\paragraph{Action on Local Operators.}
As we discussed above, a 0-form symmetry can act on a 0-dimensional point-like local operator. This action can be understood as a consequence of moving the codimension-1 topological operator $U_g(\Sigma_{d-1})$ associated to the 0-form symmetry past the local operator.

Generalizing this phenomenon, the first point to note is that a $p\ge 1$-form symmetry does not act on any local operators, as one can simply deform the corresponding codimension-$(p+1)$ topological operator $U_g(\Sigma_{d-p-1})$ past a local operator $\cO(x)$ along a codimension-$p$ submanifold $\Sigma_{d-p}$ that does not intersect the point $x$.

\paragraph{Action on Extended Operators.}
Let us now study whether higher-form symmetries can act on extended operators that are defined on a $q\ge1$-dimensional submanifold $M_q$ of spacetime. In fact, by a version of the above argument, a $p$-form symmetry cannot act on a $q$-dimensional operator if $q<p$, as in that case we can choose $\Sigma_{d-p}$ such that it does not intersect $M_q$. Thus, a $p$-form symmetry can only act on operators of dimension
\be
q\ge p.
\ee
The simplest case to study is $q=p$, where we have the following statement:

\begin{state}[Charges of $p$-Dimensional Operators Under $p$-Form Symmetry]{qhfs}
$p$-dimensional operators transform in representations of the $p$-form symmetry group $G^{(p)}$.
\end{state}

\begin{figure}
\centering
\scalebox{1.1}{
\begin{tikzpicture}
\draw [thick](-1,2) -- (-1,-1);
\draw [thick,red](-2.5,-1) -- (-1.125,0.625) node (v1) {};
\draw [thick,red](-0.875,0.875) node (v2) {} -- (0,2);
\node[red] at (0,2.5) {$U_g$};
\node at (-1,-1.5) {$\cO(M_p)$};
\node at (1.25,0.5) {=};
\draw [thick,red](3,-1) -- (5.5,2);
\draw [thick](3.5,2) -- (3.5,-0.125);
\draw [thick](3.5,-0.625) -- (3.5,-1);
\draw [red,fill=red] (3.5,0.5) ellipse (0.05 and 0.05);
\node at (3.5,-1.5) {$\cO(M_p)$};
\node[red] at (5.5,2.5) {$U_g$};
\node[red] at (2.75,0.5) {$\cO(x)$};
\node at (6.75,0.5) {=};
\begin{scope}[shift={(6,0)}]
\draw [thick,red](3,-1) -- (5.5,2);
\draw [thick](3.5,2) -- (3.5,-0.125);
\draw [thick](3.5,-0.625) -- (3.5,-1);
\node at (3.5,-1.5) {$\cO(M_p)$};
\node[red] at (5.5,2.5) {$U_g$};
\end{scope}
\node at (8.1,0.5) {$\phi(g)~\times$};
\end{tikzpicture}
}
\caption{Moving a $p$-form symmetry topological operator $U_g$ across a $p$-dimensional operator $\cO(M_p)$ leaves behind a topological local operator $\cO(x)$ living on $\cO(M_p)$ which evaluates to a number $\phi(g)\in\bC^\times$.}
\label{pfsa}
\end{figure}
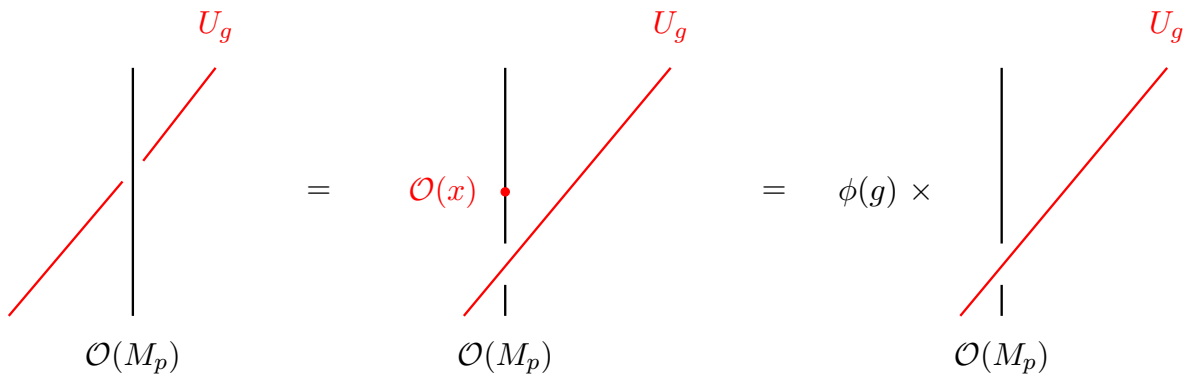

Let us discuss this action in more detail. Consider a $p\ge1$-dimensional extended operator $\cO(M_p)$ placed along a $p$-dimensional submanifold $M_p$ of spacetime. We can assume that $\cO(M_p)$ is an \textit{irreducible} $p$-dimensional operator, i.e.\ there are no topological local operators that can be inserted at a point $x\in M_p$ except multiples of the identity local operator\footnote{This means that the $p$-dimensional operator $\cO(M_p)$ cannot be expressed as a sum of two other $p$-dimensional operators. Such operators are also known as \textit{simple} $p$-dimensional operators.}. Now, note that deforming $U_g(\Sigma_{d-p-1})$ across $\cO(M_p)$ leaves behind a topological local operator $\cO(x)$ at the intersection point $x$ of $M_p$ and $\Sigma_{d-p}$;
\be
U_g(\Sigma_{d-p-1})\cO(M_p)=\cO(x)\cO(M_p)U_g(\Sigma'_{d-p-1})\,.
\ee
See figure \ref{pfsa}. Since the only possible topological local operators along $M_p$ are multiples of identity, we can replace $\cO(x)$ by a non-zero number in the above equation
\be\label{pa}
U_g(\Sigma_{d-p-1})\cO(M_p)=\phi(g)\times\cO(M_p)U_g(\Sigma'_{d-p-1}),\qquad \phi(g)\in\bC^\times = \bC-\{0\}\,.
\ee
Because of the fusion rule (\ref{fr}), the numbers $\phi(g)$ have to satisfy
\be
\phi(g)\phi(g')=\phi(gg')\,,
\ee
i.e.\ the numbers $\phi(g)$ furnish a one-dimensional representation of the $p$-form symmetry group $G^{(p)}$, and in particular the numbers $\phi(g)$ must be phase factors
\be
\phi(g)\in U(1)\subset\bC^\times\,.
\ee
Since $G^{(p)}$ is an abelian group, its one-dimensional representations are precisely its irreducible representations.

Thus, we learn that an irreducible $p$-dimensional operator $\cO(M_p)$ transforms in an irreducible representation of the $p$-form symmetry group $G^{(p)}$, thus justifying the statement \ref{qhfs}.

\begin{tech}[Remark on Higher-Charges]{}
Here, following \cite{Gaiotto:2014kfa}, we have only discussed the action of $p$-form symmetries on operators of dimension $q=p$. More generally, $p$-form symmetries also act on operators of dimension $q>p$. Such actions were studied recently in the literature in \cite{Bhardwaj:2023wzd,Bartsch:2023pzl}. Following the terminology introduced there, we say that $q$-dimensional operators transform in \textbf{$\bm q$-charges} under the $p$-form symmetry group $G^{(p)}$.

\vspace{3mm}

\ni This is true also for 0-form symmetries, in which case we have the statement
\be
\text{$q$-charges of $G^{(0)}$} = \text{$(q+1)$-representations of the group $G^{(0)}$},
\ee
where $(q+1)$-representations are categorical generalizations of the familiar ordinary representations of the group $G^{(0)}$. In this terminology, 1-representations are the ordinary representations of $G^{(0)}$ and we recover the well-known fact that local operators transform in ordinary representations of the 0-form symmetry group $G^{(0)}$. A description of higher-representations ($q\ge1$) is outside the scope of these notes, but we refer an interested reader to the references \cite{Bhardwaj:2022lsg,Bartsch:2022mpm,Bhardwaj:2022kot,Bartsch:2022ytj,Bhardwaj:2023wzd,Bartsch:2023pzl} for an introduction to higher-representations at a physics level of rigor.

\vspace{3mm}

\ni Similarly, for $p\ge1$-form symmetries, we have the statement
\be
\text{$q$-charges of $G^{(p)}$} = \text{$(q+1)$-representations of the $(p+1)$-group $\bG^{(p+1)}_{G^{(p)}}$},
\ee
where the above $(p+1)$-group is uniquely determined by the group $G^{(p)}$. Let us note that higher-groups are categorical generalizations of ordinary groups. We refer the reader to \cite{Bhardwaj:2023wzd,Bartsch:2023pzl} for more details (see also \cite{Bhardwaj:2023ayw,Bartsch:2023wvv} for generalization of higher-charges to non-invertible symmetries).
\end{tech}

\paragraph{Action Via Linking.}
An equivalent way of describing the action (\ref{pa}) is
\be\label{pa2}
U_g\left(S^{d-p-1}\right)\cO(M_p)=\phi_g\times\cO(M_p)\,,
\ee
where on the LHS we insert the operator $U$ along a small $(d-p-1)$-dimensional sphere $S^{d-p-1}$ linking a point $x\in M_p$ as shown in the figure \ref{pfsal}. 

\begin{figure}
\centering
\scalebox{1.1}{
\begin{tikzpicture}
        \draw[blue, thick] circle (1 and 0.5);
        \draw[line width=1.3mm, white] (-0.1,0.5) -- (0.1,0.5);
        \draw[thick] (0,2) -- (0,-0.4);
        \draw[thick] (0,-0.6) -- (0, -2);
        \draw[thick] (4.5,2) -- (4.5,-2);
        \node[] (n4) at (2,0) {$=$};
\node[blue] at (-2.25,0) {$U_g(S^{d-p-1})$};
\node at (0,-2.5) {$\cO(M_p)$};
\node at (4.5,-2.5) {$\cO(M_p)$};
\node at (3.25,0) {$\phi(g)~\times$};
\end{tikzpicture}
}
\caption{The linking action of a $p\ge1$-form symmetry. The symmetry generator $U_g$ inserted on $S^{d-p-1}$ acts on the extended $p$-dimensional operator $\cO$ inserted on $M_p$.}
\label{pfsal}
\end{figure}

\paragraph{Charges and Pontryagin Dual Group.}
Irreducible representations of an abelian group $G^{(p)}$ are equivalently homomorphisms
\be
\phi:~G^{(p)}\to U(1)
\ee
Such homomorphisms are often referred to as \textbf{characters}, and themselves form an abelian group, with the product operation being
\be
(\phi\phi')(g)=\phi(g)\times\phi'(g)\in U(1)\,.
\ee
The resulting group is denoted
\be
\wh G^{(p)}=\{\text{Group formed by homomorphisms $G^{(p)}\to U(1)$}\}\,,
\ee
and is referred to as the \textbf{Pontryagin dual group} of $G^{(p)}$.

Phrased in this language, the charge carried by a (irreducible) $p$-dimensional operator under a $p$-form symmetry group $G^{(p)}$ is an element of the Pontryagin dual group $\wh G^{(p)}$. Let us discuss some examples:
\bit
\item For $G^{(p)}=U(1)$, we have $\wh G^{(p)}=\Z$, namely the group formed by integers under addition. 
If we represent the elements of $G^{(p)}$ as
\be
g=e^{i\alpha}\,,\qquad\alpha\in[0,2\pi)
\ee
the possible homomorphisms are
\be
\phi(g)=g^q=e^{iq\alpha}\in U(1),\qquad q\in\Z\,.
\ee
\item For $G^{(p)}=\Z_N$, we have $\wh G^{(p)}=\Z_N$, namely the group formed by integers modulo $N$ under addition. If we represent the elements of $G^{(p)}$ as
\be
g=e^{\frac{2\pi i \alpha}N},\qquad \alpha\in\{0,1,\cdots N-1\}\,,
\ee
the possible homomorphisms are
\be
\phi(g)=g^q=e^{\frac{2\pi i q\alpha}N},\qquad q\in\{0,1,\cdots N-1\}\,.
\ee
\item Generalizing the previous example, if $\G p$ is a \textit{finite} abelian group, then we have
\be
\whG p\cong\G p
\ee
This can actually be derived as a consequence of the previous example. For a finite abelian group, we have
\be
\G p\cong\prod_{i=1}^n\Z_{N_i},\qquad N_i\in\mathbb{N}
\ee
and consequently
\be
\whG p\cong\prod_{i=1}^n\wh\Z_{N_i}\cong\prod_{i=1}^n\Z_{N_i}
\ee
\eit

\begin{tech}[Double Pontryagin Duality]{}
An important property of Pontryagin duals that we will use later is that taking the Pontryagin dual twice is equivalent to not taking the Pontryagin dual at all. More precisely, there exists a canonical isomorphism
\be
\wh{\wh G^{(p)}}\cong G^{(p)}
\ee
for any group $G^{(p)}$. Indeed, an element $g\in G^{(p)}$ defines a homomorphism
\be
h_g:~\wh G^{(p)}\to U(1)
\ee
taking the form
\be
h_g(\phi)=\phi(g)\in U(1),\qquad \phi\in\wh G^{(p)}\,.
\ee
\end{tech}

\paragraph{Special Case: $U(1)$ $p$-Form Symmetry.}
Generalizing example \ref{U10}, the continuity equation for the Noether current of a $U(1)$ $p$-form symmetry is modified in the presence of a $p$-dimensional operator of $\cO(M_p)$ of charge $q\in\Z$;
\be\label{Nm}
\cO(M_p)dj^{d-p-1}=q\delta^{d-p}(M_p)\cO(M_p)\,,
\ee
where $\delta^{d-p}(M_p)$ is the $(d-p)$-form associated to delta function on $M_p$.

Using this fact and following steps similar to those in derivation (\ref{da}), we can compute the linking action (\ref{pa2}) of $p$-form symmetry with $\cO(M_p)$ to be
\be
U_g\left(S^{d-p-1}\right)\cO(M_p)=\exp\left(iq\alpha\right)\cO(M_p)\,.
\ee

\begin{example}[Higher-Form Charges in the Maxwell Theory]{hfcMw}
Let us now return to the pure $U(1)$ gauge theory we started studying in example \ref{hfsM}. As we discussed there, the theory has two $U(1)$ higher-form symmetries, namely the electric 1-form symmetry and the magnetic $(d-3)$-form symmetry.

The operators charged under the electric 1-form symmetry are \textbf{Wilson line operators} $\cW_q$, which when inserted along a line $L$ in spacetime can be expressed as
\be
\cW_q(L)=\exp\left(2\pi iq\int_L A\right)\,.
\ee
Different Wilson lines are distinguished by a parameter $q\in \Z$ often referred to as the ``electric charge'' of the Wilson line. The operator $\cW_q(L)$ can be understood as describing an infinitely heavy probe particle of electric charge $q$ under the $U(1)$ gauge group, whose worldline is $L$.

In the presence of such a Wilson line, the equations of motion are modified as
\be
(d\star F)\cW_q(L)=q\delta^{d-1}(L)\cW_q(L)\,,
\ee
which when compared with (\ref{Nm}) implies that the Wilson line $\cW_q(L)$ has charge $q$ under the $U(1)$ electric 1-form symmetry. In other words, the electric charge of the Wilson line is the same as its 1-form charge. Thus, the integral over a $(d-2)$ manifold of $\star F$ linking a Wilson line measures the electric charge of that Wilson line: this can be seen as a statement of Gauss's law.

The operators charged under the magnetic $(d-3)$-form symmetry are \textbf{'t Hooft operators}, which are codimension-3 operators in spacetime. These operators are also referred to as \textbf{monopole operators}, especially in $d=3$. The different 't Hooft operators $\cH_q(M_{d-3})$ are distinguished by a parameter $q\in \Z$, often referred to as the ``magnetic charge'' of the 't Hooft operator. The operator $\cH_q(M_{d-3})$ can be understood as describing an infinitely heavy probe magnetic monopole of magnetic charge $q$ under the $U(1)$ gauge group, whose worldvolume is $M_{d-3}$.

The 't Hooft operator $\cH_q(M_{d-3})$ is defined by excising $M_{d-3}$ from the spacetime and requiring the gauge fields in the path integral to have a boundary condition near $M_{d-3}$ such that we have
\be\label{U1th}
\int_{S^2} F=q\,,
\ee
on a small sphere $S^2$ in spacetime which links with $M_{d-3}$. The Bianchi identity is modified in the presence of $\cH_q(M_{d-3})$;
\be
(dF)\cH_q(M_{d-3})=q\delta^{3}(M_{d-3})\cH_q(M_{d-3})\,,
\ee
which when compared with (\ref{Nm}) implies that the 't Hooft operator $\cH_q(M_{d-3})$ has charge $q$ under the $U(1)$ magnetic $(d-3)$-form symmetry. In other words, the ``magnetic charge'' of the 't Hooft operator is the same as its $(d-3)$-form charge.
\end{example}

\subsection{Discrete (Higher-Form) Gauge Theories}\label{hfgt}
In the previous subsection, we discussed electric and magnetic higher-form symmetries arising in the Maxwell theory. Both of these higher-form symmetries are continuous. In this subsection, we discuss a class of theories exhibiting discrete higher-form symmetries.

The class of theories we study are discrete higher-form gauge theories, which are examples of topological quantum field theories (TQFTs). A discrete higher-form gauge theory in $d$ spacetime dimensions has an action
\be\label{dgS}
2\pi \int a_p\cup_\eta\delta b_{d-p-1}\,,
\ee
where
\bit
\item $a_p$ is a discrete version of a $p$-form field on spacetime, 
\item $b_{d-p-1}$ is a discrete version of a $(d-p-1)$-form field on spacetime,
\item $\delta$ is a discrete version of exterior derivative, and
\item $\cup_\eta$ is discrete version of the wedge product, which is known as the \textit{cup product}.
\eit
For $p=1$, we have a discrete analogue $a_1$ of the usual gauge connection, and hence the corresponding discrete higher-form gauge theories are simply referred to as \textbf{discrete gauge theories}.

\paragraph{Cochain Description of Discrete Gauge Fields.}
To describe these discrete gauge fields concretely, one uses the language of cochains defined with respect to a triangulation on spacetime. In this language, the bullet points above become:
\bit
\item $a_p$ is a $G^{(p)}$-valued $p$-cochain, which means that $a_p$ assigns an element of the abelian group $G^{(p)}$ to every $p$-simplex in the triangulation. Note that for $p=0$, taking $\G p=\G0$ to be an abelian group is an assumption we are making in this subsection. Discrete gauge theories for finite non-abelian groups do not in general admit a description of the form (\ref{dgS}).
\item $b_{d-p-1}$ is a $\wh G^{(p)}$-valued $(d-p-1)$-cochain, which means that $b_{d-p-1}$ assigns an element of the abelian group $\wh G^{(p)}$ to every $(d-p-1)$-simplex in the triangulation.
\item $\delta$ is a differential that maps $q$-cochains to $(q+1)$-cochains valued in the same abelian group
\be
\delta c_q(v_0,v_1,\cdots,v_{q+1})=\sum_{i=0}^{q+1}(-1)^i c_q(v_0,v_1,\cdots,v_{i-1},v_{i+1},v_{i+2},\cdots, v_{q+1})
\ee
where on the LHS we have $\delta c_q$ evaluated on a $(q+1)$-simplex described by vertices $v_0,v_1,\cdots,v_{q+1}$, and on the RHS we have $c_q$ evaluated on a sub-$q$-simplices of the $(q+1)$-simplex $(v_0,v_1,\cdots,v_{q+1})$. Each such sub-$q$-simplex is obtained by deleting a vertex $v_i$ from vertices $v_0,v_1,\cdots,v_{q+1}$.
\item The cup product $\cup_\eta$ in general takes a $G$ valued $q$-cochain $c_q$ and an $H$ valued $r$-cochain $d_r$, along with a bi-homomorphism
\be
\eta:~G\times H\to K
\ee
to produce a $K$ valued $(q+r)$-cochain $c_q\cup_\eta d_r$, whose evaluation on a $(q+r)$-simplex is
\be
c_q\cup_\eta d_r(v_0,v_1,\cdots,v_{q+r})=\eta\Big(c_q(v_0,v_1,\cdots,v_q)\in G,~d_r(v_q,v_{q+1},\cdots,v_{q+r})\in H\Big)\in K\,.
\ee
The cup product used in the action (\ref{dgS}) uses the natural bi-homomorphism
\be
\eta:~G^{(p)}\times \wh G^{(p)}\to \R/\Z\,,
\ee
taking the form
\be
\eta(g,\wh g)=\alpha\in\R/\Z,\qquad g\in G^{(p)},~\wh g\in\wh G^{(p)}\,,
\ee
where $\alpha$ is defined via
\be
\wh g(g)=e^{2\pi i\alpha}\in U(1)\,,
\ee
using the fact that $\wh g$ is itself a homomorphism
\be
\wh g:~G^{(p)}\to U(1)\,.
\ee
\eit

\begin{tech}[Standard Cup Product]{}
For
\be
G=H=K=\Z_n\,,
\ee
one often uses a \textbf{standard cup product}, denoted simply as $\cup$, which takes the form
\be
c_q\cup d_r(v_0,v_1,\cdots,v_{q+r})=c_q(v_0,v_1,\cdots,v_q)d_r(v_q,v_{q+1},\cdots,v_{q+r})\,,
\ee
where on the RHS we have used the multiplication on
\be
\Z_n=\Z/n\Z\,,
\ee
descending from the usual multiplication on $\Z$. 
\end{tech}

\paragraph{Gauge Transformations.}
The discrete gauge field $a_p$ is a gauge field with gauge transformations acting as
\be
a_p\to a_p+\delta a_{p-1}\,,
\ee
where $a_{p-1}$ is an arbitrary $G^{(p)}$-valued $(p-1)$-cochain.

Similarly, the discrete gauge field $b_{d-p-1}$ is a gauge field with gauge transformations acting as
\be
b_{d-p-1}\to b_{d-p-1}+\delta b_{d-p-2}\,,
\ee
where $b_{d-p-2}$ is an arbitrary $\wh G^{(p)}$ valued $(d-p-2)$-cochain.

\paragraph{Equations of Motion.}
The equations of motion are straightforward to compute from the action (\ref{dgS});
\be
\delta a_p=0,\qquad \delta b_{d-p-1}=0\,,
\ee
meaning that $a_p$ is a $p$-cocycle and $b_{d-p-1}$ is a $(d-p-1)$-cocycle.

\paragraph{Higher-Form Symmetries.}
The above equations of motion allow us to treat $a_p$ and $b_{d-p-1}$ as discrete versions of higher-form Noether currents. That is, we can construct topological operators by integrating $a_p$ and $b_{d-p-1}$. For example, using the $G^{(p)}$ valued $p$-cocycle $a_p$, we can construct topological $p$-dimensional operators labeled by elements of $\wh G^{(p)}$
\be\label{Uhg}
U_{\wh g}(M_p)=\int_{M_p}\wh g(a_p),\qquad \wh g\in\wh G^{(p)}\,,
\ee
where $\wh g(a_p)$ is a $U(1)$ valued $p$-cocycle whose evaluation on a $p$-simplex is
\be
\big(\wh g(a_p)\big)(v_0,v_1,\cdots,v_p)=\wh g\big(a_p(v_0,v_1,\cdots,v_p)\in G^{(p)}\big)\in U(1)\,.
\ee
The above integral is then expressed as
\be
\int_{M_p}\wh g(a_p)=\prod_{S_p\in M_p} \big(\wh g(a_p)\big)(S_p)\,,
\ee
where $S_p$ are simplices in the triangulation of the submanifold $M_p$ and $\big(\wh g(a_p)\big)(S_p)$ is the evaluation of $U(1)$ valued $p$-cochain $\wh g(a_p)$ on $S_p$. The operators $U_{\wh g}$ can be seen to be topological by using a discrete version of Stokes' theorem and the equation of motion $\delta a_p=0$. These operators $U_{\wh g}$ are sometimes referred to as \textbf{Wilson operators for the gauge field $\bm{a_p}$}.

Similarly, using the $\wh G^{(p)}$ valued $(d-p-1)$-cocycle $b_{d-p-1}$, we can construct topological $(d-p-1)$-dimensional operators labeled by elements of $G^{(p)}$
\be\label{Ug}
U_{g}(M_{d-p-1})=\int_{M_{d-p-1}}b_{d-p-1}(g),\qquad g\in G^{(p)}\,,
\ee
where $b_{d-p-1}(g)$ is a $U(1)$ valued $(d-p-1)$-cocycle whose evaluation on a $(d-p-1)$-simplex is
\be
\big(b_{d-p-1}(g)\big)(v_0,v_1,\cdots,v_{d-p-1})=\left(b_{d-p-1}(v_0,v_1,\cdots,v_{d-p-1})\in\wh G^{(p)}\right)(g)\in U(1)\,.
\ee
These operators are topological as a consequence of the equation of motion $\delta b_{d-p-1}=0$. These operators $U_{g}$ are sometimes referred to as \textbf{Wilson operators for the gauge field $\bm{b_{d-p-1}}$}.

In total, the discrete higher-form gauge theory (\ref{dgS}) has two types of higher-form symmetries:
\bit
\item A $p$-form symmetry group,
\be
G^{(p)}\,,
\ee
formed by codimension-$(p+1)$ topological operators (\ref{Ug}).
\item A $(d-p-1)$-form symmetry group,
\be
G^{(d-p-1)}=\wh G^{(p)}\,,
\ee
formed by $p$-dimensional topological operators (\ref{Uhg}).
\eit
The two types of higher-form symmetries are sometimes referred to as electric and magnetic symmetries to distinguish them.

\paragraph{Charged Operators.}
A $p$-dimensional topological operator $U_{\wh g}$ carries a charge
\be
Q(U_{\wh g})=\wh g\in\wh G^{(p)}
\ee
under the $G^{(p)}$ $p$-form symmetry generated by topological operators $U_g$. This can be seen as a consequence of the fact that the equation of motion for $b_{d-p-1}$ is modified in the presence of $U_{\wh g}(M_p)$ as
\be\label{dgco}
U_{\wh g}(M_p)\delta b_{d-p-1}=\delta_{\wh g}^{d-p}(M_p)U_{\wh g}(M_p)\,,
\ee
where $\delta_{\wh g}^{d-p}(M_p)$ is a discrete analog of delta function localized on $M_p$. It is a $\wh G^{(p)}$-valued $(d-p)$-cochain having the property that
\be
\int\delta_{\wh g}^{d-p}(M_{p})\cup_\eta\Lambda_{p}=\int_{M_{p}}\wh g(\Lambda_{p})\in\R/\Z
\ee
for any $\G p$-valued $p$-cochain $\Lambda_p$.

Similarly, a $(d-p-1)$-dimensional topological operator $U_{g}$ carries a charge
\be
Q(U_{g})=g\in\wh G^{(d-p-1)}=G^{(p)}
\ee
under the $G^{(d-p-1)}=\wh G^{(p)}$ $(d-p-1)$-form symmetry generated by topological operators $U_{\wh g}$. This can be seen as a consequence of the fact that the equation of motion for $a_{p}$ is modified in the presence of $U_{g}(M_{d-p-1})$ as
\be
U_{g}(M_{d-p-1})\delta a_{p}=\delta^{p+1}_g(M_{d-p-1})U_{g}(M_{d-p-1})\,,
\ee
where $\delta^{p+1}_g(M_{d-p-1})$ is a $G^{(p)}$ valued $(p+1)$-cochain with the property
\be
\int\Lambda_{d-p-1}\cup_\eta\delta^{p+1}_g(M_{d-p-1})=\int_{M_{d-p-1}}\Lambda_{d-p-1}(g)\in\R/\Z\,,
\ee
for any $\whG p$-valued $(d-p-1)$-cochain $\Lambda_{d-p-1}$.

\section{Higher-Form Symmetries of Gauge Theories}\label{hfg}
\subsection{Screening of Extended Operators}\label{screen}
In this subsection, we discuss a powerful and simple method of understanding higher-form symmetries, which we will use in the following subsections to compute higher-form symmetries of gauge theories. This method goes under the name of screening and has been successfully used recently in various strongly coupled setups to deduce higher-form symmetries \cite{Morrison:2020ool,Albertini:2020mdx,Bhardwaj:2020phs,Closset:2020scj,DelZotto:2020esg,Bhardwaj:2021pfz,Bhardwaj:2021zrt,Bhardwaj:2021mzl,Closset:2021lwy,DelZotto:2022fnw,Cvetic:2022imb,Hubner:2022kxr,Bhardwaj:2022dyt,Bhardwaj:2023zix,Hosseini:2021ged}.

We begin with the definition of screening:

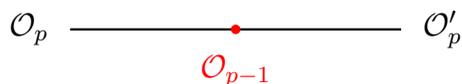
\begin{figure}
\centering
\scalebox{1.1}{
\begin{tikzpicture}
\draw [thick](-2,0) -- (0,0);
\draw [thick](0,0) -- (2,0);
\draw [red,fill=red] (0,0) ellipse (0.05 and 0.05);
\node at (-2.5,0) {$\cO_p$};
\node []at (2.5,0) {$\cO'_p$};
\node[red] at (0,-0.5) {$\cO_{p-1}$};
\end{tikzpicture}
}
\caption{The existence of above configuration implies that $\cO_p$ and $\cO'_p$ are related by screening.}
\label{scree}
\end{figure}

\begin{note}[Screening]{}
We say that a $p\ge1$-dimensional operator $\cO_p$ can be \textbf{screened} to another $p$-dimensional operator $\cO'_p$ if we can insert a $(p-1)$-dimensional operator $\cO_{p-1}$ between $\cO_p$ and $\cO'_p$. See figure \ref{scree}.

\vspace{3mm}

\ni If an operator $\cO_p$ can be screened to the identity $p$-dimensional operator, i.e.\ if there exists an operator $\cO_{p-1}$ that can be used to end $\cO_p$, then we simply say that $\cO_p$ can be \textbf{completely screened}.
\end{note}

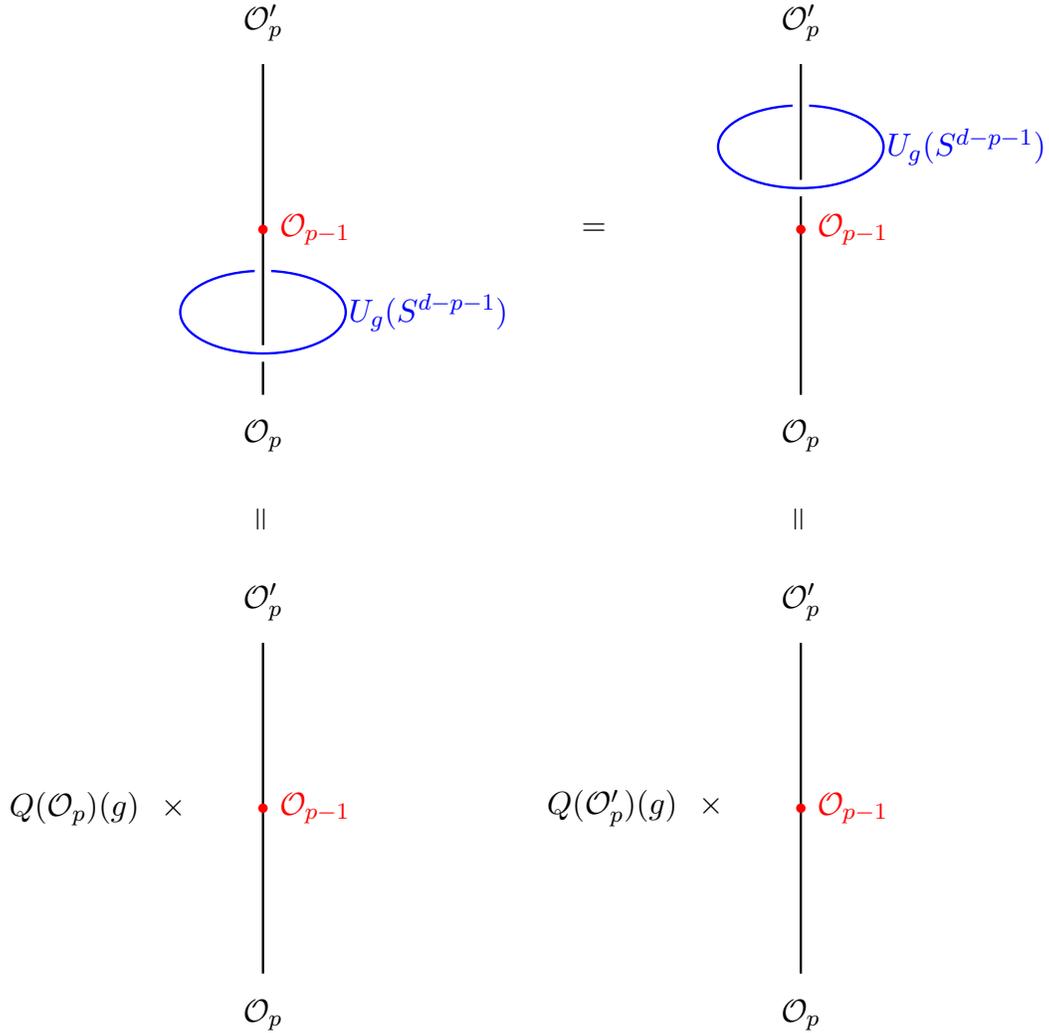
\begin{figure}
\centering
\scalebox{1.1}{
\begin{tikzpicture}
\begin{scope}[rotate=90]
\begin{scope}[shift={(-1,0)}]
\draw[blue, thick,rotate=90] circle (1 and 0.5);
\begin{scope}[rotate=90, shift={(0,-1)}]
\draw[line width=1.3mm, white] (-0.1,0.5) -- (0.1,0.5);
\end{scope}
\end{scope}
\draw [thick](-2,0) -- (-1.6,0) (-1.4,0) -- (0,0);
\draw [thick](0,0) -- (2,0);
\draw [red,fill=red] (0,0) ellipse (0.05 and 0.05);
\node at (-2.5,0) {$\cO_p$};
\node []at (2.5,0) {$\cO'_p$};
\node[red] at (0,-0.5) {~~$\cO_{p-1}$};
\node[blue] at (-1,-2) {$U_g(S^{d-p-1})$};
\end{scope}
\node at (4,0) {=};
\begin{scope}[shift={(6.5,0)},rotate=90]
\draw [thick](-2,0) -- (-1.6,0) (-1.6,0) -- (0.4,0);
\begin{scope}[shift={(1,0)}]
\draw[blue, thick,rotate=90] circle (1 and 0.5);
\begin{scope}[rotate=90, shift={(0,-1)}]
\draw[line width=1.3mm, white] (-0.1,0.5) -- (0.1,0.5);
\end{scope}
\end{scope}
\draw [thick](0.6,0) -- (2,0);
\draw [red,fill=red] (0,0) ellipse (0.05 and 0.05);
\node at (-2.5,0) {$\cO_p$};
\node [] at (2.5,0) {$\cO'_p$};
\node[red] at (0,-0.5) {~~$\cO_{p-1}$};
\node[blue] at (1,-2) {$U_g(S^{d-p-1})$};
\end{scope}
\node[rotate=90] at (0,-3.5) {=};
\begin{scope}[shift={(0,-7)},rotate=90]
\draw [thick](-2,0) -- (0,0);
\draw [thick](0,0) -- (2,0);
\draw [red,fill=red] (0,0) ellipse (0.05 and 0.05);
\node at (-2.5,0) {$\cO_p$};
\node [] at (2.5,0) {$\cO'_p$};
\node[red] at (0,-0.5) {~~$\cO_{p-1}$};
\end{scope}
\node at (4.5,-7) {$Q(\cO'_p)(g)~~\times$};
\node [rotate=90] at (6.5,-3.5) {=};
\begin{scope}[shift={(6.5,-7)},rotate=90]
\draw [thick](-2,0) -- (0,0);
\draw [thick](0,0) -- (2,0);
\draw [red,fill=red] (0,0) ellipse (0.05 and 0.05);
\node at (-2.5,0) {$\cO_p$};
\node [] at (2.5,0) {$\cO'_p$};
\node[red] at (0,-0.5) {~~$\cO_{p-1}$};
\end{scope}
\node at (-2,-7) {$Q(\cO_p)(g)~~\times$};
\end{tikzpicture}
}
\caption{On the top we have two configurations related simply by topological deformation of the topological $p$-form symmetry operator $U_g$. On the bottom we have contracted $U_g$ on $\cO_p$ and $\cO'_p$ respectively leading to phases $Q(\cO_p)(g)$ and $Q(\cO'_p)(g)$, where $Q(\cO_p),Q(\cO'_p)\in\whG p$. Since $g\in\G p$ is an arbitrary choice, this implies (\ref{Qeq}).}
\label{screeQ}
\end{figure}

\paragraph{Screening Implies Same $p$-Form Charges.}
The above definition is relevant because if $\cO_p$ can be screened to $\cO'_p$, then the two operators must have the same charges under a $p$-form symmetry group $G^{(p)}$
\be\label{Qeq}
Q(\cO_p)=Q(\cO'_p)\in\wh G^{(p)}\,.
\ee
This is justified in figure \ref{screeQ}. If an operator $\cO_p$ can be completely screened, then it is uncharged under $G^{(p)}$
\be
Q(\cO_p)=1\in\wh G^{(p)}\,.
\ee

\paragraph{$p$-Form Symmetries from Absence of Screening.}
Conversely, if $\cO_p$ and $\cO'_p$ are not related by screenings, then we can define a $p$-form symmetry characterized by the property that $\cO_p$ and $\cO'_p$ carry different charges under this $p$-form symmetry. 

More concretely, one can employ the following strategy. Define an equivalence relation on the set of $p$-dimensional operators
\be
\cO_p\sim\cO'_p
\ee
if $\cO_p$ and $\cO'_p$ are related by screenings. Let 
\be\label{ecd}
\bD_p
\ee
be the set of equivalence classes of $p$-dimensional operators obtained after imposing the above equivalence relation. This set acquires a product structure from the OPE of $p$-dimensional operators, converting it into a ring. In many cases of interest, the product structure is more restricted and $\bD_p$ is actually an abelian group. We assume this is the case in what follows.

From the above discussion, we must have $p$-form symmetries whose charges distinguish different elements of $\bD_p$. Such $p$-form symmetries form the group
\be
G^{(p)}=\wh\bD_p\,.
\ee
The possible charges under this $p$-form symmetry group are elements of
\be
\wh G^{(p)}=\bD_p\,.
\ee
That is, the charge of a $p$-dimensional operator $\cO_p$ under $G^{(p)}$ is
\be
Q(\cO_p)=[\cO_p]\in\bD_p\,,
\ee
where $[\cO_p]$ is the equivalence class in which $\cO_p$ lies.

\begin{example}[Revisiting the Maxwell Theory]{}
We can deduce the existence of electric and magnetic higher-form symmetries of a pure $U(1)$ gauge theory using the screening arguments discussed above. When there is no electric or magnetic matter content, there are no screenings relating two different Wilson operators or two different 't Hooft operators. Thus the equivalence classes of line and codimension-3 operators are
\be
\bD_1=\Z,\qquad \bD_{d-3}=\Z\,.
\ee
Taking Pontryagin duals, we recover the electric 1-form symmetry and magnetic $(d-3)$-form symmetry groups
\be
G^{(1)}=U(1),\qquad G^{(d-3)}=U(1)
\ee
that we discussed from a different point of view earlier.
\end{example}

\subsection{Abelian Gauge Theories}\label{AbG}
In this subsection, we use arguments based on screening to determine the higher-form symmetry groups that arise in gauge theories with abelian gauge groups.

\subsubsection{Maxwell Theory with Single Matter Field}
\paragraph{Setup.}
Consider a $d$-dimensional gauge theory with $U(1)$ gauge group along with a matter field $\phi$ with charge $q\in\Z$ under the $U(1)$ gauge group\footnote{The unit charge under the $U(1)$ gauge group is set by the smallest positive charge carried by the Wilson line operators that will be used to probe the theory.}. Note that $\phi$ may have any transformation properties (scalar, spinor, etc.) under the spacetime Lorentz group, and it may be a massive or a massless field.

We saw earlier that for pure Maxwell theory without the matter field $\phi$ we have a $U(1)$ 1-form symmetry. We will now see that the introduction of $\phi$ breaks this $U(1)$ 1-form symmetry to its $\Z_q$ subgroup. That is, this theory has electric 1-form symmetry group
\be
G^{(1)}=\Z_q\,.
\ee

\paragraph{Constructing Non-Genuine Local Operators Using Matter Field.}
The key point is that the matter field gives rise to a non-genuine local operator $\phi(x)$ that screens a charge $q$ Wilson line $\cW_q$ as shown in figure \ref{Wqend}.

\begin{figure}
\centering
\scalebox{1.1}{
\begin{tikzpicture}
\draw [thick](-2,0) -- (0,0);
\draw [red,fill=red] (0,0) ellipse (0.05 and 0.05);
\node at (-2.5,0) {$\cW_q$};
\node[red] at (0.5,0) {~~$\phi(x)$};
\end{tikzpicture}
}
\caption{A matter field $\phi$ of charge $q$ gives rise to a non-genuine local operator living at the end of a Wilson line of charge $q$.}
\label{Wqend}
\end{figure}

\begin{note}[Non-Genuine Operators]{}
A non-genuine $q$-dimensional operator is an operator that is attached to a collection of $p>q$-dimensional operators.
\end{note}

An insertion of the field $\phi(x)$ cannot define a genuine local operator, because such an operator will not be gauge invariant. Under a gauge transformation
\be
A(x) \to A(x) - \frac{d \theta(x)}{2\pi}\,,
\ee
the insertion $\phi(x)$ transforms as
\be
\phi(x) \to e^{i q \theta(x)}\phi(x) \,.
\ee
To make this insertion gauge invariant we can attach it to a charge $q$ Wilson line operator $\cW_q(L)$ inserted along a line $L$ that ends at the point $x$
\be
\partial L=x\,.
\ee
This is because the gauge variation of $\cW_q(L)$ is
\be
\cW_q(L)=\exp\left(2\pi i q\int_L A\right)\to \exp\left(-i q\int_{\partial L} d\theta\right)\cW_q(L)= e^{-i q\theta(x)}\cW_q(L)\,.
\ee
Thus combining the two together we obtain a gauge-invariant configuration
\be\label{LLC}
\phi(x)\cW_q(L)\,.
\ee
In other words we have constructed a non-genuine local operator $\phi$ that lives at the end of charge $q$ Wilson line $\cW_q$.

\paragraph{Screenings.}
Similarly, by inserting powers $\phi^n$ of the field $\phi$, we can construct non-genuine local operators relating Wilson lines
\be
\cW_p\sim\cW_{p+nq},\qquad \forall~p\in\Z\,.
\ee
This imposes screenings between Wilson line operators, and the group of equivalence classes of unscreened line operators defined around (\ref{ecd}) can be expressed as
\be\label{WLe}
\bD_1=\frac{\Z}{q\Z}\equiv \Z_q\,,
\ee
where the $\Z$ factor in the numerator is the group of Wilson lines before screening, and the factor $q\Z$ in the denominator is the group of completely screened Wilson lines.

\paragraph{Electric 1-Form Symmetry.}
According to the general discussion of section \ref{screen}, the Pontryagin dual of the group (\ref{WLe}) provides a 1-form symmetry group of the theory
\be
G^{(1)}=\wh\bD_1\cong\Z_q\,.
\ee
The topological codimension-2 operators constituting this 1-form symmetry are a subset of the operators described in (\ref{Ue}) given by
\be
U^{(e)}_n(\Sigma_{d-2})=\exp\left(\frac{2\pi i n}q\int_{\Sigma_{d-2}} \star F \right),\qquad n\in\{0,1,\cdots q-1\}\,.
\ee
The other codimension-2 operators in (\ref{Ue}) are not topological anymore as they they have a non-trivial linking with the charge $q$ Wilson line $\cW_q$.

\begin{example}[Higgsing $SU(2)$ to $U(1)$]{H}
Let us begin with an $SU(2)$ gauge theory at some high-enough energy scale. We will refer to this theory as the UV theory, even though it may not be UV complete (e.g.\ in $d>4$). The $SU(2)$ gauge theory under consideration has a scalar field $\Phi$ transforming in the adjoint representation of $SU(2)$. Giving a vacuum expectation value (vev) $\langle\Phi\rangle$ to the matter field breaks the $SU(2)$ gauge group to its maximal torus
\be\label{hm}
U(1)\subset SU(2)\,,
\ee
and below the scale set by $\langle\Phi\rangle$ we have a gauge theory with $U(1)$ gauge group. Again, we will refer to this theory as the IR theory, even though it may not be the deep IR.

As we will discuss later, the UV $SU(2)$ theory has an electric 1-form symmetry
\be
G^{(1)}=\Z_2\,.
\ee
This can be matched with the electric 1-form symmetry of the IR $U(1)$ theory because of the presence of W-boson produced by the Higgs mechanism (\ref{hm}), which carries a charge
\be
q_\Phi=2
\ee
under the IR $U(1)$ gauge group.
\end{example}

\begin{example}[Higgsing $SO(3)$ to $U(1)$]{}
We can replace the $SU(2)$ gauge group by the $SO(3)$ gauge group, where
\be
SO(3)=SU(2)/\Z_2\,.
\ee
Below the scale set by $\langle\Phi\rangle$, the $SO(3)$ gauge group is broken to its maximal torus
\be\label{hm2}
U(1)'\subset SO(3)\,,
\ee
where $U(1)'$ is related to the $U(1)$ group appearing in (\ref{hm}) as
\be
U(1)'=\frac{U(1)}{\Z_2}\,.
\ee
Consequently the charge of the W-boson is
\be
q_\Phi=1
\ee
under the IR $U(1)'$ gauge group and IR 1-form symmetry group is thus
\be
G^{(1)}=1\,,
\ee
which is consistent with the fact that the UV $SO(3)$ gauge theory has a trivial 1-form symmetry, as will be discussed later.
\end{example}

\begin{example}[Higgsing $U(1)$ to $\Z_q$]{}
Let us now invert the situation of the above two examples: instead of realizing a $U(1)$ gauge theory in the IR, we begin with a $U(1)$ gauge theory in the UV. Let us also include a charge $q$ scalar field $\phi$, and consider a vacuum with non-zero vev
\be
\langle\phi\rangle\neq0\,.
\ee
This higgses the $U(1)$ gauge group down to
\be
\Z_q\subset U(1)\,,
\ee
as $\Z_q$ leaves $\phi$ invariant. Thus we would expect the IR theory to be a $\Z_q$ gauge theory of the form
\be\label{dZq}
\frac{2\pi}q\int a_1\cup \delta b_{d-2}\,.
\ee
This can be seen quite explicitly as follows. Decompose $\phi$ as 
\be
\phi = \rho e^{i \alpha}\,,
\ee
and ignore the radial mode and replace it by its vev 
\be
\langle\rho\rangle = v
\ee
in the far IR. The Lagrangian then becomes
\begin{equation}
    \mathcal{L} = \frac{v^{2}}{2}  \left(d\alpha - qA \right) \wedge \star \left(d\alpha - qA \right)\,.
\end{equation}
Dualizing the circle valued scalar field $\alpha$ introduces a $(d-2)$-form field $B$ with gauge symmetry $B \rightarrow B + d\Lambda$. The Lagrangian in terms of this dual field is the well-known $BF$-theory, which reads:
\begin{equation}
    \mathcal{L} = \frac{q}{2 \pi} B \wedge dA\,.
\end{equation}
This is in fact a continuous presentation of the $\Z_q$ gauge theory in (\ref{dZq}).

As discussed in section \ref{hfgt}, the $\Z_q$ gauge theory has a $\Z_q$ 1-form symmetry, thus matching the UV 1-form symmetry of the $U(1)$ gauge theory.
\end{example}

\paragraph{Magnetic $(d-3)$-Form Symmetry.}
The magnetic $(d-3)$-form symmetry remains as
\be
G^{(d-3)}=U(1)
\ee
even after the addition of the charged matter field $\phi$.

\subsubsection{Maxwell Theory with Multiple Matter Fields}
Now consider the addition of multiple matter fields, 
\be
\phi_i,\qquad 1\le i\le m\,,
\ee
of gauge charges $q_i$ to the pure Maxwell theory. Using products of $\phi_i$ we can construct non-genuine local operators that implement screenings
\be
\cW_p\sim\cW_{p+nq}
\ee
where the minimal completely screened charge $q$ is obtained by taking the \textbf{greatest common divisor (gcd)} of the charges of all the matter fields
\be
q=\text{gcd}(q_1,q_2,\cdots,q_m)\,,
\ee
and consequently the group of unscreened Wilson lines is
\be
\bD_1=\Z/q\Z\equiv\Z_q\,,
\ee
whose Pontryagin dual is the electric 1-form symmetry group
\be
G^{(1)}=\wh\bD_1\cong \Z_q\,.
\ee
The magnetic $(d-3)$-form symmetry once again is still
\be
G^{(d-3)}=U(1)\,.
\ee

\begin{example}[Higgsing $SU(2)$ with Fundamental Matter to $U(1)$]{}
Add a new matter field $\psi$ to the UV theory in the example (\ref{H}) transforming in the fundamental representation (spin-$\frac12$ representation) of the $SU(2)$ gauge group. The matter field $\psi$ has charge
\be
q_\psi=1
\ee
under the $U(1)$ maximal torus of $SU(2)$. Thus the IR 1-form symmetry ceases to exist
\be
G^{(1)}=\wh \Z_{\text{gcd}(q_\Phi,q_\psi)}=0\,.
\ee
This is consistent with the UV 1-form symmetry which, as will be discussed later (see the discussion around \eqref{eq:su2withfundamental}), is also trivial after the addition of $\psi$.
\end{example}

\subsubsection{Multiple Abelian Gauge Fields}
We can generalize the above computations of electric 1-form symmetries to an abelian gauge theory with $r$ abelian gauge fields and gauge group
\be
\cG=\prod_{a=1}^r U(1)_a=U(1)^r\,.
\ee
The matter fields are $\phi_i$ for $1\le i\le m$, whose gauge charges $q_{i,a}$ can be organized into a $m\times r$ charge matrix $Q$.

Wilson line operators span all possible charges for the gauge group $U(1)^r$ and hence are labeled by elements of the abelian group $\Z^r$. The various charges of a matter field $\phi_i$ specify a vector $Q_i\in\Z^r$. Let us define a subgroup $Q_{\text{matter}}$ of $\Z^r$ by taking a span over integers of all $Q_i$
\be
Q_{\text{matter}}:=\text{Span}_\Z\{Q_1,Q_2,\cdots,Q_m\}\,.
\ee
Then, the abelian group of unscreened Wilson lines is obtained by modding out $\Z^r$ by $Q_{\text{matter}}$
\be\label{ES}
\bD_1=\frac{\Z^r}{Q_{\text{matter}}}\,,
\ee
whose Pontryagin dual is the electric 1-form symmetry group
\be
G^{(1)}=\wh\bD_1\,.
\ee
A practical way of computing the right hand side of (\ref{ES}) is using \textbf{Smith normal form (SNF)} of the charge matrix $Q$, which contains a diagonal matrix inside it with all other entries being zero;
\be
\text{diag}(q^D_1,q^D_2,\cdots,q^D_{\text{min}(m,r)})\subset\text{SNF}(Q)\,,
\ee
such that each $q^D_i$ is a non-negative integer. 

\begin{tech}[Smith Normal Form]{}
To compute the Smith normal form of an integer matrix $M$, we perform a sequence of the following moves\footnote{In \texttt{Mathematica} the appropriate command is \texttt{SmithDecomposition}.}:
\bit
\item Add an integer multiple of a row $R_i$ to another row $R_j$,
\item Add an integer multiple of a column $C_i$ to another column $C_j$,
\item Multiply a row by a minus sign,
\item Multiply a column by a minus sign,
\item Exchange two rows,
\item Exchange two columns,
\eit
until we reach an integer matrix $\text{SNF}(M)$ in which there is at most a single non-zero entry in each row and column, and any non-zero entry is positive.
\end{tech}

We then have
\be
\bD_1=\prod_{i=1}^{\text{min}(m,r)}\Z_{q^D_i}\,,
\ee
where $\Z_0:=\Z$ in the above equation. Its Pontryagin dual, namely the 1-form symmetry group, is
\be\label{gag1}
\G1=\prod_{i=1}^{\text{min}(m,r)}\wh\Z_{q^D_i}\,,
\ee
where we have
\be\label{wnot}
\ba
\wh\Z_{q^D_i}&\cong\Z_{q^D_i}, \qquad q^D_i>0,\\
\wh\Z_0&\cong U(1)\,.
\ea
\ee
The magnetic $(d-3)$-form symmetry on the other hand is, as usual, unaffected by the matter content
\be
\G{d-3}=U(1)^r\,.
\ee

\begin{example}[Higgsing $SU(3)$ to $U(1)\times U(1)$]{}
Slightly generalizing the example \ref{H}, consider a UV theory with $SU(3)$ gauge theory and a scalar field $\Phi$ in adjoint representation. Setting a generic vev $\langle\Phi\rangle$ leads to an IR theory with a gauge group that is the maximal torus of $SU(3)$
\be
U(1)_1\times U(1)_2\subset SU(3)\,.
\ee
The W-bosons have charges
\be
q_1=(2,-1),\qquad q_2=(-1,2)
\ee
under $U(1)_1\times U(1)_2$. The charge matrix is thus
\be
Q=\begin{pmatrix}
2&-1\\
-1&2
\end{pmatrix}\,.
\ee
whose Smith Normal Form is
\be
\text{SNF}(Q)=\begin{pmatrix}
3&0\\
0&1
\end{pmatrix}\,.
\ee
From this we compute the IR 1-form symmetry group to be
\be
G^{(1)}=\wh\Z_3\times\wh\Z_1=\Z_3\times\Z_1=\Z_3\,,
\ee
which matches the UV 1-form symmetry group that will be discussed later (see discussion around \eqref{eq:1fsofsun}).
\end{example}

\subsubsection{Including Magnetic Matter}
We can also add magnetically charged $(d-3)$-dimensional dynamical excitations, 
\be
M_i,\qquad 1\le i\le m\,,
\ee
of finite tension to an abelian gauge theory. Let $M_i$ have magnetic charges $p_{i,a}\in\Z$, which can be organized into an $m\times r$ charge matrix $P$ whose entries are $p_{i,a}$. Such abelian gauge theories with magnetic excitations arise as low energy effective theories describing Coulomb phases of supersymmetric CFTs and of more general supersymmetric QFTs.

These magnetic excitations lead to complete screening of 't Hooft operators $\cH_i$ of magnetic charges 
\be
p_{i,a},\qquad 1\le a\le r\,.
\ee
The group of unscreened 't Hooft operators is computed as
\be
\bD_{d-3}=\prod_{i=1}^{\text{min}(m,r)}\Z_{p^D_i}\,,
\ee
where $p^D_i$ are the entries inside the Smith Normal Form of $P$
\be
\text{diag}(p^D_1,p^D_2,\cdots,p^D_{\text{min}(m,r)})\subset\text{SNF}(P)\,.
\ee
The Pontryagin dual of $\bD_{d-3}$ is the magnetic $(d-3)$-form symmetry group
\be
\G{d-3}=\wh\bD_{d-3}=\prod_{i=1}^{\text{min}(m,r)}\wh\Z_{p^D_i}\,.
\ee
We remind the reader of our notation (\ref{wnot}).

The addition of magnetic excitations does not impact the electric 1-form symmetry group (\ref{gag1}).

\begin{example}[Higgsing $SU(2)$ to $U(1)$]{}
Consider again the example \ref{H}. As will be discussed later, the magnetic $(d-3)$-form symmetry group of the UV theory is trivial;
\be
G^{(d-3)}=1\,.
\ee
This is reproduced in the IR theory by noting that the 't Hooft-Polyakov monopole solution of the UV $SU(2)$ theory becomes a dynamical magnetic excitation $M$ of the IR $U(1)$ theory. As discussed around equation (\ref{coch}), monopole configurations in $SU(2)$ gauge theory are associated to loops in the group manifold of $SU(2)$. The loop associated to the 't Hooft Polyakov monopole winds once around the maximal torus $U(1)$, meaning that the IR magnetic charge of $M$ is
\be
p=1\,,
\ee
as a consequence of which all the IR 't Hooft line operators are completely screened.
\end{example}

\begin{example}[Higgsing $SO(3)$ to $U(1)$]{}
Now change the UV gauge group to $SO(3)$ in the previous example. As will be discussed later, the magnetic $(d-3)$-form symmetry group of the UV theory is now
\be
G^{(d-3)}=\Z_2\,.
\ee
To reproduce this in the IR theory, note that the IR magnetic charge of $M$ is now
\be
p=2\,,
\ee
because a loop winding once around the maximal torus $U(1)\subset SU(2)$ winds twice around the maximal torus
\be
U(1)'=\frac{U(1)}{\Z_2}\subset SO(3)\,.
\ee
\end{example}

\subsubsection{Including Dyonic Matter}
In spacetime dimension $d=4$, a generalization of the above scenario opens up, as electric and magnetically charged operators have the same dimensions. Consider a $4d$ abelian gauge theory with $U(1)^r$ gauge group. The 't Hooft operators are line operators and combine together with Wilson line operators to form general dyonic \textbf{'t Hooft-Wilson line operators}. These can be screened by dynamical particles in the theory which may also now be dyonic, i.e.\ carry both electric and magnetic charges. Let $1\le i\le m$ parametrize all the different particles, with electric charges $q_{i,a}$ and magnetic charges $p_{i,a}$. Such abelian gauge theories with dyonic excitations arise as low energy effective theories describing Coulomb phases of 4d supersymmetric CFTs and of more general 4d supersymmetric QFTs.

These dyonic excitations lead to complete screening of 't Hooft-Wilson line operators $\cD_i$ of electric and magnetic charges 
\be
\ba
&q_{i,a},\qquad 1\le a\le r\,,\\
&p_{i,a},\qquad 1\le a\le r\,,
\ea
\ee
respectively. 

Using the electric and magnetic charges of dynamical particles we define an $m\times 2r$ charge matrix $C$ whose $i$-th row is the charge vector
\be
C_i=(q_{i,1},q_{i,2},\cdots,q_{i,r},p_{i,1},p_{i,2},\cdots,p_{i,r})\,.
\ee
We then compute
\be
\text{diag}(c^D_1,c^D_2,\cdots,c^D_{\text{min}(m,2r)})\subset\text{SNF}(C)\,,
\ee
and the abelian group of unscreened 't Hooft-Wilson line operators is
\be
\bD_1=\prod_{i=1}^{\text{min}(m,2r)}\Z_{c^D_i}\,.
\ee
Its Pontryagin dual is a \textbf{dyonic 1-form symmetry group}
\be
\G1=\wh\bD_1=\prod_{i=1}^{\text{min}(m,2r)}\wh\Z_{c^D_i}\,.
\ee

\begin{example}[Dyonic Point in Seiberg-Witten Model]{}
Consider 4d $\cN=2$ pure $SU(2)$ gauge theory. We will see later that this theory has a 1-form symmetry
\be
\G1=\Z_2\,.
\ee
As is well-known from \cite{Seiberg:1994rs}, this theory has an interesting Coulomb branch of vacua, which contains two special points, namely the monopole and dyon points. Consider the IR theory at the dyon point, which is an $\cN=2$ SQED, which has $U(1)$ gauge group and a matter field $\phi$ with electric charge
\be
q_\phi=1\,.
\ee
Additionally we have a massive BPS particle that is a dyon from the point of view of SQED having electric and magnetic charges
\be
q=1, \quad p=2\,.
\ee
Collecting these charges together, we have the dyonic charge matrix
\be
C=\begin{pmatrix}
1&0\\
1&2
\end{pmatrix}\,,
\ee
whose Smith Normal Form is
\be
\text{SNF}(C)=\begin{pmatrix}
1&0\\
0&2
\end{pmatrix}\,,
\ee
implying that the IR 1-form symmetry group is also
\be
G^{(1)}=\Z_2\,.
\ee
\end{example}

\subsection{Non-Abelian Gauge Theories}
In this subsection, we use arguments based on screening to determine higher-form symmetry groups of gauge theories with non-abelian gauge groups.

\subsubsection{Pure $SU(2)$ Yang-Mills Theory}
We begin by considering a $d$-dimensional gauge theory with gauge group
\be\label{GSU2}
\cG=SU(2)\,,
\ee
and no matter fields.

\paragraph{Wilson Line Operators.}
The Wilson line operators are parametrized by irreducible representations of the gauge group (\ref{GSU2})
\be
\cW_j=\text{Tr}_{R_j}\cP e^{2\pi i\int A}\,,
\ee
where Tr$_{R_j}$ denotes trace in the spin-$j$ irreducible representation $R_j$, where $j$ is a half-integer
\be
j\in\half\Z\,.
\ee

\paragraph{Constructing Non-Genuine Local Operators Using Field Strength.}
Recall that the field strength $F$ transforms in the adjoint representation of the gauge group, which is the spin-1 representation
\be
j=1\,.
\ee
This implies that an insertion $F(x)$ of the field strength cannot provide a genuine local operator as it won't be gauge invariant. Instead, we can attach $F(x)$ to the Wilson line $\cW_{1}$ in the adjoint representation to obtain a gauge-invariant configuration
\be
F(x)\cW_{1}(L)\,,
\ee
where $L$ is a line that ends at the point $x$, i.e.
\be
\partial L=x\,,
\ee
much like in figure \ref{Wqend}. Thus, the field strength $F$ provides a non-genuine local operator that completely screens the adjoint Wilson line $\cW_{1}$.

\paragraph{Screenings of Wilson Lines.}
Similarly, by taking products of the field strength we can construct non-genuine local operators that completely screen any integer-spin Wilson line
\be
\cW_j,\qquad j\in\Z\,.
\ee
More generally, we can construct non-genuine local operators relating any Wilson lines whose spins differ by an integer
\be
\cW_j\sim\cW_{j'},\qquad j'-j\in\Z\,.
\ee
Thus the group of unscreened Wilson lines is
\be
\bD_1=\Z_2\,.
\ee
Concretely, the trivial Wilson line and spin$-\half$ Wilson lines can be chosen as representatives of the two non-trivial equivalence classes of line operators forming the group $\whG1=\bD_1$. 

\paragraph{Electric 1-Form Symmetry.}
Taking the Pontryagin dual of $\bD_1$, we obtain the electric 1-form symmetry group of $SU(2)$ pure Yang-Mills theory
\be\label{esu2}
\G1=\wh\bD_1=\Z_2\,.
\ee

\paragraph{'t Hooft Operators.}
The 't Hooft operators for a non-abelian gauge group $\cG$ are simply obtained from 't Hooft operators for $U(1)$ discussed around (\ref{U1th}). This is explained in the box below.

\begin{tech}['t Hooft Operators and Cocharacters]{}
Recall that a $U(1)$ 't Hooft operator $\cH$ has the property that on a small sphere $S^2$ linking $\cH$ we have
\be
\int_{S^2} F=m\in\Z
\ee
where $F$ is the field strength for $U(1)$ and $m$ is the magnetic charge of $\cH$. Now the idea is to promote $\cH$ to a 't Hooft operator $\cH_\phi$ for gauge group $\cG$ (which is assumed to be a connected Lie group) by letting the above $U(1)$ lie inside $\cG$. Concretely this is done as follows. 

We begin by picking a homomorphism
\be\label{coch}
\phi:~U(1)\to\cG
\ee
referred to as a \textbf{cocharacter}. This defines a subgroup
\be
U(1)_\phi=\text{Image}(\phi)\subseteq\cG
\ee
Moreover we obtain an integer
\be
m_\phi\in\Z
\ee
describing the \textit{winding number} of the map $\phi$ around $U(1)_\phi$.

Given a cocharacter $\phi$, we obtain a 't Hooft operator $\cH_\phi$ for a $\cG$ gauge theory. An insertion
\be
\cH_\phi(M_{d-3})\,,
\ee
of the 't Hooft operator $\cH_\phi$ along a codimension-3 submanifold $M_{d-3}$ of spacetime is defined by excising $M_{d-3}$ from spacetime and requiring that the $\cG$ gauge fields in the path integral have a boundary condition near $M_{d-3}$ such that we have
\be\label{mcc}
\int_{S^2} F_\phi=m_\phi
\ee
on a small sphere $S^2$ linking $M_{d-3}$, where $F_\phi$ is the field strength for $U(1)_\phi$.
\end{tech}

In the context of the current discussion, the various 't Hooft operators are thus parametrized by cocharacters
\be
\phi:~U(1)\to\cG=SU(2)\,.
\ee

\paragraph{Screenings of 't Hooft Operators.}
As in the 't Hooft-Polyakov monopole solution, we can produce dynamical solutions for $SU(2)$ gauge fields having the property (\ref{mcc}). This can be done for any choice of $\phi$, as the only requirement is that the loop in the group manifold for $SU(2)$ defined by $\phi$
be shrinkable to a trivial loop. Since the first homotopy group for $SU(2)$ is trivial,
\be\label{p1su2}
\pi_1(SU(2))=1\,,
\ee
every such loop is indeed shrinkable.

The creation operators for these solutions provide non-genuine codimension-4 operators completely screening the codimension-3 't Hooft operators. 
Thus, all 't Hooft operators can be completely screened, leading to
\be
\bD_{d-3}=1\,.
\ee

\paragraph{Magnetic $(d-3)$-Form Symmetry.}
Since there are no unscreened 't Hooft operators, there is no magnetic $(d-3)$-form symmetry
\be\label{msu2}
\G{d-3}=\wh\bD_{d-3}=1\,.
\ee

\paragraph{Topological Codimension-2 Operators.}
Above we deduced the existence of a non-trivial electric 1-form symmetry (\ref{esu2}). The topological codimension-2 operator $U$ generating this symmetry is understood as a solitonic defect\footnote{This terminology was introduced in \cite{Bhardwaj:2022dyt} for local and extended operators that generate a solitonic configuration of fields around them.} inducing an
\be
SO(3)=SU(2)/\Z_2
\ee
vortex configuration in the $SU(2)$ theory that cannot be lifted to an $SU(2)$ vortex configuration.

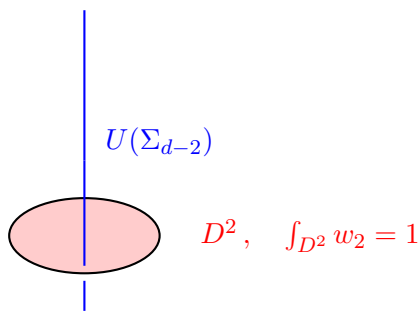
\begin{figure}
\centering
\scalebox{1}{
\begin{tikzpicture}
\begin{scope}[rotate=90]
\begin{scope}[shift={(-1,0)}]
\draw[fill=red!20, pattern color=red, thick, rotate=90] circle (1 and 0.5);
\begin{scope}[rotate=90, shift={(0,-1)}]
\end{scope}
\end{scope}
\draw [thick, blue](-2,0) -- (-1.6,0) (-1.4,0) -- (0,0);
\draw [thick, blue](0,0) -- (2,0);
\node[blue] at (0.25,-1) {$U(\Sigma_{d-2})$};
\node[red] at (-1,-3) {$D^2\,, \quad \int_{D^2} w_2 = 1$};
\end{scope}
\end{tikzpicture}
}
\caption{Codimension-2 topological operator $U$ generating the $\Z_2$ electric 1-form symmetry, inserted along $\Sigma_{d-2}$. On a small 2-dimensional disk $D_2$ intersecting $\Sigma_{d-2}$, we have a non-trivial second Stiefel-Whitney class $w_2$ turned on, which obstructs the lifting of $SO(3)$ gauge fields to $SU(2)$ gauge fields.}
\label{fig:diracstringin3d}
\end{figure}

In more detail, consider an insertion,
\be
U(\Sigma_{d-2})\,,
\ee
of the operator $U$ along a codimension-2 submanifold $\Sigma_{d-2}$ of spacetime. Then the gauge fields near $\Sigma_{d-2}$ are valued in $SO(3)$ such that we have
\be\label{OC}
\int_{D_2} w_2=1\in\Z/2\Z\equiv\Z_2
\ee
a small disk $D^2$ intersecting $\Sigma_{d-2}$ at a point, and $w_2$ is the \textit{second Stiefel-Whitney class} (see below) for $SO(3)$ bundles. Non-triviality of $w_2$ means that the $SO(3)$ gauge fields cannot be lifted to $SU(2)$ gauge fields near $\Sigma_{d-2}$.
See figure \ref{fig:diracstringin3d}.

The operator $U$ is known as a \textbf{Gukov-Witten operator} \cite{Gukov:2006jk,Gukov:2008sn} and can also be understood as a \textbf{Dirac string} for an $SO(3)$ monopole configuration that cannot be lifted to an $SU(2)$ monopole configuration. As it is a Dirac string, it can be placed anywhere, meaning that $U$ is topological. And the operator $U$ is not trivial because it can be detected by the Wilson line operator in the fundamental representation of $SU(2)$.

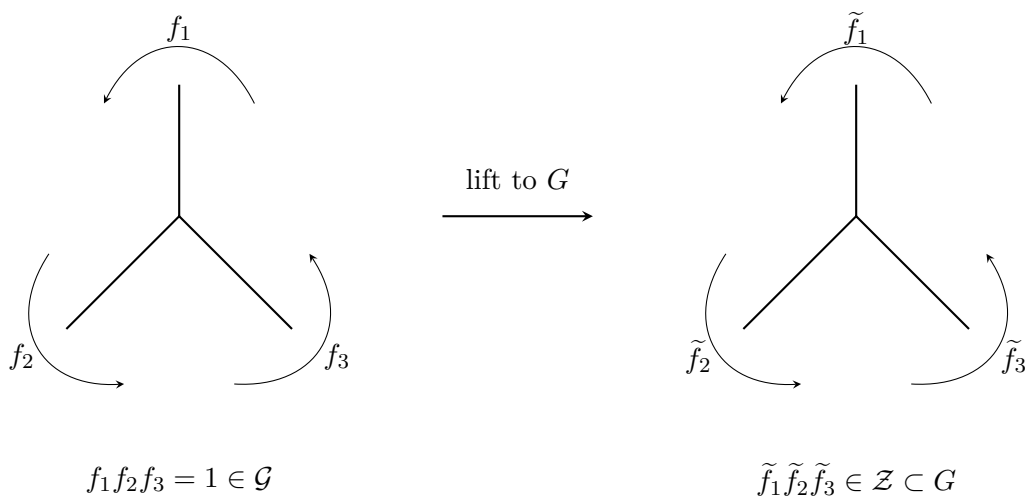
\begin{figure}
\centering
\scalebox{1}{
\begin{tikzpicture}
\draw [thick](0,1.25) -- (0,-0.5) (0,-0.5) -- (-1.5,-2) (0,-0.5) -- (1.5,-2);
\draw [-stealth](1,1) .. controls (0.5,2) and (-0.5,2) .. (-1,1);
\node at (0,2) {$f_1$};
\begin{scope}[rotate=120, shift={(-1,1)}]
\draw [-stealth](1,1) .. controls (0.5,2) and (-0.5,2) .. (-1,1);
\node at (0,2) {$f_2$};
\end{scope}
\begin{scope}[rotate=240, shift={(1,1)}]
\draw [-stealth](1,1) .. controls (0.5,2) and (-0.5,2) .. (-1,1);
\node at (0,2) {$f_3$};
\end{scope}
\node at (0,-4) {$f_1f_2f_3=1\in\cG$};
\draw [thick,-stealth](3.5,-0.5) -- (5.5,-0.5);
\node at (4.5,0) {lift to $G$};
\begin{scope}[shift={(9,0)}]
\draw [thick](0,1.25) -- (0,-0.5) (0,-0.5) -- (-1.5,-2) (0,-0.5) -- (1.5,-2);
\draw [-stealth](1,1) .. controls (0.5,2) and (-0.5,2) .. (-1,1);
\node at (0,2) {$\widetilde{f}_1$};
\begin{scope}[rotate=120, shift={(-1,1)}]
\draw [-stealth](1,1) .. controls (0.5,2) and (-0.5,2) .. (-1,1);
\node at (0,2) {$\widetilde{f}_2$};
\end{scope}
\begin{scope}[rotate=240, shift={(1,1)}]
\draw [-stealth](1,1) .. controls (0.5,2) and (-0.5,2) .. (-1,1);
\node at (0,2) {$\widetilde{f}_3$};
\end{scope}
\node at (0,-4) {$\widetilde{f}_1\widetilde{f}_2\widetilde{f}_3\in\cZ\subset G$};
\end{scope}
\end{tikzpicture}
}
\caption{Solid lines represent codimension-1 loci implementing transition functions $f_i$, which come together along a codimension-2 locus. As one goes in a circle around the codimension-2 locus, one picks up $f_1f_2f_3$ which must be equal to the the identity in $\cG$ to have a consistent $\cG$ bundle. Picking lifts $\wt f_i$ of the transition functions to $G$ may destroy this consistency condition and in general we have $\wt f_1\wt f_2\wt f_3\in\cZ$, which specifies the 2-cochain $w_2$ capturing the obstruction of lifting the $\cG$ bundle to a $G$ bundle.}
\label{obstc}
\end{figure}

\begin{tech}[Obstruction Classes]{}
Consider a principal bundle for a Lie group
\be
\cG=G/\cZ\,,
\ee
where $\cZ$ is a subgroup of the center $Z(G)$ of another Lie group $G$. A principal bundle for $\cG$ can be described in terms of transition functions valued in $\cG$ on codimension-1 loci. These codimension-1 loci come together and form codimension-2 junctions. Consider a lift of these $\cG$-valued transition functions to $G$-valued transition functions. Before the lift, the product of transition functions around a codimension-2 junction is 1. After the lift, the product of transition functions around a codimension-2 junction will be a lift of 1, or in other words an element in $\cZ$. See figure \ref{obstc}. By Poincaré duality, this defines a $\cZ$-valued 2-cochain $w_2$. See the discussion around (\ref{PD}) for more details on Poincaré duality.

The cohomology class of $w_2$ is known as the obstruction class for lifting the $\cG$ bundle under study to a $G$ bundle. It is independent of the various choices made above like the choice of lift.

For
\be
\cG=SO(3),\qquad G=SU(2),\qquad \cZ=\Z_2\,,
\ee
the obstruction class is also known as the \textbf{second Stiefel-Whitney class}.
\end{tech}

\subsubsection{Pure $SO(3)$ Yang-Mills Theory}
Let us now consider a $d$-dimensional gauge theory with gauge group
\be
\label{GSO3}
\cG=SO(3)=\frac{SU(2)}{\Z_2}
\ee
and no matter fields. The $\Z_2$ subgroup of $SU(2)$ being modded out is the center of $SU(2)$.

The difference between this case and the case of pure $SU(2)$ Yang-Mills theory discussed above seems rather minute, given that the only difference between them is whether a discrete $\Z_2$ is included in the gauge group or not. However, we will see here that this difference crucially modifies the electric and magnetic higher-form symmetries.

\paragraph{Wilson Line Operators.}
The Wilson line operators are parametrized by irreducible representations of the gauge group $SO(3)$, of which there are actually fewer than in the case of $SU(2)$. A given spin-$j$ representation $R_j$ of $SU(2)$ descends to a representation of $SO(3)$ only if the center,
\be
\Z_2\subset SU(2)\,,
\ee
acts trivially on $R_j$. The $\Z_2$ center acts as a sign on every vector in the representation $R_j$, with the sign being $(-1)^{2j}$. We can thus express the action of the center as
\be
R_j\to(-1)^{2j}R_j\,.
\ee
Thus, only the integer spin representations of $SU(2)$ are representations of $SO(3)$
\be
\text{$R_j$ is an $SO(3)$ representation}\iff j\in\Z\,.
\ee

\paragraph{Electric 1-Form Symmetry.}
As discussed above, all integer spin Wilson lines can be completely screened by products of the field strength. Thus there is no electric 1-form symmetry, i.e.
\be
\G1=1\,,
\ee
in contrast with (\ref{esu2}).

\paragraph{'t Hooft Operators.}
't Hooft operators now correspond to cocharacters
\be
\phi:~U(1)\to\cG=SO(3)\,.
\ee
Now, not all of the 't Hooft operators can be completely screened because we have a nontrivial first homotopy group
\be\label{p1so3}
\pi_1(SO(3))=\Z_2\,.
\ee
\begin{tech}[First Homotopy Group of $SO(3)$]{}
We can deduce (\ref{p1so3}) from (\ref{p1su2}) as follows. Consider a path in $SU(2)$ relating the two elements of the $\Z_2$ center of $SU(2)$. Since the end points of this path are identified after quotienting out $\Z_2$, this path descends to a loop on $SO(3)$. Such a loop is not contractible in $SO(3)$ as the corresponding path in $SU(2)$ is not contractible, simply because one is not allowed to move the end points of a path. However, twice of this path is a loop in $SU(2)$ and hence contractible. We have thus deduced (\ref{p1so3}).
\end{tech}
Thus, the group of unscreened 't Hooft operators is
\be
\bD_{d-3}=\Z_2\,.
\ee

\paragraph{Magnetic $(d-3)$-Form Symmetry.}
The magnetic $(d-3)$-form symmetry is thus
\be
\G{d-3}=\wh\bD_{d-3}=\Z_2\,,
\ee
in contrast with (\ref{msu2}).

\paragraph{Topological Surface Operators.}
The topological surface operator generating the magnetic $(d-3)$-form symmetry can be expressed as
\be\label{topm}
U(\Sigma_2)=\exp\left(i\pi\int_{\Sigma_2} w_2\right)\,,
\ee
where $w_2$ is the second Stiefel-Whitney class of $SO(3)$ gauge bundles.

\subsubsection{Pure Yang-Mills Theory with General Gauge Group}
Consider a pure $d$-dimensional gauge theory based on a non-abelian gauge algebra which is simple, i.e.
\be\label{gval}
\fg\in\{\su(n),\so(n),\sp(n),\fe_6,\fe_7,\fe_8,\ff_4,\fg_2\}\,.
\ee
The gauge group is taken to be a connected Lie group $\cG$ such that its Lie algebra is $\fg$
\be
\text{Lie}(\cG)=\fg\,.
\ee
Such a group can be expressed as
\be
\cG=\frac{G}{\cZ}\,,
\ee
where:
\bit
\item $G$ is the simply connected group having
\be
\text{Lie}(G)=\fg\,,
\ee
which takes the form
\be\label{Gval}
G\in\{SU(n),\Spin(n),\text{Sp}(n),E_6,E_7,E_8,F_4,G_2\}
\ee
respectively according to the form (\ref{gval}) taken by $\fg$. The fact that $G$ is simply connected means that its first homotopy group is trivial
\be\label{p1G}
\pi_1(G)=1\,.
\ee
\item $\cZ$ is a subgroup of the center $Z(G)$ of $G$
\be
\cZ\subseteq Z(G)\,.
\ee
The centers for various groups in (\ref{Gval}) are compiled in table \ref{tab:repsandcentralcharges}, where various possibilities for $\cG$ are also discussed.
\eit
Let us note that the center $Z(\cG)$ of the gauge group $\cG$ can be expressed as
\be\label{ZG}
Z(\cG)=\frac{Z(G)}{\cZ}
\ee
At the present moment, we do not include any matter fields.

\begin{table}[]
    \centering
    \begin{tabular}{|c|c|c|c|}
    \hline
         Simply Connected Group & Center & Representation & Charge of Representation  \\
         \hline
         \hline
         $SU(n)$ & $\mathbb{Z}_n$  & Fundamental, $\bm{F}$ & $1 \mod n$ \\
         && Antisymmetric, $\bm{\Lambda}^2$ & $2 \mod n$ \\
          && Symmetric, $\bm{S}^2$ & $2 \mod n$ \\
         \hline
         $\text{Sp}(n)$ & $\mathbb{Z}_2$ & Fundamental, $\bm{F}$ & $1 \mod 2$ \\
         \hline
         $\Spin(2n+1)$ & $\mathbb{Z}_2$ & Vector, $\bm{F}$ & $0 \mod 2$ \\
         &&Irreducible Spinor, $\bm{S}$ & $1 \mod 2$ \\
         \hline
         $\Spin(4n+2)$ & $\mathbb{Z}_4$ &Vector, $\bm{F}$& $2 \mod 4$ \\ 
         && Irreducible Spinor, $\bm{S}$ & $1 \mod 4$ \\
         && Irreducible Co-spinor, $\bm{C}$ & $3 \mod 4$ \\
         \hline
         $\Spin(4n)$ & $\mathbb{Z}_2 \times \mathbb{Z}_2$ &Vector, $\bm{F}$& $(1 \mod 2, 1\mod 2)$ \\ 
         && Irreducible Spinor, $\bm{S}$ & $(1 \mod 2,0 \mod 2)$ \\
         && Irreducible Co-spinor, $\bm{C}$ & $(0 \mod 2, 1 \mod 2)$ \\
         \hline
         $E_6$ & $\Z_3$ & Fundamental, $\bm{27}$ & $1 \mod 3$ \\
         \hline
         $E_7$ & $\Z_2$ & Fundamental, $\bm{56}$ & $1 \mod 2$ \\
         \hline
    \end{tabular}
    \caption{Here we list the simply connected groups having non-trivial centers. For each group, we list some representations and the charges of these representations under the center. For $E_8$, $F_4$ and $G_2$, the center is trivial.}
    \label{tab:repsandcentralcharges}
\end{table}

\paragraph{Electric 1-Form Symmetry.}
The Wilson line operators $\cW_R$ are labeled by arbitrary irreducible representations $R$ of the gauge group $\cG$. These can be identified with the irreducible representations of the simply connected group $G$ (or equivalently irreducible representations of the gauge algebra $\fg$) having the property that the subgroup
\be
\cZ\subseteq Z(G)\subset G
\ee
acts trivially.

Screenings induced by non-genuine local operators constructed out of products of the field strength $F$ relate
\be
\cW_R\sim\cW_{R'}
\ee
if $R$ and $R'$ have the same charges under the center $Z(\cG)$
\be
q(R)=q(R')\in\wh Z(\cG)\,.
\ee
Thus, the group of unscreened Wilson lines is
\be
\bD_1=\wh Z(\cG)
\ee
implying that the 1-form symmetry group is
\be
\G1= Z(\cG)\,,
\ee

\paragraph{Magnetic $(d-3)$-Form Symmetry.}
The 't Hooft operators $\cH_\phi$ are labeled by cocharacters
\be
\phi:~U(1)\to\cG\,.
\ee
As before, screenings relate
\be
\cH_\phi\sim\cH_{\phi'}
\ee
if the loops in $\cG$ specified by $\phi$ and $\phi'$ inside the group manifold of $\cG$ are related by a homotopy.
Thus, the group of unscreened 't Hooft operators is
\be
\bD_{d-3}=\pi_1(\cG)=\cZ\,.
\ee
The fact that $\pi_1(\cG)=\cZ$ follows from (\ref{p1G}) because the quotient by $\cZ$ of $G$ converts homotopically distinct paths in $G$ into loops in $\cG$. Taking the Pontryagin dual, we have the $(d-3)$-form symmetry group
\be\label{mag}
\G{d-3}=\widehat{ \pi_1(\cG)}=\wh\cZ\,.
\ee 

\paragraph{Topological Operators.}
The topological codimension-2 operators generating electric 1-form symmetries are
\be
U^{(e)}_g,\qquad g\in Z(\cG)\,,
\ee
which can be understood as Dirac strings inducing vortex configurations for the group
\be
G_{\text{min}}=\frac{\cG}{Z(\cG)}
\ee
that cannot be lifted to vortex configurations for $\cG$. That is, around an insertion $U^{(e)}_g(\Sigma_{d-2})$ we have
\be
\int_{D^2} w_2^{\text{min}}=g\,,
\ee
where $w_2^{\text{min}}$ is the obstruction class for lifting $G_{\text{min}}$ bundles to $\cG$ bundles. See the discussion after (\ref{OC}) for more details on obstruction classes.

On the other hand, the topological surface operators generating magnetic $(d-3)$-form symmetries are
\be\label{mago}
U^{(m)}_h(\Sigma_2)=\int_{\Sigma_2} h(w_2),\qquad h\in\wh\cZ\,,
\ee
where $w_2$ is the obstruction class for lifting $\cG$ gauge bundles to $G$ bundles.

\subsubsection{Including Matter Fields}\label{namat}
Let us now include matter fields (having any spacetime transformation properties and any mass) in the above setup
\be
\phi_i, \qquad 1\le i\le m\,,
\ee
transforming in irreducible representations $R_i$ of the gauge group $\cG$.

\paragraph{Electric 1-Form Symmetry.}
Before including matter content, we have the group $\wh Z(\cG)$ of unscreened Wilson lines. The inclusion of matter fields induces additional screenings. For example, Wilson lines $\cW_{R_i}$ for representations $R_i$ can be completely screened even though we have
\be
q(R_i)\neq0\in\wh Z(\cG)\,.
\ee
In general, integer linear combinations\footnote{Here we are regarding $\wh Z(\cG)$ as an additive group.} of the charges
\be
q(R_i)\in \wh Z(\cG),\qquad 1\le i\le m
\ee
generate a subgroup
\be
Q_{\text{matter}}\subseteq\wh Z(\cG)\,,
\ee
and the group of unscreened Wilson lines is
\be\label{g1na0}
\bD_1=\frac{\wh Z(\cG)}{Q_{\text{matter}}}\,.
\ee
The electric 1-form symmetry group is
\be\label{g1na1}
\G1=\wh\bD_1=\wh{\frac{\wh Z(\cG)}{Q_{\text{matter}}}}\,.
\ee
Equivalently, the only codimension-2 operators which remain topological are the ones corresponding to elements in $Z(\cG)$ under which the completely screened Wilson lines
\be
Q_{\text{matter}}\subseteq\wh Z(\cG)
\ee
are uncharged. Thus, we can express $\G1$ as
\be\label{g1na2}
\G1=\text{Subgroup of $Z(\cG)$ leaving all $R_i$ invariant}\,.
\ee

\paragraph{Computation of $\G1$.}
For any possible gauge group
\be
\cG\neq\Spin(4n),\qquad n\ge2\,,
\ee
the center is
\be
Z(\cG)=\Z_g,\qquad g\ge1\,,
\ee
and we can express $\G1$ as
\be
\G1=\Z_{\text{gcd}(g,q_1,q_2,\cdots q_m)}\,,
\ee
where
\be
q_i\equiv q(R_i)\in\wh Z(\cG)=\Z_g=\{0,1,\cdots, g-1\}\,.
\ee

\begin{example}[$\su(n)$ Gauge Theories with Adjoint and Fundamental Matter]{}
In various examples of section \ref{AbG}, we quoted results about electric 1-form symmetries of various gauge theories. Here we provide a derivation of those results.

First consider $SU(2)$ gauge theory with an adjoint matter field $\Phi$. As the adjoint matter has trivial charge under center
\be
Z(SU(2))=\Z_2\,,
\ee
the electric 1-form symmetry group is
\be
\G1=Z(SU(2))=\Z_2\,.
\ee
For $SO(3)$ gauge theory we have
\be
Z(SO(3))=Z(SU(2))/\Z_2=1\,,
\ee
and hence we have
\be
\G1=1\,,
\ee
which is not impacted by the addition of matter field $\Phi$.

Now add also a fundamental matter field $\psi$ along with the adjoint field $\Phi$ to the $SU(2)$ gauge theory. The fundamental representation has a non-trivial charge under $Z(SU(2))$, and hence the electric 1-form symmetry group is broken to
\be\label{eq:su2withfundamental}
\G1=1\,.
\ee
Let us now extend the gauge group from $SU(2)$ to $SU(n)$ and consider only the matter field $\Phi$ in adjoint of $SU(n)$. Again the adjoint representation is uncharged under the center
\be
Z(SU(n))=\Z_n\,,
\ee
and hence the electric 1-form symmetry group is
\be\label{eq:1fsofsun}
\G1=Z(SU(n))=\Z_n\,.
\ee
\end{example}

\paragraph{Magnetic $(d-3)$-Form Symmetry.}
The inclusion of matter does not impact the magnetic symmetry, which continues to be given by (\ref{mag}).

\paragraph{Topological Operators.}
The topological codimension-2 operators generating electric 1-form symmetries are 
\be
U^{(e)}_g,\qquad g\in\G1\,,
\ee
which can be understood as Dirac strings inducing vortex configurations for the group
\be
G_{\text{min}}=\frac{\cG}{\G1}
\ee
that cannot be lifted to vortex configurations for $\cG$. The group $G_{\text{min}}$ is the minimal connected Lie group with Lie algebra $\fg$ that allows the representations $R_i$ of the matter fields $\phi_i$. Around an insertion $U^{(e)}_g(\Sigma_{d-2})$ we have
\be
\int_{D^2} w_2^{\text{min}}=g\,,
\ee
where $w_2^{\text{min}}$ is the obstruction class for lifting $G_{\text{min}}$ bundles to $\cG$ bundles.

The topological surface operators generating magnetic $(d-3)$-form symmetries continue to be (\ref{mago}).

\subsubsection{Multiple Gauge Group Factors}\label{gna}
We can now easily generalize the above arguments to determine electric and magnetic higher-form symmetries of a gauge theory with a general non-abelian gauge algebra
\be
\fg=\bigoplus_a\fg_a\,,
\ee
where each $\fg_a$ is simple and thus valued in (\ref{gval}). Such theories include \textbf{quiver gauge theories} which play an important role in various areas of research in quantum field theory.

\paragraph{Gauge Group.}
The simply connected group associated to $\fg$ is
\be
G=\prod_a G_a\,,
\ee
where $G_a$ is the simply connected group associated to each $\fg_a$.

We want to consider a gauge group that is a connected Lie group $\cG$ such that its Lie algebra is $\fg$
\be
\text{Lie}(\cG)=\fg\,.
\ee
Just as before, such a group can be expressed as
\be
\cG=\frac{G}{\cZ}\,,
\ee
where $\cZ$ is a subgroup of the center $Z(G)$ of $G$
\be
\cZ\subseteq Z(G)=\prod_a Z(G_a)\,.
\ee
The center $Z(\cG)$ of the gauge group $\cG$ can be expressed as in (\ref{ZG}).

\paragraph{Matter Fields.}
We include matter fields (having any spacetime transformation properties and any mass)
\be
\phi_i, \qquad 1\le i\le m\,,
\ee
transforming in irreducible representations $R_i$ of the gauge group $\cG$.
We can express the representations as tensor products
\be
R_i=\bigotimes_{a}R_{i,a}\,,
\ee
where
\be
R_{i,a}=\text{An irreducible representation of $G_a$}\,.
\ee
However, there are constraints on $R_{i,a}$ for $R_i$ to be a representation of $\cG$, as in order for this to be possible, the subgroup $\cZ$ must act trivially on $R_i$.

\paragraph{Higher-Form Symmetries.}
The analysis of electric and magnetic 1-form symmetries is similar to as above. Finally the electric 1-form symmetry can be expressed either as (\ref{g1na1}) or (\ref{g1na2}), and the magnetic $(d-3)$-form symmetry as (\ref{mag}). The topological operators for these higher-form symmetries are the same as described in section \ref{namat}.

The computation of (\ref{g1na0}) can be easily done using Smith Normal Form. First of all, we can express $\wh Z(\cG)$ as
\be
\wh Z(\cG)=\prod_{\alpha=1}^p\Z_{n_\alpha}\,,
\ee
and each charge can thus be expressed as
\be
q(R_i)=(q_{1,i},\cdots,q_{\alpha,i},\cdots,q_{p,i}),\qquad q_{\alpha,i}\in\{0,1,\cdots,n_\alpha-1\}\,.
\ee
From this define a $p\times m$ matrix $Q$ whose $(\alpha,i)$-th entry is $q_{\alpha,i}$, using which we define a $2p\times (m+p)$ matrix
\be
Y=(Z|Q)\,,
\ee
where $Z$ is a diagonal $p\times p$ matrix whose $\alpha$-th diagonal entry is $n_\alpha$.
Then we have
\be
\G1=\prod_{i=1}^{\text{min}(2p,m+p)}\wh\Z_{q^D_i}\,,
\ee
where $q^D_i$ are defined via
\be
\text{diag}(q^D_1,q^D_2,\cdots,q^D_{\text{min}(2p,m+p)})\subset\text{SNF}(Y)\,.
\ee

\begin{example}[An Orthosymplectic Quiver Gauge Theory]{osp}
Consider a $d$-dimensional quiver gauge theory with gauge group
\be
\cG=\Spin(4n+2)\times\text{Sp}(m)\,,
\ee
and matter fields in bifundamental representation
\be\label{bif}
R=V\ot F\,,
\ee
of $\cG$, where $V$ is the vector representation of $\Spin(4n+2)$ and $F$ is the fundamental representation of $\text{Sp}(m)$. We have
\be
\wh Z(\cG)=\wh Z(\Spin(4n+2))\times\wh Z(\text{Sp}(m))=\Z_4\times\Z_2\,,
\ee
and the charge of $R$ is
\be
q(R)=(2,1)\,,
\ee
from which we construct the matrix
\be
Y=\begin{pmatrix}
4&0&2\\
0&2&1
\end{pmatrix}\,,
\ee
whose Smith Normal Form is
\be
\text{SNF}(Y)=\begin{pmatrix}
0&4&0\\
0&0&1
\end{pmatrix}\,,
\ee
implying that the electric 1-form symmetry of the quiver gauge theory is
\be
\G1=\Z_4\,.
\ee
The magnetic $(d-3)$-form symmetry is trivial because $\cG$ is simply connected.
\end{example}

\subsection{Discrete Theta Angle in 4d Gauge Theories}
In $d=4$ spacetime dimensions, there are more kinds of non-abelian gauge theories that carry discrete theta angles \cite{Aharony:2013hda}. These in general can have dyonic 1-form symmetries. The aim of this subsection is to describe these 1-form symmetries.

\subsubsection{Pure $SO(3)_-$ Yang-Mills Theory}\label{SO3-}
In the previous subsection we discussed pure $SU(2)$ and $SO(3)$ Yang-Mills theories in general $d$, which are both based on the gauge algebra $\su(2)$. In $d=4$, there is another pure Yang-Mills theory based on the $\su(2)$ gauge algebra \cite{Gaiotto:2010be}, which has
\be
\cG=SO(3)
\ee
gauge group, and is referred to as the $SO(3)_-$ theory \cite{Aharony:2013hda}. Correspondingly, the usual $SO(3)$ theory discussed above is sometimes also referred to as the $SO(3)_+$ theory.

\paragraph{Line Operators.}
Let us begin by recalling the discussion of line operators in the $SO(3)_+$ theory. First of all, we have purely Wilson line operators
\be\label{Wj}
\cW_j,\qquad j\in\Z\,.
\ee
Additionally, we have line operators that induce $SO(3)$ monopole configurations on a sphere $S^2$ linking them. Consider a monopole configuration described by a cocharacter
\be
\phi:~U(1)\to SO(3)\,.
\ee
The various line operators inducing the monopole configuration $\phi$ are labeled by integer spin representations of $\su(2)$. Let us denote them as
\be
\cD_{j,\phi},\qquad j\in\Z\,.
\ee
In particular, we have a line operator for the trivial representation $j=0$ of $\su(2)$ that we call a purely 't Hooft line operator and denote as
\be
\cH_\phi\equiv\cD_{0,\phi}\,.
\ee
Note that $\cD_{j,\phi}$ arises inside $\cW_j\cH_\phi$ the OPE of $\cW_j$ and $\cH_\phi$, a fact that we express by identifying the two
\be\label{+}
\cD_{j,\phi}\equiv\cW_j\cH_\phi\,.
\ee

\begin{tech}[Notation for OPE]{}
Often in the literature, additive notation is used for the OPE, and the resulting line operator is instead expressed as $\cW+\cH$. In this text, we use a product notation as in equation (\ref{+}). We do so in order to avoid confusion between a fusion product and a direct sum of lines. A correlation function of the fusion product $L_1\ot L_2$ of two lines $L_1$ and $L_2$ is
\be
\langle L_1L_2\rangle\,,
\ee
but a correlation function of a direct sum $L_1\oplus L_2$ is
\be
\langle L_1\rangle + \langle L_2\rangle\,.
\ee
\end{tech}

For the $SO(3)_-$ theory, we have the same set (\ref{Wj}) of purely Wilson line operators. However, the spectrum of line operators inducing monopole configurations is different. If $\phi$ can be lifted to a cocharacter for $SU(2)$, i.e.\ if the associated $SO(3)$ monopole configuration can be lifted to an $SU(2)$ monopole configuration, the dyonic lines are the same as in the $SO(3)_+$ theory
\be
\cD_{j,\phi},\qquad j\in\Z\,.
\ee
In particular, we still have purely 't Hooft operators
\be
\cH_\phi\,,
\ee
for such cocharacters $\phi$. However, if $\phi$ cannot be lifted to a cocharacter for $SU(2)$, i.e.\ if the associated $SO(3)$ monopole configuration cannot be lifted to an $SU(2)$ monopole configuration, the dyonic operators must carry half-integer spin representations of $\su(2)$ and we denote them as
\be
\cD_{j,\phi} \,,\qquad j\not\in\Z \,.
\ee
In particular, we no longer have purely 't Hooft operators for such cocharacters $\phi$.
We may formally write such line operators as
\be\label{-}
\cD_{j,\phi}\equiv\cW_j\cH_\phi\,,\qquad j\not\in\Z \,,
\ee
though it should be noted that neither $\cW_j$ nor $\cH_\phi$ is a genuine line operator individually\footnote{As we will discuss later, both $\cW_j$ for half-integer $j$ and $\cH_\phi$ for non-$SU(2)$ $\phi$ exist as non-genuine line operators in the $SO(3)_-$ theory.}. However, their product $\cD_{j,\phi}$ is a genuine line operator.

\paragraph{Witten Effect.}
An interesting way of understanding the existence of the $SO(3)_-$ theory with the above spectrum of line operators is to begin with the $SO(3)_+$ theory and changing the usual (as opposed to discrete) theta angle as
\be\label{tt}
\theta\to\theta+2\pi \,.
\ee
As is well-known, such a transformation on the theta angle induces the Witten effect, according to which a magnetically charged line operator acquires an additional electric charge, which converts the line operators (\ref{+}) of the $SO(3)_+$ theory to the line operators (\ref{-}) if $\phi$ cannot be lifted to an $SU(2)$ cocharacter. Since the line operators (\ref{-}) do not exist (as genuine operators) in the $SU(2)$ and $SO(3)_+$ theory, we conclude that we have landed on a new theory, which we have dubbed as the $SO(3)_-$ theory.

\paragraph{Screenings.}
The screenings of line operators in the $SO(3)_-$ theory are analogous to the screenings in the $SO(3)_+$ theory. All the purely Wilson and purely 't Hooft operators are completely screened. On the other hand, the 't Hooft-Wilson operators (\ref{-}) are also all related by screenings
\be
\cD_{j,\phi}\sim \cD_{j',\phi'}\,,
\ee
but are not completely screened. Thus the $SO(3)_-$ theory has a group
\be
\bD_1=\Z_2\,,
\ee
of unscreened line operators, implying that the theory has a 1-form symmetry group
\be
\G1=\wh\bD_1=\Z_2 \,.
\ee
The 't Hooft-Wilson line operators (\ref{-}) are non-trivially charged under $\G1$, while the purely Wilson and purely 't Hooft Wilson operators are not charged under it. Thus, we may refer to $\G1$ as a \textbf{dyonic 1-form symmetry group}.

\paragraph{Discrete Theta Angle.}
The action for the $SO(3)_-$ theory contains the term
\be\label{Pw}
\pi\int\frac{\cP(w_2)}2 \,,
\ee
where $\cP(w_2)$ is a $\Z_4$ valued degree-4 class obtained by taking Pontryagin square of the second Stiefel-Whitney class $w_2$ of $SO(3)$ gauge bundles. It can be expressed as
\be
\cP(w_2)=\wt w_2\cup \wt w_2-\delta\wt w_2\cup_1\wt w_2 \,,
\ee
where $\wt w_2$ is a $\Z_4$ valued 2-cochain that is a lift of the $\Z_2$ valued 2-cocycle $w_2$, and $\cup_1$ is a higher cup product for whose definition we refer the reader to appendix B of \cite{Gaiotto:2015zta}. The half in (\ref{Pw}) is allowed by the extra assumption that we are studying 4d gauge theories only on spacetimes admitting spin structures, as in this case $\cP(w_2)$ is an even class, i.e.\ its integrals take values only in $\{0,2\}\subset\Z_4=\{0,1,2,3\}$.

Just like the ordinary theta angle term is a square of the field strength, the term (\ref{Pw}) is a square of a dynamical degree-2 field $w_2$, which can be interpreted as a \textit{fractional field strength} in an $SO(3)$ gauge theory. For this reason, (\ref{Pw}) is referred to as a \textbf{discrete theta angle}.

As we discussed above, the term (\ref{Pw}) can be understood as a fractional contribution to the action arising by implementing the transformation (\ref{tt}).

\subsubsection{General Non-Abelian 4d Gauge Theory}
Now consider a general 4d $\cG$ gauge theory of the type considered in section (\ref{gna}). We add in a discrete theta angle, specified by a homomorphism
\be
\gamma:~\cZ\to\wh\cZ\,.
\ee
The corresponding theory is called $\cG_\gamma$ gauge theory. We will discuss the form of the corresponding term in the Lagrangian of the gauge theory in what follows.

\paragraph{'t Hooft-Wilson Line Operators.}
Consider a line operator inducing a monopole configuration for $\cG$, with the property that
\be
\int_{S^2}w_2=z\in\cZ\,,
\ee
on an $S^2$ linking the line. Such a line operator must transform in an irreducible representation $R$ of the gauge algebra $\fg$, having charge
\be
q(R)=\gamma(z)\in\wh\cZ
\ee
under $\cZ$.
Note that if $\gamma(z)=0$ then $R$ is a representation of the gauge group $\cG$, but if $\gamma(z)\neq0$ then $R$ is not a representation of $\cG$.

\paragraph{1-Form Symmetry.}
Including just the magnetic screenings and the screenings by the field strength, we reduce the group of unscreened line operators to
\be
\bD_\gamma\subset\cZ\times\wh Z(G) \,.
\ee
This is made up of elements
\be\label{eq:unscreenedwl}
(z,\wt\gamma(z))\in\cZ\times\wh Z(G) \,,
\ee
where $z\in\cZ$ and $\wt\gamma(z)\in\wh Z(G)$ is any lift of $\gamma(z)\in\wh\cZ$ under the surjective homomorphism
\be
\wh i:~\wh Z(G)\to\wh\cZ \,,
\ee
Pontryagin dual to the injective homomorphism
\be
i:\cZ\to Z(G) \,,
\ee
corresponding to the fact that $\cZ$ is a subgroup of $Z(G)$.

Substituting $z=0$, we see that purely Wilson lines form a subgroup
\be
\wh Z(\cG)\subseteq\bD_\gamma \,.
\ee
This follows from the fact that $\widehat{Z}(\cG) = \widehat{Z}(G) / \widehat{\cZ}$, so that the possible lifts into $\widehat{Z}(G)$ of the trivial element $0 \in \widehat{\cZ}$ are indeed given by elements of $\widehat{Z}(\cG)$.

The matter content completely screens part of this subgroup
\be
Q_{\text{matter}}\subseteq\wh Z(\cG)\subseteq\bD_\gamma\,,
\ee
and thus the group of unscreened 't Hooft-Wilson lines is
\be
\bD_1=\frac{\bD_\gamma}{Q_{\text{matter}}} \,,
\ee
implying that the dyonic 1-form symmetry group is
\be
\G1=\wh\bD_1=\wh{\frac{\bD_\gamma}{Q_{\text{matter}}}}\,.
\ee

\paragraph{Discrete Theta Term.}
The extra discrete theta angle term in the action of a $\cG_\gamma$ theory can be expressed as
\be
2\pi\int\cP_\sigma(w_2)\,,
\ee
where $w_2$ is the $\cZ$ valued degree-2 obstruction class for lifting $\cG$ gauge bundles to $G$ bundles and $\cP_\sigma(w_2)$ is an $\R/\Z$ valued degree-4 class obtained by applying a Pontryagin square operation on $w_2$. The Pontryagin square operation is specified by choosing a quadratic function \cite{Kapustin:2013qsa}
\be
\sigma:~\cZ\to\R/\Z \,,
\ee
refining the bi-homomorphism
\be
\eta:~\cZ\times\cZ\to\R/\Z
\ee
dual to the homomorphism $\gamma$:
\be
\eta(z,z')\equiv(\gamma(z))(z')\in U(1)\cong \R/\Z \,.
\ee
where we have used the fact that $\gamma(z)\in\wh\cZ$ which naturally contracts with $z'\in\cZ$ to provide a $U(1)$ element.

\begin{tech}[Quadratic Refinement]{}
Beginning with a bi-homomorphism, $\eta: \cZ \times \cZ \to \mathbb{R}/\Z$, a quadratic refinement $\sigma: \cZ \to \mathbb{R}/\Z$ is a quadratic function which obeys
\be
\eta(z_1,z_2) = \sigma(z_1 + z_2) - \sigma(z_1) - \sigma(z_2) \,.
\ee
\end{tech}

\ni Different choices of $\sigma$ define the same Pontryagin square operation on 4d spacetimes admitting spin structures.

\begin{example}[An Orthosymplectic Gauge Theory with Discrete Theta Angle]{}
We consider a theory similar to the theory considered in (\ref{osp}). We have a 4d gauge theory with gauge algebra
\be
\fg=\so(4n+2)\oplus\sp(m)\,.
\ee
The associated simply connected group is
\be
G=\Spin(4n+2)\times\text{Sp}(m)\,,
\ee
whose center is
\be
Z(G)=Z(\Spin(4n+2))\times Z(\text{Sp}(m))=\Z_4\times\Z_2\,.
\ee
We take the \textbf{gauge group} to be
\be
\cG=\text{SO}(4n+2)\times\text{Sp}(m)\,,
\ee
for which we have
\be
\cZ=\Z_2\,.
\ee
We additionally add matter fields in bifundamental representation (\ref{bif}) and turn on a discrete theta angle corresponding to the non-trivial homomorphism
\be
\gamma:~\Z_2\to\Z_2\,.
\ee
Let us compute the 1-form symmetry group. We have
\be
\cZ\times\wh Z(G)=\Z_2\times\Z_4\times\Z_2\,.
\ee
The group of unscreened line operators is constructed as follows via \eqref{eq:unscreenedwl}. Denote elements $z \in \cZ=\Z_2$ as $z \in \{0,1 \}$.  The surjective homomorphism $\wh i$ is
\be
\{ \{ (0,0), (2,0), (0,1), (2,1) \} \to \wh 0 \,, \{ (1,0), (3,0), (1,1), (3,1) \} \to \wh 1 \} \,.
\ee
We should construct all pairs $(z, \wt \gamma(z))$. Let's begin with $z=0 \in \Z_2$. We are looking for uplifts of $\wh 0$, which are $\{(0,0),(2,0),(0,1),(2,1)\} \in \wh Z(G) = \Z_4 \times \Z_2$. Similarly, beginning with $z=1\in\Z_2$, we are looking for uplifts of $\wh 1$, which are $\{ (1,0), (3,0), (1,1), (3,1) \} \in \wh Z(G) = \Z_4 \times \Z_2$. We thus obtain
\be
\ba
\bD_\gamma&=\{(0,0,0),(0,2,0),(1,1,0),(1,3,0),(0,0,1),(0,2,1),(1,1,1),(1,3,1)\}\,,\\
&=\Z_4\times\Z_2\subset\cZ\times\wh Z(G)=\Z_2\times\Z_4\times\Z_2\,.
\ea
\ee
Let us take the generator for $\Z_4$ factor of $\bD_\gamma$ to be $(1,1,1)$ and the generator for $\Z_2$ factor of $\bD_\gamma$ to be $(0,2,1)$. The charge provided by matter content is $(0,2,1)$ which implies that
\be
\bD_1=\Z_4\,,
\ee
and hence the 1-form symmetry group is
\be
\G1=\Z_4\,.
\ee
\end{example}

\section{Properties of Higher-Form Symmetries}\label{hfp}
In the previous sections, we introduced higher-form symmetries and discussed many examples of quantum field theories in various dimensions admitting such symmetries. We also described computations of higher-form symmetries in many general classes of gauge theories via screening. 

In this section, we describe various properties of higher-form symmetries. In particular, we will discuss how to couple a theory to backgrounds for such symmetries. This is intricately connected to 't Hooft anomalies, which are RG flow invariants and also characterize obstructions to the gauging of higher-form symmetries. These properties have been utilized for various physical applications following the appearance of \cite{Gaiotto:2014kfa}. See \cite{Aharony:2016jvv,Hsin:2016blu,Seiberg:2016rsg,Cordova:2017kue,Cordova:2017vab,Gaiotto:2017tne,Komargodski:2017keh,Gaiotto:2017yup,Benini:2017dus,Hsin:2018vcg,Hsin:2019gvb,Bhardwaj:2016clt,Bhardwaj:2020ymp,Lee:2020ojw,Bergman:2020ifi,Hsieh:2020jpj,Cordova:2018qvg,Gukov:2020btk,Das:2022auy,Iqbal:2021tui,Hofman:2017vwr,Apruzzi:2021nmk,GarciaEtxebarria:2019caf,Cordova:2019bsd,Closset:2020afy,vanBeest:2022fss,Eckhard:2019jgg,Bah:2020uev,Apruzzi:2020zot,Buican:2021xhs,Bhardwaj:2016dtk,Iqbal:2021rkn,Braun:2021sex,Closset:2021lhd,Lee:2021obi,Bah:2021brs,Beratto:2021xmn,Comi:2023lfm,Creutzig:2021ext,DelZotto:2022ras,Moradi:2023dan} for a sample of these works (in no particular order).

\subsection{Background Fields}
We begin with the discussion of background fields. The discussion is slightly different depending on whether the higher-form symmetry under discussion is discrete or continuous.

\paragraph{Revisiting Continuous 0-Form Symmetries.}
Recall that a background field\footnote{Often one adds an extra adjective `gauge' and refers to the `background field' as `background gauge field'. We will drop this adjective here in order to avoid confusion with the closely related notion of a gauge field. Throughout this text, a background field is a fixed, nondynamical, field on spacetime that is not summed in the path integral, while a gauge field is a dynamical field being summed over in the path integral.} for a continuous 0-form symmetry group $\G0$ is a connection $A$ on a $\G0$ bundle on spacetime, which locally can be described as a 1-form field on spacetime. Using the information of the connection, we can compute the field strength $F$, which is a closed 2-form field on spacetime.

In particular, for
\be
\G0= U(1)\,,
\ee
we have
\be
F=dA\,.
\ee
The small gauge transformations for the background field are
\be
A\to A+d\Lambda\,,
\ee
for 0-forms $\Lambda$ on spacetime. We also have large gauge transformations which shift the holonomies $\oint A$ by integers.

For a theory with $\G0$ 0-form symmetry, the correlation functions are functions of the background field $A$. One denotes the partition function in the presence of a background field $A$ as $Z[A]$.

The presence of a background field modifies the action by a coupling
\be
2\pi\int A \wedge j^{d-1}\,,
\ee
where $j^{d-1}$ is the Noether current for the $\G 0=U(1)$ $0$-form symmetry.

\paragraph{Continuous $p$-Form Symmetries.}
Since $p\ge1$-form symmetries are abelian, a continuous $p\ge1$-form symmetry group must be of the form
\be
\G p=U(1)^n,\qquad n>0\,.
\ee
We can thus restrict the discussion to a 
\be\label{ppU1}
\G p=U(1)\,,
\ee
$p$-form symmetry, without loss of generality. In this case, we have

\begin{note}[Background Field for Continuous $p$-Form Symmetry]{}
A background field for a $\G p=U(1)$ $p$-form symmetry is a $(p+1)$-form $B_{p+1}$ on spacetime, whose field strength
\be
F_{p+2}=dB_{p+1}\,,
\ee
is a closed $(p+2)$-form on spacetime.
\end{note}
The small gauge transformations for the background field are
\be
B_{p+1}\to B_{p+1}+d\Lambda_p\,,
\ee
for $p$-forms $\Lambda_p$ on spacetime. We also have large gauge transformations which shift by integers the holonomies $\oint B_{p+1}$ on $(p+1)$-dimensional closed submanifolds of spacetime.

In a theory with (\ref{ppU1}) symmetry, correlation functions are a function of the background field $B_{p+1}$. In particular, the partition function in the presence of $B_{p+1}$ is expressed as
\be
Z[B_{p+1}]\,.
\ee
The presence of a background field modifies the action by a coupling
\be
2\pi\int B_{p+1}\wedge j^{d-p-1}\,,
\ee
where $j^{d-p-1}$ is the Noether current for the $\G p=U(1)$ $p$-form symmetry.

\begin{example}[Maxwell Theory]{}
We return to the example of Maxwell theory introduced in example \ref{hfsM}. As discussed there, this theory admits an electric
\be
\G1=U(1)\,,
\ee
1-form symmetry, and a magnetic
\be
\G{d-3}=U(1)\,,
\ee
$(d-3)$-form symmetry. The background fields for these symmetries are
\be
B^e_2,\qquad B^m_{d-2}\,,
\ee
respectively. From the knowledge of the Noether currents, we see that in the presence of $B^e_2$, the action contains the term
\be\label{eb}
2\pi \int B^e_2\wedge \star F\,,
\ee
while in the presence of $B^m_{d-2}$, the action contains the term
\be\label{mb}
2\pi \int B^m_{d-2}\wedge F\,.
\ee
\end{example}

\paragraph{Discrete $p$-Form Symmetries.}
Now we consider the case of a discrete $p\ge0$-form symmetry group $\G p$. We need a discrete version of differential forms. As discussed in section \ref{hfgt}, this is provided by discrete cochains on spacetime.

\begin{note}[Background Field for Discrete $p$-Form Symmetry]{}
A background field for a discrete $\G p$ $p$-form symmetry is a $\G p$ valued $(p+1)$-cochain $B_{p+1}$ on spacetime.
\end{note}

\ni If background fields for other symmetries are not turned on, then $B_{p+1}$ is a cocycle, i.e.\ we have
\be
\delta B_{p+1}=0\,.
\ee
However, if background fields for other symmetries are turned on, then $B_{p+1}$ may not be a cocycle. This happens if $\G p$ participates in a non-trivial group extension or a higher-group symmetry. We discuss such aspects later.

The gauge transformations for the background field are
\be
B_{p+1}\to B_{p+1}+\delta\Lambda_p\,,
\ee
for $\G p$ valued $p$-cochains $\Lambda_p$ on spacetime.

\begin{example}[Discrete Gauge Theories]{}
Consider a discrete (higher-form) gauge theory of the form considered in section \ref{hfgt}. It has an electric $\G p$ $p$-form symmetry and a magnetic $\whG p$ $(d-p-1)$-form symmetry, both of which are associated to discrete versions of Noether currents. Using them, we can describe a coupling to background fields $B^e_{p+1}$ and $B^m_{d-p}$ as the following terms in the action
\be\label{dgtB}
2\pi\int\left(B^e_{p+1}\cup_\eta b_{d-p-1}+a_p\cup_\eta B^m_{d-p}\right)\,,
\ee
where recall that $a_p$ and $b_{d-p-1}$ are dynamical gauge fields.
\end{example}

\paragraph{Electric 1-Form Symmetry of Non-Abelian Gauge Theory}
Consider a non-abelian gauge theory with a connected gauge group $\cG$. As we have discussed above, its electric 1-form symmetry is a subgroup of the center of the gauge group
\be
\G1\subseteq Z(\cG)\,.
\ee
In the presence of a background $B^e_2$ for the 1-form symmetry is turned on, the $\cG$ gauge theory path integral sums over gauge fields living in all those bundles for the group
\be
\cG/\G1\,,
\ee
for which the obstruction of lifting to a $\cG$ bundle is
\be
w_2=B_2^e\,.
\ee

\begin{example}[Pure $SU(2)$ Yang-Mills Theory]{SU2B}
As a concrete example, consider pure $SU(2)$ Yang-Mills theory, which has
\be
\G1=Z(SU(2))=\Z_2\,,
\ee
1-form symmetry. In the presence of background $B_2^e$, the path integral sums over
\be
SU(2)/\Z_2=SO(3)\,,
\ee
gauge bundles with Stiefel-Whitney class $w_2$ fixed to be
\be
w_2=B_2^e\,.
\ee
\end{example}

\paragraph{Magnetic $(d-3)$-Form Symmetry of Non-Abelian Gauge Theory}
Recall that a non-abelian gauge theory with a connected gauge group $\cG$ has a magnetic $(d-3)$-form symmetry given by
\be
\G{d-3}=\wh\cZ\,,
\ee
where $\cZ$ is a subgroup of the center $Z(G)$ of the simply connected group $G$ in terms of which the gauge group $\cG$ is specified as
\be
\cG=G/\cZ\,.
\ee
Such a gauge theory has a $\cZ$-valued dynamical field $w_2$, namely the obstruction class for lifting $\cG$ bundles to $G$ bundles. In the presence of a $\wh\cZ$-valued background field $B^m_{d-2}$ for the magnetic symmetry, the action acquires a coupling
\be\label{Mba}
2\pi\int B^m_{d-2}\cup_\eta w_2\,,
\ee
where the cup product utilizes the natural bi-homomorphism
\be
\eta:~\wh\cZ\times\cZ\to\R/\Z\,.
\ee

\begin{example}[Pure $SO(3)$ Yang-Mills Theory]{SO3B}
As a concrete example, consider pure $SO(3)$ Yang-Mills theory, which has
\be
\G{d-3}=\Z_2\,.
\ee
In this case the background field turns on the coupling
\be
2\pi \int B^m_{d-2}\cup_\eta w_2=\pi\int B_{d-2}^m\cup w_2\,,
\ee
where $w_2$ is the second Stiefel-Whitney class, and on the RHS we have the standard cup product between two $\Z_2$-valued cochains.
\end{example}

\paragraph{Background Field as a Network of Topological Defects.}
For a discrete $\G p$ $p\ge0$-form symmetry, the background field $B_{p+1}$ can be described in terms of a network of codimension-$(p+1)$ topological defects comprising the $\G p$ symmetry. This uses a version of Poincaré duality.

\begin{tech}[Operators vs. Defects]{}
In these lecture notes we will use equivalently the terms \textbf{operator} and \textbf{defect}. This is customary in the literature on relativistic QFTs.

This is due to the following reason. A defect may be interpreted as introducing additional degrees of freedom into a theory that are localized along a submanifold of space, in turn modifying the Hilbert space. Now consider a $p$-dimensional operator $\cO$ inserted along a submanifold $M_p$ of spacetime. We may now choose a time direction such that $M_p$ does not entirely lie inside a time slice. In such a situation, the intersection of $M_p$ with the time slice may be viewed as a defect as it in general modifies the Hilbert space at that time slice. 
\end{tech}

\begin{figure}
\centering
\scalebox{1}{
\begin{tikzpicture}
\draw [thick](0,1.25) -- (0,-0.5) (0,-0.5) -- (-2.5,-3) (0,-0.5) -- (2.5,-3);
\draw [thick,red](-1.5,0) -- (0,-2) -- (1.5,0) --cycle;
\draw [thick](-2.5,-3) -- (2.5,-3);
\draw [thick,red](0,-2) -- (0,-4.5);
\end{tikzpicture}
}
\caption{Local pieces of two triangulations in $d=2$, shown in red and black, that are dual to each other. Note that 2-simplices (or faces) of one of the triangulations corresponds to 0-simplices (or vertices) of the other triangulation, in the sense that each face of a triangulation contains a single vertex of the other triangulation. Similarly, the 1-simplices (or edges) are in a one-to-one correspondence, in the sense that an edge of a triangulation intersects precisely one edge of the other triangulation.}
\label{dtri}
\end{figure}
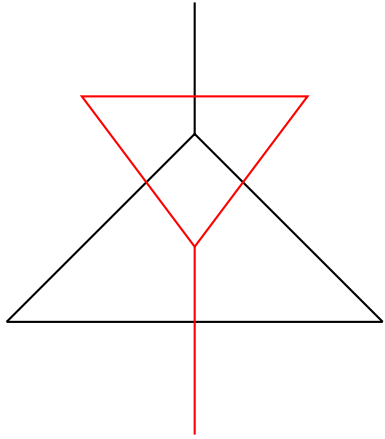

As discussed in section \ref{hfgt}, the cochain $B_{p+1}$ is defined in terms of a triangulation $\cT$ of spacetime. The topological defects live on a dual triangulation $\wh\cT$, which is defined as follows:
\bit
\item A vertex (or 0-simplex) of $\wh\cT$ corresponds to a $d$-simplex of $\cT$, and is placed inside the $d$-simplex.
\item An edge (or 1-simplex) $\wh s_1$ of $\wh\cT$ corresponds to a $(d-1)$-simplex $s_{d-1}$ of $\cT$, and connects the two vertices of $\wh\cT$ associated to the two $d$-simplices of $\cT$ sharing the $(d-1)$-simplex $s_{d-1}$. The edge $\wh s_1$ intersects once the $(d-1)$-simplex $s_{d-1}$.
\item Continuing in the above fashion, an $i$-simplex $\wh s_i$ of $\wh\cT$ corresponds to a $(d-i)$-simplex $s_{d-i}$ of $\cT$, intersecting it at one point. The boundary of $s_i$ is comprised of $(i-1)$-simplices of $\wh\cT$ associated to $(d-i+1)$-simplices of $\cT$ whose boundary contains $s_{d-i}$.
\eit
See figure \ref{dtri} for a depiction of the two triangulations in $d=2$.

Now, the network of topological defects is constructed as follows. For a $(p+1)$-simplex $s_{p+1}$ of $\cT$ we have
\be\label{PD}
B_{p+1}(s_{p+1})=g\in\G p\,.
\ee
Given this, insert the codimension-$(p+1)$ topological defect $U_g$ along the dual $(d-p-1)$-simplex $\wh s_{d-p-1}$ of $\wh\cT$.

The partition function $Z[B_{p+1}]$ in the presence of a background field is then the same as the correlation function of the corresponding network of topological defects. This is easy to demonstrate in a situation when we have a discrete Noether current $j_{d-p-1}$ (which may not always be the case), as in such a situation the topological codimension-$(p+1)$ operators constituting the $\G p$ $p$-form symmetry can be expressed as
\be
U_g=\int g(j_{d-p-1}),\qquad g\in\G p\,,
\ee
where $j_{d-p-1}$ which is a dynamical $\whG p$-valued $(d-p-1)$-cochain. As discussed above, in such a case, turning on a background field $B_{p+1}$ corresponds to adding to the action a coupling
\be
2\pi\int B_{p+1}\cup_\eta j_{d-p-1}\,.
\ee
By Poincaré duality, computing partition function in the presence of this term is precisely equivalent to computing the correlation function of a network of $U_g$ defects placed on the Poincaré dual codimension-$(p+1)$ chain to the cochain $B_{p+1}$.

\paragraph{Gauge Transformations as Rearrangements of Topological Defects.}
Just like background fields describe networks of topological defects on spacetime, gauge transformations of background fields describe deformations of the topological defects including local rearrangements of the network of topological defects. Examples will be discussed in the next subsection, see figure \ref{taaq}.

\subsection{'t Hooft Anomalies and SPT Phases}
\subsubsection{Basics of 't Hooft Anomalies}\label{basA}
\paragraph{Pure 't Hooft Anomalies.}
A pure 't Hooft anomaly of a $p$-form symmetry arises when both of the below points are true
\bit
\item Gauge transformations of background field $B_{p+1}$ do not leave the partition function invariant
\be
Z[B_{p+1}]\neq Z[B^g_{p+1}]\,,
\ee
where $B^g_{p+1}$ is gauge related to $B_{p+1}$, i.e.\
\be\label{Bg}
\ba
B^g_{p+1}&=B_{p+1}+d\Lambda_p,\qquad\text{$\G p$ continuous}\,,\\
B^g_{p+1}&=B_{p+1}+\delta\Lambda_p,\qquad\text{$\G p$ discrete}\,.
\ea
\ee
\item Even after adding a counter-term, which is a function of $B_p$, to the action, it is still impossible to have $Z[B_{p+1}]= Z[B^g_{p+1}]$.
\eit

\paragraph{Mixed 't Hooft Anomalies.}
We can also have situations, where we have
\be
Z[B_{p+1}]= Z[B^g_{p+1}]\,,
\ee
but if we also turn on a background for  another $q$-form symmetry, then the partition functions are not invariant under gauge transformations
\be
Z[B_{p+1},B_{q+1}]\neq Z[B^g_{p+1}, B_{q+1}]\,,
\ee
in such a way that cannot be fixed by adding a counterterm, which is now a function of both $B_{p+1}$ and $B_{q+1}$.
In such situation, we say that there is a mixed 't Hooft anomaly between the $p$-form and $q$-form symmetries.

There can also be mixed 't Hooft anomalies involving more than two symmetries, whose definition is a straightforward generalization of the above.

\begin{tech}[Other Types of Anomalies]{}
't Hooft anomalies are closely related to two other types of anomalies, but should not be confused with them:
\bit
\item {\bf ABJ Anomalies:} These arise when we have a transformation which is a symmetry of the classical Lagrangian, but fails to be a symmetry of the full quantum partition function (which also includes the measure of the path integral). Such a transformation is \textit{not} a global symmetry of the QFT. One often says that such a transformation is a symmetry of the classical theory, but is afflicted with an ABJ anomaly, due to which it is not a symmetry of the quantum-mechanical theory. In the context of 4d gauge theories with continuous gauge groups and continuous 0-form symmetries, ABJ anomalies are characterized by triangle diagrams having gauge currents on two vertices, and a global current on the third vertex.
\item {\bf Gauge Anomalies:} These arise when a gauge transformation of dynamical gauge fields does not leave its contribution to the path integral invariant. In such a situation, the resulting gauge theory is ill-defined. In the context of 4d gauge theories with continuous gauge groups, the triangle diagrams for these anomalies have gauge currents on all three vertices.
\item {\bf 't Hooft Anomalies:} In the context of 4d gauge theories with continuous gauge groups, the triangle diagrams for these anomalies have global currents on all three vertices.
\eit
\end{tech}

\begin{example}[Maxwell Theory]{AMT}
Consider $d$-dimensional pure $U(1)$ gauge theory in the presence of non-trivial backgrounds for both electric and magnetic higher-form symmetries, which involves adding the two terms (\ref{eb}) and (\ref{mb}) into the action. Performing a gauge transformation for the electric background
\be
B_2^e\to B_2^e+d\Lambda_1^e\,,
\ee
modifies the action
\be\label{Svar}
S\to S+2\pi\int\Lambda_1^e\wedge d\star F\,.
\ee
The equation of motion for $\star F$ is modified in the presence of magnetic background to be
\be
d\star F=-dB^m_{d-2}\,.
\ee
Substituting this into (\ref{Svar}), we see that the partition function is modified as
\be
Z\big[B_2^e+d\Lambda_1^e,B_{d-2}^m\big]=\exp\left(-2\pi i\int\Lambda_1^e\wedge dB^m_{d-2}\right)\times Z\big[B_2^e,B_{d-2}^m\big]\,.
\ee
Thus, we see that there is a mixed 't Hooft anomaly between the electric and magnetic higher-form symmetries of the Maxwell theory.

We can also see the anomaly if we interchange the role of electric and magnetic pieces in the above computation. Performing a magnetic gauge transformation
\be
B_{d-2}^m\to B_{d-2}^m+d\Lambda_{d-3}^m\,,
\ee
in the presence of an electric background leads to the following anomalous variation of the partition function
\be
Z\big[B_2^e,B_{d-2}^m+d\Lambda_{d-3}^m\big]=\exp\left(2\pi i\int dB^e_2\wedge\Lambda_{d-3}^m \right)\times Z\big[B_2^e,B_{d-2}^m\big]\,.
\ee
\end{example}

\begin{figure}
\centering
\scalebox{1.1}{
\begin{tikzpicture}
\draw [thick,blue](4.5,2) -- (4.5,-1);
\draw [thick,red](3,-1) -- (4.375,0.625) node (v1) {};
\draw [thick,red](4.625,0.875) node (v2) {} -- (5.5,2);
\node[red] at (5.5,2.5) {$U_g$};
\node[blue] at (4.5,-1.5) {$U_{g'}$};
\node at (6.5,0.5) {$\neq$};
\begin{scope}[shift={(4.5,0)}]
\draw [thick,red](3,-1) -- (5.5,2);
\draw [thick,blue](3.5,2) -- (3.5,-0.125);
\draw [thick,blue](3.5,-0.625) -- (3.5,-1);
\node[blue] at (3.5,-1.5) {$U_{g'}$};
\node[red] at (5.5,2.5) {$U_g$};
\end{scope}
\draw [thick,blue](5,-3.5) -- (1.5,-7) (2.5,-6) -- (3.5,-7) (3.5,-5) -- (5.5,-7);
\node at (6.5,-5) {$\neq$};
\begin{scope}[shift={(6,-0.5)}]
\draw [thick,blue](5,-3) -- (1.5,-6.5) (4.5,-5.5) -- (3.5,-6.5) (3.5,-4.5) -- (5.5,-6.5);
\end{scope}
\node[blue] at (3.5,-7.5) {$U_{g'}$};
\node[blue] at (5.5,-7.5) {$U_{g''}$};
\node[blue] at (9.5,-7.5) {$U_{g'}$};
\node[blue] at (11.5,-7.5) {$U_{g''}$};
\node[blue] at (7.5,-7.5) {$U_{g}$};
\node[blue] at (1.5,-7.5) {$U_{g}$};
\node[blue] at (2.691,-5.2559) {$U_{gg'}$};
\node[blue] at (10.4439,-5.2481) {$U_{g'g''}$};
\node[blue] at (5,-3) {$U_{gg'g''}$};
\node[blue] at (11,-3) {$U_{gg'g''}$};
\end{tikzpicture}
}
\caption{Two examples of 't Hooft anomalies. In the top case, $U_{g'}$ is charged under $U_g$, as moving $U_g$ across $U_{g'}$ changes correlation functions. In the bottom case, one may analogously say that the junction between $U_{gg'}$ and $U_{g''}$ is charged under $U_{g'}$.}
\label{taaq}
\end{figure}
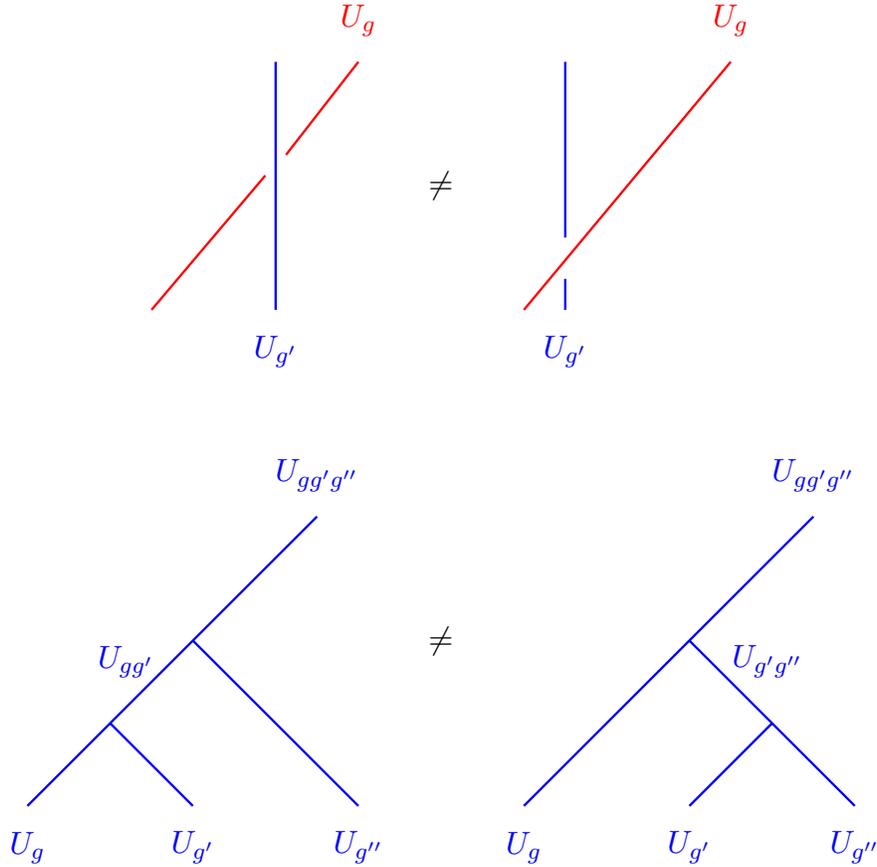

\paragraph{Description in terms of Topological Operators.}
Consider now discrete symmetries. Above we discussed that gauge transformations of background fields are equivalent to deformations of topological defects generating these symmetries. Thus, a 't Hooft anomaly arises when such deformations are not topological and change the correlation functions. This is only possible when the topological defects or their junctions cross each other. See figure \ref{taaq}. Equivalently, we have the statement

\begin{state}['t Hooft Anomalies as Topological Defects Carrying Charges]{}
A 't Hooft anomaly for a set of discrete symmetries arises when topological defects for these symmetries, or junctions between such topological defects, carry non-trivial charges under the symmetry. See figure \ref{taaq}.
\end{state}

\ni Addition of counterterms corresponds to redefining the topological operators and their junctions.

\begin{example}[Discrete Gauge Theories]{DHFGT}
Consider a discrete (higher-form) gauge theory with backgrounds for both electric and magnetic symmetries turned on as in (\ref{dgtB}). Performing a discrete version of the computation done in example \ref{AMT}, we find a mixed 't Hooft anomaly between the two symmetries. Performing a gauge transformation
\be
B^e_{p+1}\to B^e_{p+1}+\delta\Lambda^e_p\,,
\ee
in the presence of $B^m_{d-p}$ modifies the partition function as
\be
Z\big[B^e_{p+1}+\delta\Lambda^e_p,B^m_{d-p}\big]=\exp\left(2\pi i \, s(p)\int\Lambda^e_p\cup_\eta B^m_{d-p}\right)\times Z\big[B_{p+1}^e,B_{d-p}^m\big]\,,
\ee
where we have defined 
\be
s(n):=(-1)^n,\qquad n\in\Z\,.
\ee
Similarly, performing a gauge transformation
\be\label{mgau}
B^m_{d-p}\to B^m_{d-p}+\delta\Lambda^m_{d-p-1}\,,
\ee
in the presence of $B^e_p$ modifies the partition function as
\be
Z\big[B^e_{p+1},B^m_{d-p}+\delta\Lambda^m_{d-p-1}\big]=\exp\left(-2\pi i\int B^e_{p+1}\cup_\eta\Lambda^m_{d-p-1}\right)\times Z\big[B_{p+1}^e,B_{d-p}^m\big]\,.
\ee
This 't Hooft anomaly has to do with the fact that the topological operators constituting the $\G p$ electric $p$-form symmetry are charged under the $\whG p$ magnetic $(d-p-1)$-form symmetry, and vice versa. The gauge transformation (\ref{mgau}) corresponds to moving the topological operators for the magnetic symmetry such that they sweep out a $(p+1)$-chain Poincaré dual to the cochain $\Lambda^m_{d-p-1}$. Due to the presence of background $B_{p+1}^e$, we have a network of topological operators for the electric symmetry inserted inside spacetime. Thus, under the above deformation, the magnetic operators cross the electric ones generating a phase described by intersections between the chain occupied by the electric operators with the chain along which the magnetic operators are deformed. By Poincaré duality, the product of these phases is computed as
\be
\exp\left(-2\pi i\int B^e_{p+1}\cup_\eta\Lambda^m_{d-p-1}\right)\,.
\ee
which is the anomalous variation of the partition function discussed above.
\end{example}

\paragraph{'t Hooft Anomalies of Discrete 0-Form Symmetries in 2d.}
Using the above description of the 't Hooft anomaly in terms of movements of topological defects, we can completely characterize possible 't Hooft anomalies of 0-form symmetries in $d=2$.

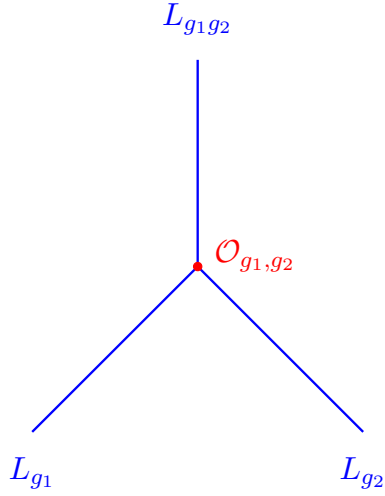
\begin{figure}
\centering
\scalebox{1.1}{
\begin{tikzpicture}
\draw [thick,blue](3.5,-5) -- (1.5,-7)  (3.5,-5) -- (5.5,-7);
\node[blue] at (5.5,-7.5) {$L_{g_2}$};
\node[blue] at (1.5,-7.5) {$L_{g_1}$};
\node[blue] at (3.5,-2) {$L_{g_1g_2}$};
\draw [thick,blue](3.5,-2.5) -- (3.5,-5);
\draw [red,fill=red] (3.5,-5) ellipse (0.05 and 0.05);
\node[red] at (4.1961,-4.8714) {$\cO_{g_1,g_2}$};
\end{tikzpicture}
}
\caption{A configuration of topological local and line operators associated to a 0-form symmetry in 2d.}
\label{LOg}
\end{figure}

Consider a 0-form symmetry described by a discrete group $\G0$. That is, we have topological line operators
\be
L_g,\qquad g\in\G0\,.
\ee
Additionally choose topological local operators
\be\label{jlo}
\cO_{g_1,g_2}\,,
\ee
at a junction where line operators $L_{g_1}$ and $L_{g_2}$ come together and transform into the line operator $L_{g_1g_2}$. See figure \ref{LOg}.

\begin{figure}
\centering
\scalebox{1.1}{
\begin{tikzpicture}
\draw [thick,blue](5,-3.5) -- (1.5,-7) (2.5,-6) -- (3.5,-7) (3.5,-5) -- (5.5,-7);
\begin{scope}[shift={(8,-0.5)}]
\draw [thick,blue](5,-3) -- (1.5,-6.5) (4.5,-5.5) -- (3.5,-6.5) (3.5,-4.5) -- (5.5,-6.5);
\end{scope}
\node[blue] at (3.5,-7.5) {$L_{g_2}$};
\node[blue] at (5.5,-7.5) {$L_{g_3}$};
\node[blue] at (11.5,-7.5) {$L_{g_2}$};
\node[blue] at (13.5,-7.5) {$L_{g_3}$};
\node[blue] at (9.5,-7.5) {$L_{g_1}$};
\node[blue] at (1.5,-7.5) {$L_{g_1}$};
\node[blue] at (2.5385,-5.1991) {$L_{g_1g_2}$};
\node[blue] at (12.4439,-5.2481) {$L_{g_2g_3}$};
\node[blue] at (5,-3) {$L_{g_1g_2g_3}$};
\node[blue] at (13,-3) {$L_{g_1g_2g_3}$};
\draw [red,fill=red] (3.5,-5) ellipse (0.05 and 0.05);
\node[red] at (1.8479,-5.7795) {$\cO_{g_1,g_2}$};
\draw [red,fill=red] (2.5,-6) ellipse (0.05 and 0.05);
\node[red] at (4.5,-5) {$\cO_{g_1g_2,g_3}$};
\draw [red,fill=red] (12.5,-6) ellipse (0.05 and 0.05);
\draw [red,fill=red] (11.5,-5) ellipse (0.05 and 0.05);
\node[red] at (13.1921,-5.8132) {$\cO_{g_2,g_3}$};
\node[red] at (10.7203,-4.7259) {$\cO_{g_1,g_2g_3}$};
\node at (6.5,-5.5) {=};
\node at (8.25,-5.5) {$\omega(g_1,g_2,g_3)~\times$};
\end{tikzpicture}
}
\caption{Performing an F-move changes a correlator by a 3-group-cochain $\omega$.}
\label{Asso}
\end{figure}
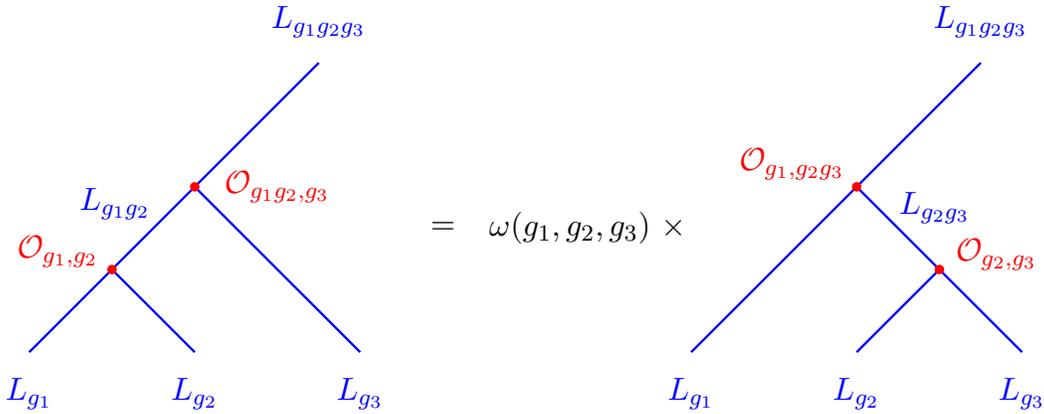

Consider a movement of topological lines shown in figure \ref{Asso}, which is sometimes referred to as an \textbf{associator} or as an \textbf{F-move}. Such a move might modify the correlation function by a complex factor
\be
\omega(g_1,g_2,g_3)\in\bC^\times\,,
\ee
which characterizes the 't Hooft anomaly of the $\G0$ 0-form symmetry once counterterms have been taken into account. These numbers can be collected into a function
\be
\omega:~(\G0)^3\to\bC^\times\,,
\ee
Such a function is known as a $\bC^\times$-valued 3-group-cochain on $\G0$.

\begin{note}[Group-Cochains]{}
Let $M$ be a group. An $M$-valued $p$-group-cochain $\alpha_p$ on $\G0$ is a function
\be
\alpha:~(\G 0)^p\to M\,.
\ee
\end{note}

\begin{figure}
\centering
\scalebox{1}{
\begin{tikzpicture}
\draw [thick,blue](4.75,-3.75) -- (1.5,-7) (2.25,-6.25) -- (3,-7) (3.75,-4.75) -- (6,-7);
\node[blue] at (3,-7.5) {$L_{g_2}$};
\node[blue] at (4.5,-7.5) {$L_{g_3}$};
\node[blue] at (1.5,-7.5) {$L_{g_1}$};
\node[blue] at (4.75,-3.25) {$L_{g_1g_2g_3g_4}$};
\draw [red,fill=red] (3.75,-4.75) ellipse (0.05 and 0.05);
\draw [red,fill=red] (2.25,-6.25) ellipse (0.05 and 0.05);
\node[blue] at (6.25,-7.5) {$L_{g_4}$};
\draw [thick,blue](3,-5.5) -- (4.5,-7);
\draw [red,fill=red] (3,-5.5) ellipse (0.05 and 0.05);
\draw [thick,-stealth](3.5,-8) -- (1.5,-9.5);
\begin{scope}[shift={(-4.5,-6.5)}]
\draw [thick,blue](4.75,-3.75) -- (1.5,-7) (3.75,-6.25) -- (3,-7) (3.75,-4.75) -- (6,-7);
\node[blue] at (3,-7.5) {$L_{g_2}$};
\node[blue] at (4.5,-7.5) {$L_{g_3}$};
\node[blue] at (1.5,-7.5) {$L_{g_1}$};
\node[blue] at (4.75,-3.25) {$L_{g_1g_2g_3g_4}$};
\draw [red,fill=red] (3.75,-4.75) ellipse (0.05 and 0.05);
\draw [red,fill=red] (3.75,-6.25) ellipse (0.05 and 0.05);
\node[blue] at (6.25,-7.5) {$L_{g_4}$};
\draw [thick,blue](3,-5.5) -- (4.5,-7);
\draw [red,fill=red] (3,-5.5) ellipse (0.05 and 0.05);
\end{scope}
\draw [thick,-stealth](-0.5,-14.5) -- (-0.5,-16);
\begin{scope}[shift={(-4.5,-13.5)}]
\draw [thick,blue](4.75,-3.75) -- (1.5,-7) (4.5,-5.5) -- (3,-7) (3.75,-4.75) -- (6,-7);
\node[blue] at (3,-7.5) {$L_{g_2}$};
\node[blue] at (4.5,-7.5) {$L_{g_3}$};
\node[blue] at (1.5,-7.5) {$L_{g_1}$};
\node[blue] at (4.75,-3.25) {$L_{g_1g_2g_3g_4}$};
\draw [red,fill=red] (3.75,-4.75) ellipse (0.05 and 0.05);
\draw [red,fill=red] (4.5,-5.5) ellipse (0.05 and 0.05);
\node[blue] at (6.25,-7.5) {$L_{g_4}$};
\draw [thick,blue](3.75,-6.25) -- (4.5,-7);
\draw [red,fill=red] (3.75,-6.25) ellipse (0.05 and 0.05);
\end{scope}
\draw [thick,-stealth](4.75,-8) -- (6.75,-9.5);
\begin{scope}[shift={(3.5,-6.5)}]
\draw [thick,blue](4.75,-3.75) -- (1.5,-7) (2.25,-6.25) -- (3,-7) (3.75,-4.75) -- (6,-7);
\node[blue] at (3,-7.5) {$L_{g_2}$};
\node[blue] at (4.5,-7.5) {$L_{g_3}$};
\node[blue] at (1.5,-7.5) {$L_{g_1}$};
\node[blue] at (4.75,-3.25) {$L_{g_1g_2g_3g_4}$};
\draw [red,fill=red] (3.75,-4.75) ellipse (0.05 and 0.05);
\draw [red,fill=red] (2.25,-6.25) ellipse (0.05 and 0.05);
\node[blue] at (6.25,-7.5) {$L_{g_4}$};
\draw [thick,blue](5.25,-6.25) -- (4.5,-7);
\draw [red,fill=red] (5.25,-6.25) ellipse (0.05 and 0.05);
\end{scope}
\draw [thick,-stealth](2.5,-19) -- (4.25,-19);
\draw [thick,-stealth](7.5,-14.5) -- (7.5,-16);
\begin{scope}[shift={(3.5,-13.5)}]
\draw [thick,blue](4.75,-3.75) -- (1.5,-7) (4.5,-5.5) -- (3,-7) (3.75,-4.75) -- (6,-7);
\node[blue] at (3,-7.5) {$L_{g_2}$};
\node[blue] at (4.5,-7.5) {$L_{g_3}$};
\node[blue] at (1.5,-7.5) {$L_{g_1}$};
\node[blue] at (4.75,-3.25) {$L_{g_1g_2g_3g_4}$};
\draw [red,fill=red] (3.75,-4.75) ellipse (0.05 and 0.05);
\draw [red,fill=red] (4.5,-5.5) ellipse (0.05 and 0.05);
\node[blue] at (6.25,-7.5) {$L_{g_4}$};
\draw [thick,blue](5.25,-6.25) -- (4.5,-7);
\draw [red,fill=red] (5.25,-6.25) ellipse (0.05 and 0.05);
\end{scope}
\end{tikzpicture}
}
\caption{By performing sequential F-moves, we can go between the same initial and final configurations in two different ways. Equating the $\omega$ factors produced in the two ways leads to equation (\ref{Pentaeq})}
\label{Penta}
\end{figure}
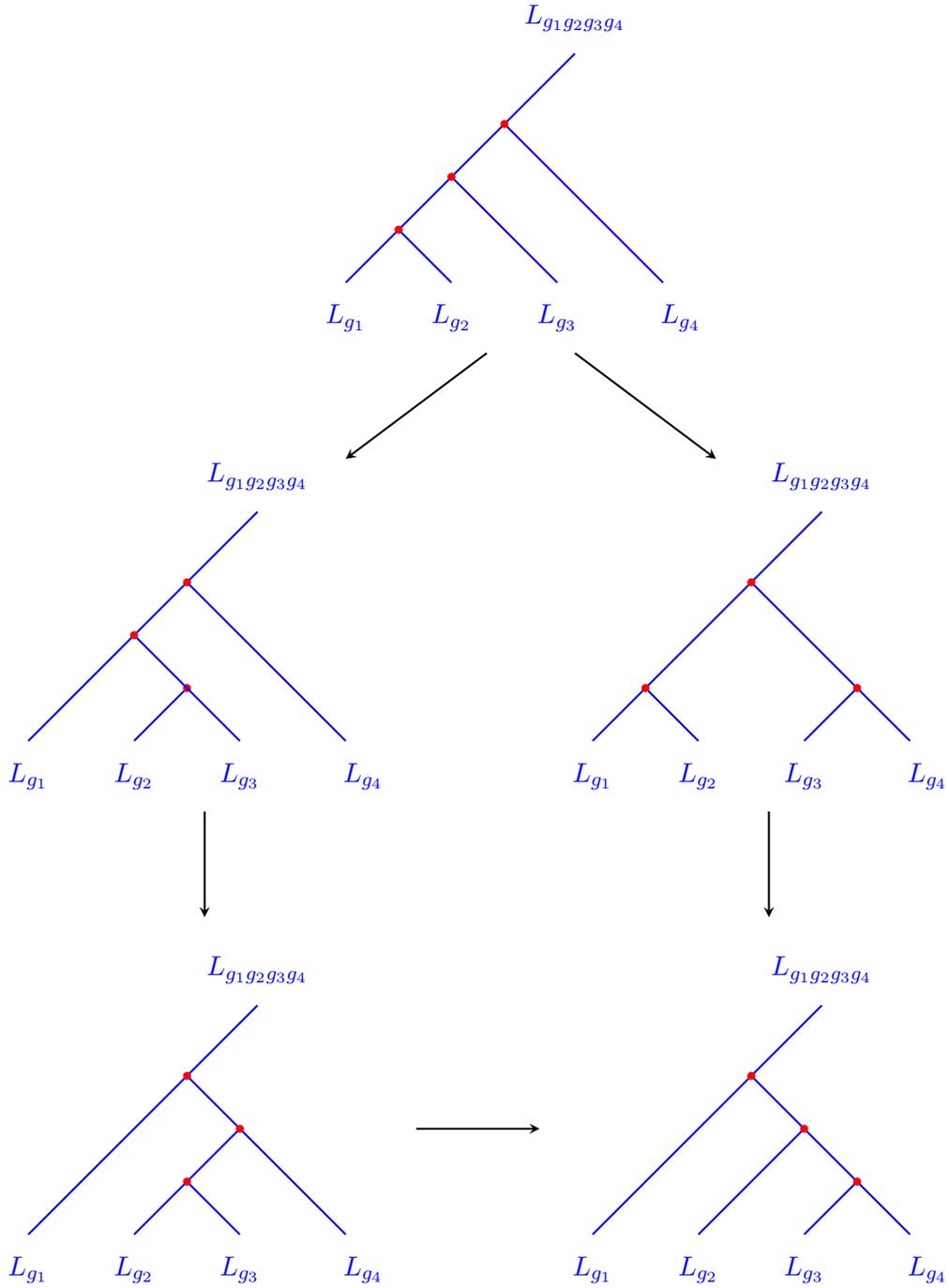
There is a consistency condition, known as the \textbf{pentagon identity}, on $\omega$ arising by equating the two sets of moves shown in figure \ref{Penta}
\be\label{Pentaeq}
\delta\omega(g_1,g_2,g_3,g_4)=\frac{\omega(g_2,g_3,g_4)\omega(g_1,g_2g_3,g_4)\omega(g_1,g_2,g_3)}{\omega(g_1g_2,g_3,g_4)\omega(g_1,g_2,g_3g_4)}=1\,,
\ee
which means that $\omega$ is a closed 3-group-cochain on $\G0$.

\begin{note}[Differential on Group-Cochains and Group Cohomology]{}
The differential $\delta$ maps $\alpha_p$ to an $M$-valued $(p+1)$-group-cochain on $\G0$
\be
\ba
\delta\alpha_p(g_1,g_2,\cdots,g_{p+1})=&\alpha_p(g_2,g_3,\cdots,g_{p+1})\alpha_p^{s(p+1)}(g_1,g_2,\cdots,g_p)\\
&\prod_{i=1}^p\alpha_p^{s(i)}(g_1,g_2,\cdots,g_{i-1},g_ig_{i+1},g_{i+2},g_{i+3}\cdots,g_{p+1})\,,
\ea
\ee
where $s(n)=(-1)^n$, i.e.\ we take $\alpha_p$ for even $i$ and $\alpha_p^{-1}$ for odd $i$.

The $p$-th cohomology group obtained using this differential is denoted
\be
H^p(\G0,M)\,,
\ee
and is referred to as the $p$-th \textbf{group cohomology} of $\G0$ valued in $M$.
\end{note}

Addition of counter-terms corresponds to changing the junction local operators as
\be
\cO(g_1,g_2)\to\beta(g_1,g_2)\cO(g_1,g_2),\qquad \beta(g_1,g_2)\in\bC^\times\,,
\ee
which modifies the associator as
\be
\omega\to\omega\times\delta\beta\,,
\ee
as the reader can check. Thus, only the cohomology class $[\omega]$ of $\omega$ determines a 't Hooft anomaly.

In conclusion, 't Hooft anomalies for a discrete 0-form symmetry group $\G0$ in 2d are characterized by elements of its third group cohomology valued in $\bC^\times$
\be
[\omega]\in H^3(\G0,\bC^\times)\,.
\ee
For later purposes let us note that
\be
H^p(\G0,\bC^\times)=H^p(\G0,U(1))\,,
\ee
so $[\omega]$ can be regarded as an element of $H^3(\G0,U(1))$.

\subsubsection{SPT Phases}\label{SPT}
't Hooft anomalies are intimately tied to the notion of SPT phases, which are very important in various diverse areas of theoretical physics. Let us begin with the definition of a symmetry protected topological (SPT) phase as often used in the hep-th literature

\begin{note}[SPT Phase]{}
A $d$-dimensional SPT phase protected by a set of $p\ge0$-form symmetries
\be
\cS=\{\G{p_1},\G{p_2},\cdots,\G{p_n}\}\,,
\ee
is an invertible $d$-dimensional topological quantum field theory (TQFT) admitting the symmetries $\cS$, such that its partition functions are trivial when all background fields are turned off
\be
Z[B_{p_1+1}=0,B_{p_2+1}=0,\cdots,B_{p_n+1}=0]=1\,.
\ee
For general background fields, the partition functions are phase factors
\be
Z[B_{p_1+1},B_{p_2+1},\cdots,B_{p_n+1}]\in U(1)\,.
\ee
\end{note}

\paragraph{Faithful vs. Non-Faithful Symmetries. }
Since the partition functions without symmetry backgrounds are all trivial, it means that an SPT phase is the same as a trivial theory, when symmetry is not taken into account. The only operators in a trivial theory are identity operators
\be
D_q^{\id}\,,
\ee
of general dimension $0\le q\le d-1$. Thus the topological operators constituting the symmetries $\G{p_i}$ must all be the same as the identity operator
\be
U_g=D_{d-p_i-1}^{\id},\qquad\forall~g\in\G{p_i},~\forall~i\,.
\ee
Thus, the $p$-form symmetries of an SPT phase are quite different from the $p$-form symmetries we have been discussing throughout the paper, in which we have been implicitly assuming that no two topological operators are the same
\be
U_g\neq U_{g'}\,,\qquad \text{if }g\neq g'\,.
\ee
We say that a symmetry is \textbf{non-faithful} if the corresponding topological operator is the same as the identity operator. In this sense, in this text we have been studying completely faithful symmetry groups, in which only the identity symmetry is non-faithful. On the other hand, the symmetry groups involved in an SPT phase are completely non-faithful, in the sense that all symmetries are non-faithful. In general, one can consider groups in which only some proper subgroups are non-faithful.

Let us emphasize that the faithfulness of a symmetry is not an RG invariant notion. Beginning with a $p$-form symmetry $\G p$ that is completely faithful in the UV, one may obtain an IR theory in which $\G p$ is completely non-faithful. This is illustrated with an example below.

\begin{example}[Faithful to Non-Faithful: 4d $\cN=1$ $SU(2)$ Super-Yang-Mills]{}
Consider minimally supersymmetric pure Yang-Mills theory in 4d with gauge group
\be
\cG=SU(2)\,.
\ee
All matter fields in this theory transform in the adjoint representation of $SU(2)$, and thus the theory has an electric 1-form symmetry
\be\label{f1s}
\G1=Z(\cG)=\Z_2\,.
\ee
This symmetry acts non-trivially on the Wilson line operator
\be
\cW_{j=\half}\,,
\ee
transforming in spin-1/2 representation of $SU(2)$. Since the identity operator must act trivially on all operators, we conclude that this $\Z_2$ 1-form symmetry must be completely faithful.

In the IR, the theory admits two degenerate vacua
\be
v_s,\qquad s\in\{0,1\}\,,
\ee
and the IR effective theory in each vacuum is an SPT phase for the $\Z_2$ 1-form symmetry. More precisely, the partition functions for these SPTs are
\be
Z[B_2]=\exp\left(\pi is\int \frac{\cP(B_2)}2\right),\qquad s\in\{0,1\}\,,
\ee
where $\cP$ is the Pontryagin square operation discussed around equation (\ref{Pw}). Thus the (\ref{f1s}) 1-form symmetry is completely non-faithful in the IR.
\end{example}

\paragraph{Information of an SPT Phase.}
According to the above discussion, the symmetry generating topological defects are all trivial for an SPT phase. This then raises the question of where the information regarding non-trivial partition functions is stored. This information is stored entirely in the junctions of these topological operators. 

Let us explain this using the example of SPT phases for a discrete $\G0$ 0-form symmetry group in $d=2$. We have topological line operators
\be
L_g=D_1^\id,\qquad \forall~g\in\G0\,.
\ee
As in (\ref{jlo}), we have to choose topological local operators $\cO_{g_1,g_2}$ at a junction of $L_{g_1}$, $L_{g_2}$ and $L_{g_1g_2}$. As the lines are all identity, the junction local operators are chosen from the set of genuine local operators of the trivial theory, which is $\bC^\times$
\be
\cO_{g_1,g_2}=\beta(g_1,g_2)\in\bC^\times\,.
\ee
Thus, the choice of junction local operators is encoded in a $\bC^\times$-valued 2-group-cochain $\beta$ on $\G0$. Demanding that the 0-form symmetry is non-anomalous implies
\be
\delta\beta=\omega=1\,,
\ee
where $\omega$ was described in figure \ref{Asso}. In other words, $\beta$ is a cocycle.

It is straightforward to see that for a given $\beta$, the partition function in the presence of a background $A_1$ for $\G0$ can be expressed as
\be\label{2dSPT}
Z[A_1]=\int A_1^*\beta\,.
\ee

\begin{tech}[Pullback of Group-Cochains]{}
Given a background field $A_1$ for a $\G0$ 0-form symmetry and an $M$-valued $p$-group-cochain $\alpha_p$, we can define an $M$-valued $p$-cochain
\be
A_1^*\alpha_p\,,
\ee
on spacetime, which is referred to as the pullback of $\alpha_p$ under the background $A_1$. It is defined as
\be
A_1^*\alpha_p(v_0,v_1,\cdots,v_p):=\alpha_p\big(A_1(v_0,v_1),A_1(v_1,v_2),\cdots,A_1(v_{p-1},v_p)\big)\,.
\ee
\end{tech}
Note that replacing
\be
\beta\to\beta+\delta\lambda\,,
\ee
for a 1-group-cochain $\lambda$, does not change $Z[A_1]$ on a closed spacetime manifold. Thus, we obtain the following statement

\begin{state}[2d SPT Phases for Discrete $\G0$ 0-Form Symmetry]{}
SPT phases $[\beta]$ for a discrete 0-form symmetry group $\G0$ in spacetime dimension $d=2$ are characterized by elements of the second group cohomology with coefficients in $U(1)$
\be
[\beta]\in H^2\left(\G0,U(1)\right)\,.
\ee
\end{state}

\ni We can easily generalize (\ref{2dSPT}) to obtain

\begin{state}[Group Cohomology Type SPT Phases Protected by $\G0$]{}
A class of $d$-dimensional SPT phases for a discrete $\G0$ 0-form symmetry is given by elements
\be
[\alpha_d]\in H^d\left(\G0,U(1)\right)\,.
\ee
The partition functions of the SPT phase can be expressed as
\be
Z[A_1]=\int A_1^*\alpha_d\,.
\ee
\end{state}

\subsubsection{Anomaly Theory and Anomaly Polynomial}
For the discussion of the anomaly theory, we consider $(d+1)$-dimensional SPT phases.

\paragraph{SPT Phases on Open Manifold.}
Note that SPT phases do not have a 't Hooft anomaly, i.e.\ we have
\be\label{nta}
Z[B_{p_i+1}+\delta\Lambda_{p_i}]=Z[B_{p_i+1}]\,,
\ee
for every closed $(d+1)$-dimensional spacetime manifold $M_{d+1}$.
However, if we consider the SPT phase on an open spacetime $M_{d+1}$ with a $d$-dimensional boundary
\be
M_d=\partial M_{d+1}\,,
\ee
then (\ref{nta}) is no longer satisfied but we have
\be\label{sba}
\frac{Z[B_{p_i+1}+\delta\Lambda_{p_i}]}{Z[B_{p_i+1}]}=Z_\partial[B_{p_i+1},\Lambda_{p_i}]\,,
\ee
where $Z_\partial[B_{p_i+1},\Lambda_{p_i}]$ is localized entirely on the boundary $M_d$ and is a function of both the background fields $B_{p_i+1}$ and the gauge transformation parameters $\Lambda_{p_i}$ along $M_d$. We can interpret $Z_\partial$ as a 't Hooft anomaly arising in trying to define the SPT phase on an open manifold. 

\begin{figure}
\centering
\scalebox{1.1}{
\begin{tikzpicture}[x=0.75pt,y=0.75pt,yscale=-1,xscale=1]
\draw [line width=0.75]    (185.22,44) -- (185.22,113.1)(188.22,44) -- (188.22,113.1) ;
\draw    (135.72,44) .. controls (125.93,55.75) and (48.32,27.4) .. (22.32,47.4) .. controls (-3.68,67.4) and (69.17,117.82) .. (79.44,130.14) .. controls (89.71,142.46) and (118.58,125.95) .. (135.72,113.1) ;
\draw    (135.72,44) -- (135.72,113.1) ;
\draw    (417.67,43) .. controls (407.88,54.75) and (330.27,26.4) .. (304.27,46.4) .. controls (278.27,66.4) and (351.12,116.82) .. (361.39,129.14) .. controls (371.66,141.46) and (400.53,124.95) .. (417.67,112.1) ;
\draw [line width=0.75]    (419.17,43) -- (419.17,112.1)(416.17,43) -- (416.17,112.1) ;

\draw (14.05,13) node [anchor=north west][inner sep=0.75pt]   [align=left] {$\displaystyle \mathcal{M}_{d+1}$};
\draw (178.05,16) node [anchor=north west][inner sep=0.75pt]   [align=left] {$\displaystyle \mathcal{M}_{d}$};
\draw (177.7,66) node [anchor=north west][inner sep=0.75pt]   [align=left] {};
\draw (51,71) node [anchor=north west][inner sep=0.75pt]   [align=left] {$\displaystyle \mathcal{I}_{d+1}$};
\draw (244.7,66) node [anchor=north west][inner sep=0.75pt]   [align=left] {=};
\draw (296,12) node [anchor=north west][inner sep=0.75pt]   [align=left] {$\displaystyle \mathcal{M}_{d+1}$};
\draw (325.95,69) node [anchor=north west][inner sep=0.75pt]   [align=left] {$\displaystyle \mathcal{I}_{d+1}$};
\draw (184,118) node [anchor=north west][inner sep=0.75pt]   [align=left] {$\displaystyle \mathfrak{T}$};
\draw (420,69) node [anchor=north west][inner sep=0.75pt]   [align=left] {$\displaystyle \mathfrak{T}$};
\draw (54,137) node [anchor=north west][inner sep=0.75pt]   [align=left] {bulk theory only};
\draw (280,136) node [anchor=north west][inner sep=0.75pt]   [align=left] {bulk and boundary theory};
\end{tikzpicture}
}
\caption{The figure shows a $(d+1)$-dimensional SPT phase $\cI_{d+1}$ on an open $(d+1)$-dimensional manifold $\cM_{d+1}$ with boundary $\cM_d$ and a $d$-dimensional theory $\fT$ whose anomaly theory is $\cI_{d+1}$. Putting $\fT$ along $\cM_d$ as shown on the RHS makes the combined system anomaly free.}
\label{SPTb}
\end{figure}
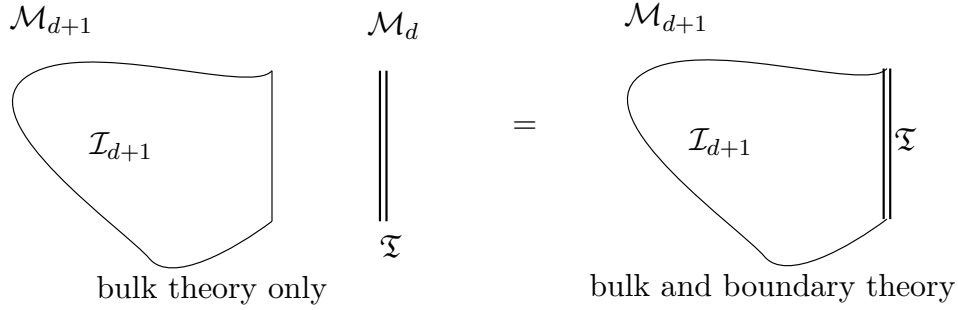

\paragraph{Anomaly Theory.}
We can now place a $d$-dimensional theory $\fT$ along $M_d$, which has the same symmetries as that of the SPT phase but carries a 't Hooft anomaly 
\be
\cI_d[B_{p_i+1},\Lambda_{p_i}]=Z^{-1}_\partial[B_{p_i+1},\Lambda_{p_i}]\,.
\ee
The combined hybrid $(d,d+1)$-dimensional system comprising of the $(d+1)$-dimensional SPT and the $d$-dimensional theory $\fT$ is anomaly free. See figure \ref{SPTb}.

\begin{note}[Anomaly Theory]{}
Given a 't Hooft anomaly
\be
\cI_d[B_{p_i+1},\Lambda_{p_i}]\,,
\ee
for $d$-dimensional theories,
the associated anomaly theory $Z[B_{p_i+1}]$ is a $(d+1)$-dimensional SPT phase for which
\be
Z_\partial[B_{p_i+1},\Lambda_{p_i}]=\cI_d^{-1}[B_{p_i+1},\Lambda_{p_i}]\,.
\ee
\end{note}

In this text, we will refer to an anomaly theory as $\cI_{d+1}$ and its partition function as
\be
\cI_{d+1}[B_{p_i+1}]\,.
\ee

\paragraph{Anomaly Polynomial.}
For a continuous symmetry, the effective action $S^{\text{eff}}_{d+1}$ of the anomaly theory $\cI_{d+1}$ defined via
\be
\cI_{d+1}[B_{p_i+1}]=\exp(2\pi iS^{\text{eff}}_{d+1})\,,
\ee
is not closed, and using it we can define a $(d+2)$-dimensional anomaly polynomial
\be
\cI_{d+2}=dS^{\text{eff}}_{d+1}\,,
\ee
as is familiar from the study of continuous 0-form symmetries.

The 't Hooft anomaly $\cI_d$, the anomaly theory $\cI_{d+1}$ and the anomaly polynomial $\cI_{d+2}$ are related to each other by a descent procedure.

\paragraph{Examples.}
Above we discussed many examples with 't Hooft anomalies. Let us discuss the corresponding anomaly theories and anomaly polynomials (if they exist):
\bit
\item For the Maxwell theory discussed in example \ref{AMT}, the anomaly theory is
\be
\cI_{d+1}=\exp\left(2\pi i\int B^e_2\wedge dB_{d-2}^m \right)\,,
\ee
and the anomaly polynomial is
\be
\cI_{d+2}=\int dB_2^e\wedge dB_{d-2}^m\equiv \int F_3^e\wedge F_{d-1}^m\,,
\ee
where $F_3^e$ and $F_{d-1}^m$ are field strengths for the electric and magnetic background fields respectively.
\item For the discrete higher-form gauge theories discussed in example \ref{DHFGT}, the anomaly theory is
\be
\cI_{d+1}=\exp\left(-2\pi is(p)\int B_{p+1}^e\cup_\eta B_{d-p}^m\right),\qquad s(p)=(-1)^p\,.
\ee
\item For discrete 0-form symmetry groups in 2d discussed in (\ref{basA}), the anomaly theory is
\be
\cI_3=\int A_1^*\omega_3\,.
\ee
See the remark after equation (\ref{2dSPT}).
\eit

\subsubsection{Mixed 't Hooft Anomaly for a Non-Abelian Gauge Theory}
In the previous subsection, we discussed background fields for electric and magnetic higher-form symmetries in $d$-dimensional non-abelian gauge theories. It turns out that, just as in the Maxwell case, a non-abelian gauge theory also admits a mixed 't Hooft anomaly between the two symmetries.

\paragraph{Three Obstructions.}
Recall that background field $B_{d-2}^m$ for the magnetic symmetry appears as a term (\ref{Mba}) in the action, where note that $w_2$ is the $\cZ$-valued obstruction class for lifting $\cG$ bundles to $G$ bundles. On the other hand, turning on a background $B_2^e$ for the electric symmetry corresponds to considering bundles for the group $\cG/\G1$ for which the $\G1$-valued obstruction class for lifting to $\cG$ bundles is the background field $B_2^e$.

Overall, we can consider lifting directly from 
\be
\frac{\cG}{\G1}=\frac{G/\cZ}{\G1}\equiv\frac{G}{\cX}\;,\qquad \cX\subseteq Z(G)\,,
\ee
bundles to $G$ bundles, for which let us denote the $\cX$-valued obstruction class to be $W_2$. As will be discussed below, it turns out that even though $W_2$ and $B_2^e$ are closed, given that there is a well-defined obstruction of lifting $G/\cX$ bundles either to $G$ bundles or to $G/\cZ$ bundles respectively, the field $w_2$ is not closed if $B_2^e$ is non-trivial. This is because $w_2$ is the obstruction for lifting $G/\cZ$ bundles to $G$ bundles, but when $B_2^e$ is non-zero, we have to consider $G/\cX$ bundles that cannot be regarded as $G/\cZ$ bundles to begin with.

\paragraph{Relating the Obstructions.}
Clearly, $\cX$ should be a combination of $\cZ$ and $\G1$, and $W_2$ should be a combination of $w_2$ and $B_2^e$. Let us describe these combinations in more detail. First of all, 
$\cZ$ is naturally a subgroup of $\cX$, thus providing an injective homomorphism
\be
i:~\cZ\to\cX\,.
\ee
The remaining part of $\cX$ after taking out $\cZ$ is $\G1$
\be
\G1=\cX/\cZ\,,
\ee
thus providing a surjective homomorphism
\be
\pi:~\cX\to\G1\,.
\ee
Combining these two homomorphisms, we obtain a short exact sequence
\be\label{ses}
0\to\cZ\to\cX\to\G1\to0\,.
\ee
The obstructions are related as
\be\label{Wrel}
W_2=i(w_2)+\wt B_2^e\,,
\ee
where $i(w_2)$ is the $\cX$-valued 2-cochain obtained by applying the homomorphism $i$ on $w_2$
\be
i(w_2)(v_0,v_1,v_2)\equiv i\big(w_2(v_0,v_1,v_2)\in\cZ\big)\in\cX\,,
\ee
and $\wt B_2^e$ is an $\cX$-valued 2-cochain which is a lift of $B_2^e$ via the homomorphism $\pi$, i.e.\ we have
\be
\pi(\wt B_2^e)=B_2^e\,.
\ee

\paragraph{Non-Closure of the Obstruction $w_2$.}
Applying the differential $\delta$ on both sides of (\ref{Wrel}), we obtain
\be
0=i(\delta w_2)+\delta\wt B_2^e\,,
\ee
since $\delta$ commutes with $i$. Now, since $\delta$ commutes with $\pi$, we have
\be
\pi(\delta\wt B_2^e)=\delta B_2^e=0\,.
\ee
Consequently, we can obtain $\delta\wt B_2^e$ by applying $i$ on a $\cZ$-valued 3-cochain that is denoted as $\Bock(B_2^e)$ and referred to as the Bockstein of $B_2^e$ with respect to the short exact sequence (\ref{ses})
\be\label{Bock}
\delta\wt B_2^e=i(\Bock(B_2^e))\,.
\ee
Substituting it back, we obtain the key relationship
\be\label{kr}
\delta w_2=-\Bock(B_2^e)\,,
\ee
describing the non-closure of $w_2$.

\begin{tech}[Associated Long Exact Sequence in Cohomology]{}
Given a short exact sequence
\be
0\to A\to B\to C\to 0\,,
\ee
of abelian groups, we obtain a long exact sequence in cohomology
\be
\begin{tikzpicture}[descr/.style={fill=white,inner sep=1.5pt}]
        \matrix (m) [
            matrix of math nodes,
            row sep=1em,
            column sep=2.5em,
            text height=1.5ex, text depth=0.25ex
        ]
        { \dots & H^{p}(M,A) & H^{p}(M,B) & H^{p}(M,C) \\
            & H^{p+1}(M,A) & H^{p+1}(M,B) & H^{p+1}(M,C) & \dots \\
        };

        \path[overlay,->, font=\scriptsize,>=latex]
        (m-1-1) edge (m-1-2)
        (m-1-2) edge (m-1-3)
        (m-1-3) edge (m-1-4)
        (m-1-4) edge[out=355,in=175,blue] node[descr,yshift=0.3ex] {Bock} (m-2-2)
        (m-2-2) edge (m-2-3)
        (m-2-3) edge (m-2-4)
        (m-2-4) edge (m-2-5)
        (m-2-5);
\end{tikzpicture}
\ee
where $M$ is any manifold. The maps in each row descend straightforwardly from maps appearing in the short exact sequence applied to cochains. The connecting map between two rows is precisely the Bockstein map discussed above for $p=2$. For general $p$ the Bockstein map is a straightforward generalization specified via
\be
\delta\wt c_p=i(\Bock(c_p))\,,
\ee
on cochains, where $c_p$ is a $C$-valued $p$-cochain.
\end{tech}

\paragraph{Computing the 't Hooft Anomaly.}
Let us now perform a gauge transformation for the magnetic background field $B^m_{d-2}$
\be
B^m_{d-2}\to B^m_{d-2}+\delta\Lambda^m_{d-3}\,.
\ee
From the coupling (\ref{Mba}), we see that the action is modified as
\be
S\to S+2\pi s(d)\int \Lambda^m_{d-3}\cup_\eta\delta w_2^m,\qquad s(d)=(-1)^d\,,
\ee
which using (\ref{kr}) implies a mixed 't Hooft anomaly between the electric and magnetic symmetries
\be
Z\big[B_2^e,B_{d-2}^m+d\Lambda_{d-3}^m\big]=\exp\left(-2\pi is(d)\int \Lambda_{d-3}^m\cup_\eta\Bock(B_2^e) \right)\times Z\big[B_2^e,B_{d-2}^m\big]\,.
\ee
The anomaly theory is thus
\be\label{ATna}
\cI_{d+1}[B_2^e,B_{d-2}^m]=\exp\left(2\pi is(d)\int B_{d-2}^m\cup_\eta\Bock(B_2^e)\right)\,.
\ee

\begin{tech}[Vanishing of Bockstein: Split Short Exact Sequence]{}
In the expression for the anomaly (\ref{ATna}), only the group cohomology class of $\Bock(B_2^e)$ is relevant. Whenever the cohomology class vanishes
\be
[\Bock(B_2^e)]=0\,,
\ee
the anomaly also vanishes. This happens if the short exact sequence (\ref{ses}) \textbf{splits}, i.e.\ if we can find a homomorphism
\be
s:~\G1\to\cX\,,
\ee
such that
\be
\pi\circ s=1\,.
\ee
The existence of a homomorphism $s$ with the above properties is equivalent to factorization of $\cX$ as
\be
\cX=\cZ\times\cY\,,
\ee
where $\cY\cong\G1$.
\end{tech}

\begin{example}[Pure $SO(6)$ Yang-Mills Theory]{SO6}
As a simple example consider a theory with gauge group
\be
\cG=SO(6)\,,
\ee
and no matter content. The simply connected group is
\be
G=\Spin(6)=SU(4)\,,
\ee
whose center is
\be
Z(G)=\Z_4\,,
\ee
and we have
\be
\cZ=\Z_2\,,
\ee
since
\be
SO(6)=\Spin(6)/\Z_2\,.
\ee
The magnetic $(d-3)$-form symmetry is
\be
\G{d-3}=\wh\cZ=\Z_2\,,
\ee
and the electric 1-form symmetry is
\be
\G1=Z(\cG)=Z(G)/\cZ=\Z_4/\Z_2=\Z_2\,.
\ee
These groups sit in a short exact sequence
\be
\ba
&0\to\cZ\to \cX=Z(G)\to\G1\to0\\
=~~&0\to\Z_2\to \Z_4\to\Z_2\to0\,.
\ea
\ee
Turning on a $\Z_2$-valued magnetic background $B_{d-2}^m$ turns on a coupling 
\be
\pi\int B^m_{d-2}\cup w_2\,,
\ee
where $w_2$ is the $\Z_2$-valued obstruction for lifting $SO(6)$ bundles to $\Spin(6)$ bundles, and $\cup$ is the standard cup product. On the other hand, the $\Z_2$-valued background $B_2^e$ is the obstruction for lifting
\be
PSO(6)=SO(6)/\Z_2\,,
\ee
bundles to $SO(6)$ bundles. Overall, we have a $\Z_4$-valued obstruction $W_2$ for lifting $PSO(6)$ bundles to $\Spin(6)$ bundles, which can be expressed as
\be
W_2=2w_2+\wt B_2^e\,,
\ee
where the injective map
\be
i:~\Z_2\to\Z_4\,,
\ee
is just multiplication by 2, i.e.\
\be
\ba
2w_2(v_0,v_1,v_2)&=0\qquad\text{if $w_2(v_0,v_1,v_2)=0$}\,,\\
2w_2(v_0,v_1,v_2)&=2\qquad\text{if $w_2(v_0,v_1,v_2)=1$}\,,
\ea
\ee
and
\be
\ba
\wt B_2^e(v_0,v_1,v_2)&\in\{0,2\}\qquad\text{if $B_2^e(v_0,v_1,v_2)=0$}\,,\\
\wt B_2^e(v_0,v_1,v_2)&\in\{1,3\}\qquad\text{if $B_2^e(v_0,v_1,v_2)=1$}\,.
\ea
\ee
One can check that 
\be
\delta \wt B_2^e(v_0,v_1,v_2,v_3)\in\{0,2\}\,,
\ee
leading to
\be
\delta \wt B_2^e\equiv 2\Bock(B_2^e)\,,
\ee
where $\Bock(B_2^e)$ is a $\Z_2$-valued 3-cochain. The obstruction $w_2$ is non-closed as
\be
\delta w_2=-\Bock(B_2^e)\,,
\ee
and the anomaly theory (\ref{ATna}) is
\be\label{SO6A}
\cI_{d+1}[B_2^e,B_{d-2}^m]=\exp\left(\pi i\int B_{d-2}^m\cup\Bock(B_2^e)\right)\,,
\ee
as the sign $s(d)$ is irrelevant for $\Z_2$-valued quantities.
\end{example}

\subsection{Gauging}
\subsubsection{Definition}\label{Gdef}
Gauging is the procedure of promoting the background field $B_{p+1}$ for a $p$-form symmetry to a dynamical $(p+1)$-form (or $(p+1)$-cochain) gauge field $b_{p+1}$
\be\label{gauge}
B_{p+1}\to b_{p+1}\,.
\ee
Concretely this means that we sum over all possible \textit{gauge-inequivalent} backgrounds $B_{p+1}$. That is, we choose a representative $B_{p+1}$ inside each equivalence class of background fields up to background gauge transformations, and sum over all such representatives. If the original theory is $\fT$ and the $p$-form symmetry group being gauged is $\G p$, the theory obtained after gauging is referred to as the \textbf{gauged theory} and denoted as
\be
\fT/\G p\,.
\ee
The partition function of the gauged theory is
\be\label{gdef}
\ba
Z_{\fT/\G p}\propto\sum_{[B_{p+1}]}Z_\fT[B_{p+1}],\qquad\quad&\text{$\G p$ discrete}\,,\\
Z_{\fT/\G p}\propto\int \cD[B_{p+1}]~~Z_\fT[B_{p+1}],\qquad&\text{$\G p$ continuous}\,.
\ea
\ee
The above procedure is sensible only if the choice of representatives is immaterial, i.e.\ if we have
\be
Z_\fT[B_{p+1}^g]= Z_\fT[B_{p+1}]\,,
\ee
where $B_{p+1}^g$ is related to $B_{p+1}$ by a gauge transformation as in (\ref{Bg}), leading to the following statement.

\begin{state}['t Hooft Anomaly is Obstruction for Gauging]{}
A $\G p$ $p$-form symmetry group can be gauged only if it carries no 't Hooft anomaly.
\end{state}

\paragraph{Gauging Electric 1-Form Symmetries of Non-Abelian Gauge Theories.}
Using the above summing description of gauging, it is easy to understand gauging of electric 1-form symmetries of non-abelian gauge theories. Let us begin with an example and then we present the general results.

\begin{example}[From $SU(2)$ to $SO(3)$]{}
Consider $d$-dimensional pure $SU(2)$ Yang-Mills theory which has an electric 1-form symmetry
\be
\G1=\Z_2\,.
\ee
As we discussed in example \ref{SU2B}, in the presence of a background field $B_2^e$, the $SU(2)$ gauge theory sums over a subset of $SO(3)$ bundles, namely those for which the obstruction of lifting to an $SU(2)$ bundle is
\be
w_2=B_2^e\,.
\ee
Upon gauging the $\Z_2$ 1-form symmetry, we are now promoting $w_2$ to a dynamical field, and summing over all $SO(3)$ bundles irrespective of what their obstruction is. Thus, the gauged theory is $d$-dimensional pure $SO(3)$ Yang-Mills theory
\be
\fT_{SU(2)}/\Z_2^{(1)}=\fT_{SO(3)}\,.
\ee
\end{example}
We can easily generalize the above argument. Consider a $d$-dimensional non-abelian gauge theory $\fT_\cG$ having gauge group $\cG$ and some matter content. The electric 1-form symmetry group is
\be
\G1\subseteq Z(\cG)\,.
\ee
Consider gauging a subgroup
\be
\H1\subseteq\G1\,.
\ee
In the presence of a background field $B_2^e$ for $\H1$ 1-form symmetry, the $\cG$ gauge theory sums over $\cG/H$ bundles for which $B_2^e$ is the obstruction of lifting to $\cG$ bundles. Summing over all $B_2^e$ means that we are now summing over all $\cG/H$ bundles, and thus the theory $\fT_\cG/\H1$ obtained after gauging $\H1$ differs from $\fT_\cG$ only in the fact that the gauge group for $\fT_\cG/\H1$ is
\be
\cG/\H1\,.
\ee
We can denote the resulting theory as $\fT_{\cG/\H1}$. Then the result of gauging can be expressed as
\be\label{egau}
\fT_\cG/\H1=\fT_{\cG/\H1}\,.
\ee

\paragraph{Gauging Magnetic $(d-3)$-Form Symmetries of Non-Abelian Gauge Theories.}
It is similarly easy to understand gauging of magnetic $(d-3)$-form symmetries of non-abelian gauge theories. We again begin with an example and then present the general results.

\begin{example}[From $SO(3)$ to $SU(2)$]{}
Consider $d$-dimensional pure $SO(3)$ Yang-Mills theory which has a magnetic $(d-3)$-form symmetry
\be
\G{d-3}=\Z_2\,.
\ee
As we discussed in example \ref{SO3B}, in the presence of a background field $B_{d-2}^m$, the action of the $SO(3)$ gauge theory includes a term
\be
\pi\int B_{d-2}^m\cup w_2\,,
\ee
where $w_2$ is a dynamical field.
Summing over $B_{d-2}^m$, the above term acts as a Lagrange multiplier which sets
\be
w_2=0\,.
\ee
In other words, the resulting path integral sums only over $SU(2)$ gauge bundles. Thus, the gauged theory is the $d$-dimensional pure $SU(2)$ Yang-Mills theory
\be
\fT_{SO(3)}/\Z_2^{(d-3)}=\fT_{SU(2)}\,.
\ee
\end{example}
We can easily generalize the above argument. Consider a $d$-dimensional non-abelian gauge theory $\fT_\cG$ having gauge group $\cG$ and some matter content. The magnetic $(d-3)$-form symmetry is
\be
\G{d-3}=\wh\cZ\,,
\ee
where $\cZ$ relates $\cG$ to the simply connected group $G$ via
\be
\cG=G/\cZ\,.
\ee
Consider gauging a subgroup
\be
\H{d-3}\subseteq\G{d-3}\,.
\ee
Applying the Lagrange multiplier argument, we learn that this gauging procedure modifies the gauge group to
\be
G/\cY\,,
\ee
where
\be
\cY=\wh{\G{d-3}/\H{d-3}}\subseteq\cZ\,.
\ee
Thus we have
\be
\fT_{\cG}/\H{d-3}=\fT_{G/\cY}\,.
\ee

\subsubsection{Discrete Torsion}
There is a slight generalization of the above gauging procedure in which we add coefficients to the sums or integrals appearing in (\ref{gdef})
\be\label{gdt}
\ba
Z_{\fT/\G p}\propto\sum_{[B_{p+1}]}\cI_d[B_{p+1}]Z_\fT[B_{p+1}]\,,\qquad\quad&\text{$\G p$ discrete}\,,\\
Z_{\fT/\G p}\propto\int \cD[B_{p+1}]~~\cI_d[B_{p+1}]Z_\fT[B_{p+1}]\,,\qquad&\text{$\G p$ continuous}\,.
\ea
\ee
These coefficients have to be themselves  gauge invariant
\be
\cI_d[B_{p+1}^g]=\cI_d[B_{p+1}]\,.
\ee
In principle, these coefficients $\cI_d[B_{p+1}]$ could be the partition functions of any $d$-dimensional theory $\fT'$ with non-anomalous $\G p$ symmetry. In such a situation, the above gauging procedure corresponds to stacking $\fT$ with a decoupled copy of $\fT'$, leading to a total of $\G p\times\G p$ symmetry and then gauging the diagonal $\G p$ symmetry. 

However, the case of $\cI_d$ being a $d$-dimensional $\G p$ SPT phase is particularly interesting as it simply provides another way of gauging a $\G p$ $p$-form symmetry of $\fT$. This is because an SPT phase is a trivial theory (when symmetry is forgotten), and hence stacking with an SPT phase does not add any new degrees of freedom to $\fT$. In such a case, when $\cI_d$ is an SPT phase, the gauging procedure (\ref{gdt}) is referred to as gauging with a \textbf{discrete torsion} given by the SPT phase $\cI_d$.

\begin{example}[From $SU(2)$ to $SO(3)_-$]{}
Consider 4d pure $SU(2)$ Yang-Mills theory which has a 1-form symmetry
\be
\G1=\Z_2\,.
\ee
There is an SPT phase for $\Z_2$ 1-form symmetry in 4d whose partition function in the presence of a background is
\be\label{4dSPT}
\cI_4[B_2]=\exp\left(\pi i\int \frac{\cP(B_2)}2\right)\,.
\ee
Gauging the $\Z_2$ 1-form symmetry of the $SU(2)$ theory with discrete torsion $\cI_4$ leads to a gauge theory with $SO(3)$ gauge group along with the term (\ref{Pw}) in the action. This is precisely the 4d pure $SO(3)_-$ Yang-Mills theory discussed in section \ref{SO3-}, which carries a discrete theta angle. Thus, we learn that
\be
(\fT_{SU(2)}\times\cI_4)/\Z_2^{(1)}=\fT_{SO(3)_-}\,.
\ee
\end{example}

\subsubsection{Dual Symmetries Obtained After Gauging}
Consider the gauging (\ref{gauge}), which provides a $\G p$-valued gauge field $b_{p+1}$ in the gauged theory $\fT/\G p$. Just as with any gauge field we can consider Wilson operators $\cW_R$ for $b_{p+1}$ which are parametrized by irreducible representations $R$ of $\G p$. These operators are $(p+1)$-dimensional operators of $\fT/\G p$.

If $\G p$ is continuous, then $\cW_R$ are non-topological operators. However, if $\G p$ is discrete, then $\cW_R$ are topological operators since $b_{p+1}$ is necessarily flat in this case. Given that the $\cW_R$ are topological, they must generate a new symmetry in the gauged theory. We are thus led to the following statement:

\begin{state}[Dual Symmetries of Gauged Theory]{}
A gauged theory $\fT/\G p$ obtained after gauging a discrete \textit{abelian} $\G p$ $p$-form symmetry group admits a $(d-p-2)$-form symmetry given by the Pontryagin dual group
\be\label{dual}
\G{d-p-2}=\whG p\,.
\ee
Such symmetries are known as \textbf{dual symmetries} or \textbf{quantum symmetries} arising from the gauging of $\G p$. 
\end{state}

\ni Note the restriction to abelian $\G p$ in the above statement. For $p>0$, $\G p$ must be abelian, but for $p=0$, $\G0$ can be non-abelian. This has the following interesting consequence:

\begin{tech}[Non-Invertible Representation Symmetries]{}
First note that there always exists an irreducible representation $R$ of dimension bigger than one for a discrete non-abelian group $\G0$. As such it is impossible to find another representation $R^*$ of $\G0$ such that
\be
R\ot R^*=\id\,,
\ee
where on the RHS we have the trivial representation. This is because dimensions of representations are multiplicative under the tensor product.

Consequently, in a gauged theory $\fT/\G0$ we have topological Wilson line operators that are \textbf{non-invertible}. The symmetries generated by non-invertible topological operators are referred to as non-invertible symmetries, and have been a major focus of research recently \cite{Heidenreich:2021xpr,
Kaidi:2021xfk, Choi:2021kmx, Roumpedakis:2022aik,Bhardwaj:2022yxj, Choi:2022zal,Cordova:2022ieu,Choi:2022jqy,Kaidi:2022uux,Antinucci:2022eat,Bashmakov:2022jtl,Damia:2022bcd,Choi:2022rfe,Bhardwaj:2022lsg,Bartsch:2022mpm,Damia:2022rxw,Apruzzi:2022rei,Lin:2022xod,GarciaEtxebarria:2022vzq,Heckman:2022muc,Niro:2022ctq,Kaidi:2022cpf,Antinucci:2022vyk,Chen:2022cyw,Lin:2022dhv,Bashmakov:2022uek,Karasik:2022kkq,Cordova:2022fhg,GarciaEtxebarria:2022jky,Decoppet:2022dnz,Moradi:2022lqp,Runkel:2022fzi,Choi:2022fgx,Bhardwaj:2022kot, Bhardwaj:2022maz,Bartsch:2022ytj,Heckman:2022xgu,Antinucci:2022cdi,Apte:2022xtu,Delcamp:2023kew,Kaidi:2023maf,Li:2023mmw,Brennan:2023kpw,Etheredge:2023ler,Lin:2023uvm, Putrov:2023jqi, Carta:2023bqn,Koide:2023rqd, Zhang:2023wlu, Cao:2023doz, Dierigl:2023jdp, Inamura:2023qzl,Chen:2023qnv, Bashmakov:2023kwo, Choi:2023xjw,vanBeest:2023dbu,Lawrie:2023tdz,Apruzzi:2023uma,Chen:2023czk,Bah:2023ymy}. In the above situation, one says that $\fT/\G0$ has a
\be
\Rep(\G0)\,,
\ee
non-invertible $(d-2)$-form symmetry.
\end{tech}

\paragraph{Dual of Dual.}
Consider now gauging the dual symmetry (\ref{dual}). It turns out that the resulting gauged theory is the original theory $\fT$, i.e.\ we have
\be\label{dd}
\left(\fT/\G p\right)/\G{d-p-2}=\fT\,,
\ee
and the dual of the dual symmetry $\G{d-p-2}$ is the original symmetry $\G p$. 

It can be quickly seen by noting that turning on a background $B_{d-p-1}$ of the dual symmetry adds a coupling
\be
2\pi\int B_{d-p-1}\cup_\eta b_{p+1}\,.
\ee
After gauging $B_{d-p-1}$, and adding coupling to the new dual $p$-form symmetry background $B_{p+1}^{\text{new}}$, we have
\be
2\pi\int b_{d-p-1}\cup_\eta b_{p+1}+B_{p+1}^{\text{new}}\cup_\eta b_{d-p-1}\,.
\ee
Integrating out $b_{d-p-1}$ sets
\be
b_{p+1}= (-1)^p B_{p+1}^{\text{new}}\,.
\ee
Thus we indeed return, up to a sign, to the original theory $\fT$, justifying (\ref{dd}), and find that the new dual symmetry is the original $\G p$ symmetry.

\paragraph{Gauging Electric 1-Form Symmetries of Non-Abelian Gauge Theories.}
As we saw in section \ref{Gdef}, gauging electric 1-form symmetry of a gauge theory reduces the gauge group. More precisely, it implements a discrete quotient on the gauge group. A smaller gauge group has a larger magnetic $(d-3)$-form symmetry. These new magnetic symmetries can be understood as dual symmetries arising from the 1-form gauging. Let us begin with an example before describing the general results

\begin{example}[$SU(2)$ to $SO(3)$]{}
Recall that background for $\Z_2$ electric 1-form symmetry of the pure $SU(2)$ Yang-Mills theory is identified as the obstruction class 
\be
B_2=w_2 \,,
\ee
for lifting $SO(3)$ bundles to $SU(2)$ bundles. After gauging the $\Z_2$ 1-form symmetry, we obtain the pure $SO(3)$ Yang-Mills theory. The dual $\Z_2$ $(d-3)$-form symmetry is generated by topological operator
\be
\exp(i\pi\int b_2)=\exp(i\pi\int w_2)\,.
\ee
Now recall from (\ref{topm}) that the RHS is precisely the topological operator generating the magnetic $(d-3)$-form symmetry of the $SO(3)$ theory. Thus, the dual $(d-3)$-form symmetry is precisely the magnetic $(d-3)$-form symmetry we have been studying.
\end{example}

More generally, gauging an electric 1-form symmetry of a non-abelian gauge theory we have the result (\ref{egau}). The magnetic $(d-3)$-form symmetries of $\fT_\cG$ and $\fT_{\cG/\H1}$ are
\be\label{mags}
\G{d-3}(\fT_\cG)=\wh\cZ,\qquad \G{d-3}(\fT_{\cG/\H1})=\wh\cY\,,
\ee
where $\cZ$ and $\cY$ are defined via
\be
\cG=G/\cZ,\qquad \cG/\H1=G/\cY\,,
\ee
in terms of the simply connected group $G$. The two groups $\cZ$ and $\cY$ are related by a short exact sequence
\be
0\to\cZ\to\cY\to\H1\to0\,,
\ee
whose Pontryagin dual short exact sequence relates the magnetic symmetries (\ref{mags})
\be
0\to\whH1\to\wh\cY\to\wh\cZ\to0\,.
\ee
The subgroup $\whH1$ of $\G{d-3}(\fT_{\cG/\H1})=\wh\cY$ is the dual $(d-3)$-form symmetry arising from the gauging of $\H1$ 1-form symmetry.

\subsubsection{Gauging Exchanges Twisted and Charged Sectors}\label{tcex}
\paragraph{Twisted becomes Charged.}
Defects charged under the dual symmetry $\G{d-p-2}$ of $\fT/\G p$ are in twisted sectors for the symmetry $\G p$ of $\fT$.

\begin{note}[Twisted Sectors]{}
A codimension-$(p+2)$ operator $\cO$ is said to be in the twisted sector for a $p$-form symmetry, if $\cO$ is a non-genuine operator arising at an end of the codimension-$(p+1)$ topological operator associated to the $p$-form symmetry.
\end{note}

In order to see this, consider a twisted sector operator $\cO$, inserted along a submanifold $M_{d-p-2}$ of spacetime, lying at the end of a topological operator
\be
U_g,\qquad g\in\G p\,.
\ee
After gauging $\G p$, the topological operator $U_g$ implements a gauge symmetry, and hence does not act on any physical observables. Consequently, in $\fT/\G p$, the operator $U_g$ can be identified with the identity operator. One often says that $U_g$ becomes \textbf{invisible} after gauging. Thus, $\cO$ is a genuine codimension-$(p+2)$ operator of the gauged theory $\fT/\G p$. See figure \ref{invisible}.

\begin{figure}
\centering
\scalebox{1.1}{
\begin{tikzpicture}
\draw [thick,blue](-2,0) -- (0,0);
\draw [red,fill=red] (0,0) ellipse (0.05 and 0.05);
\node[blue] at (-2.5,0) {$U_g$};
\node[red] at (0.5,0) {$\cO$};
\draw [thick,-stealth](1.25,0) -- (3.25,0);
\node at (2.25,0.25) {\footnotesize{gauge $G^{(p)}$}};
\begin{scope}[shift={(4,0)}]
\draw [red,fill=red] (0,0) ellipse (0.05 and 0.05);
\node[red] at (0.5,0) {$\cO$};
\end{scope}
\end{tikzpicture}
}
\caption{A non-genuine codimension-$(p+2)$ operator $\cO$ living in twisted sector for a $\G p$ $p$-form symmetry becomes a genuine operator after gauging of $\G p$. Technically, this is true only if $\cO$ is uncharged under $\G p$, as will be discussed later.}
\label{invisible}
\end{figure}

Now we claim that $\cO$ has charge
\be\label{tcc}
g\in\G p=\wh{\whG p}=\whG{d-p-2}\,,
\ee
under the dual $\G{d-p-2}$ symmetry of $\fT/\G p$. This can be seen by noting that in $\fT$, insertion of $\cO(M_{d-p-2})$ imposes
\be\label{dB}
\delta B_{p+1}=\delta_g(M_{d-p-2})\,,
\ee
on the background field.
This is simply a translation of the fact that $\cO$ lies in the $g$-twisted sector.

After gauging $\G p$, (\ref{dB}) becomes
\be
\delta b_{p+1}=\delta_g(M_{d-p-2})\,.
\ee
Recall from the discussion around (\ref{dgco}), that this means precisely that $\cO$ has charge $g$ under $\G{d-p-2}$.

\begin{example}[$SU(2)$ to $SO(3)$: 't Hooft Line]{}
The operators charged under $\Z_2$ magnetic $(d-3)$-form symmetry of $d$-dimensional pure $SO(3)$ Yang-Mills theory are 't Hooft operators inducing $SO(3)$ monopole configurations that cannot be lifted to $SU(2)$ monopole configurations. In the pure $SU(2)$ Yang-Mills theory, the Dirac string for such 't Hooft operators is visible and can be identified with the topological codimension-2 operator generating the $\Z_2$ 1-form symmetry of the $SU(2)$ theory. Thus, such 't Hooft operators are in the twisted sector of the electric 1-form symmetry in the $SU(2)$ theory.
\end{example}

\paragraph{Charged to Twisted.}
On the other hand, operators charged under $\G p$ symmetry of $\fT$ are identified with twisted sector operators for $\G{d-p-2}$ symmetry of $\fT/\G p$. To see this, begin with twisted sector operators in $\fT/\G p$ and gauge $\G{d-p-2}$. As discussed around (\ref{dd}), this leads us back to the theory $\fT$. According to the above discussion, the twisted sector operators of $\fT/\G p$ become charged operators of $\fT$ under this process.

In more detail, an operator $\cO$ with charge
\be
\wh g\in\whG p\,,
\ee
under $\G p$ symmetry of $\fT$ lies in the $\wh g$-twisted sector of the
\be
\G{d-p-2}=\whG p\,,
\ee
symmetry of $\fT/\G p$.

\begin{example}[$SU(2)$ to $SO(3)$: Wilson Line]{}
A Wilson line
\be
\cW_j,\qquad j\not\in\Z\,,
\ee
is charged under the $\Z_2$ 1-form symmetry of $SU(2)$ theory. As we have discussed earlier, the only genuine Wilson line operators of the $SO(3)$ theory are $\cW_j$ with $j\in\Z$. All the above Wilson lines for $j\not\in\Z$ are in the twisted sector for the $\Z_2$ 1-form symmetry of the $SO(3)$ theory, attached to the magnetic 1-form symmetry generator (\ref{topm}).
\end{example}

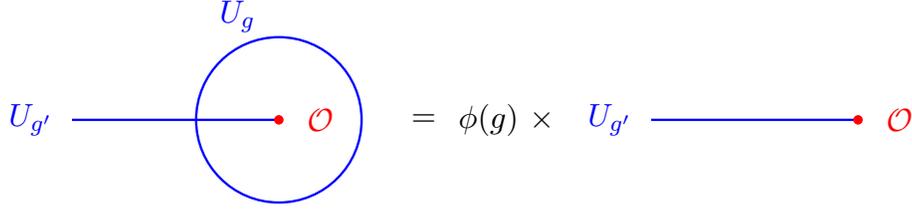
\begin{figure}
\centering
\scalebox{1.1}{
\begin{tikzpicture}
\draw [thick,blue](-2.5,0) -- (0,0);
\draw [red,fill=red] (0,0) ellipse (0.05 and 0.05);
\node[blue] at (-3,0) {$U_{g'}$};
\node[red] at (0.5,0) {$\cO$};
\draw [thick,blue] (0,0) ellipse (1 and 1);
\node[blue] at (-0.5,1.25) {$U_g$};
\node at (1.75,0) {=};
\begin{scope}[shift={(7,0)}]
\draw [thick,blue](-2.5,0) -- (0,0);
\draw [red,fill=red] (0,0) ellipse (0.05 and 0.05);
\node[blue] at (-3,0) {$U_{g'}$};
\node[red] at (0.5,0) {$\cO$};
\end{scope}
\node at (2.75,0) {$\phi(g)~\times$};
\end{tikzpicture}
}
\caption{An operator $\cO$ that is both in twisted sector for a $\G p$ $p$-form symmetry and also charged under it.}
\label{tcboth}
\end{figure}

\paragraph{Charged+Twisted to Twisted+Charged.}
Above we assumed that a twisted sector operator is not charged and a charged operator is not in the twisted sector in order to simplify the arguments. We can drop this assumption and consider operators that are both in the twisted sector and are charged as shown in figure \ref{tcboth}. Note that this is possible only if $d$ and $p$ are related as
\be
d=2p+2\,.
\ee
Let $\cO$ be an operator in $\fT$ that has the properties
\be
(q,t)=(\wh g,g)\in\whG p\times\G p\,,
\ee
where $q$ denotes the charge under $\G p$ $p$-form symmetry and $t$ denotes the twisted sector under $\G p$ $p$-form symmetry in which $\cO$ lies.
In the gauged theory $\fT/\G p$, the operator $\cO$ has properties
\be
(q,t)=(g,\wh g)\in\G p\times\whG p\,,
\ee
where now $q$ is the charge under $\G{d-p-2}=\whG p$ $(d-p-2)$-form symmetry and $t$ is the twisted sector under $\G{d-p-2}$.

Thus, gauging exchanges the values of $q$ and $t$.

\paragraph{Addition of SPT Phases.}
In the above discussion, we have been cavalier about the definition of charges of twisted sector operators. A proper definition requires making a choice of operators lying at an intersection of topological operators constituting the $p$-form symmetry.

Recall that such choices were encountered in the discussion of SPT phases. Indeed, the choices are permuted by stacking $\G p$ SPT phases with $\fT$, which changes the coupling of $\fT$ to $\G p$ symmetry. 

Let us see this in action in an example.

\begin{example}[$SU(2)$ to $SO(3)_\pm$]{}
Consider 4d pure $SU(2)$ Yang-Mills theory. The twisted sector line operators in this theory are of the form
\be
\cD_{j,\phi}:=\cW_j\cH_\phi,\qquad j\in\Z/2,~~\phi:~U(1)\to SO(3)\,,
\ee
where the cocharacter $\phi$ must not be liftable to an $SU(2)$ cocharacter. There are two options for charges of these line operators:
\ben
\item $\cD_{j,\phi}$ are charged if $j\not\in\Z$ and uncharged if $j\in\Z$\,,
\item $\cD_{j,\phi}$ are charged if $j\in\Z$ and uncharged if $j\not\in\Z$\,.
\een
These two choices are exchanged by stacking with the SPT phase (\ref{4dSPT}).

Gauging the $\Z_2$ 1-form symmetry using the coupling provided by the first choice leads to a theory in which lines
\be
\cD_{j,\phi},\qquad j\in\Z\,,
\ee
are in the untwisted sector for the dual 1-form symmetry. From the discussion in section \ref{SO3-}, we see that this is the $SO(3)_+$ theory.

On the other hand, gauging the $\Z_2$ 1-form symmetry using the coupling provided by the second choice leads to a theory in which lines
\be
\cD_{j,\phi},\qquad j\not\in\Z\,,
\ee
are in the untwisted sector for the dual 1-form symmetry. From the discussion in section \ref{SO3-}, we see that this is the $SO(3)_-$ theory.
\end{example}

\subsubsection{Gauging Exchanges Mixed 't Hooft Anomalies and Group Extensions}
Consider gauging a proper subgroup
\be
\H p\subseteq\G p\,,
\ee
of a $p$-form symmetry group $\G p$. The gauged theory contains not only a dual
\be
\H{d-p-2}=\widehat{H}^{(p)}\,,
\ee
$(d-p-2)$-form symmetry, but also a residual
\be
\K p:=\G p/\H p\,,
\ee
$p$-form symmetry. Depending on how $\H p$ is embedded inside $\G p$, the two symmetries of the gauged theory carry a mixed 't Hooft anomaly.

In order to see this, note that a background field $B_{p+1}^G$ for $\G p$ $p$-form symmetry decomposes as
\be
B_{p+1}^G=i(B_{p+1}^H)+\wt B_{p+1}^K\,,
\ee
where $B_{p+1}^H$ and $B_{p+1}^K$ are background fields for $\H p$ and $\K p$ respectively in the ungauged theory, $i$ is the injective homomorphism
\be
i:~\H p\to\G p\,,
\ee
and $\wt B_{p+1}^K$ is a lift to $\G p$ of the $\K p$-valued field $B_{p+1}^K$. From the analogous discussion around (\ref{kr}), we see that the above relationship implies
\be\label{GHK}
\delta B^H_{p+1}=-\Bock(B^K_{p+1})\,,
\ee
i.e.\ $B^H_{p+1}$ is not closed in general.

After gauging $\H p$, the relation (\ref{GHK}) becomes
\be
\delta b^H_{p+1}=-\Bock(B^K_{p+1})\,,
\ee
for the gauge field $b^H_{p+1}$. The presence of coupling
\be
2\pi\int B_{d-p-1}\cup_\eta b^H_{p+1}\,,
\ee
where $B_{d-p-1}$ is background field for $\H{d-p-2}$ implies a mixed 't Hooft anomaly in the gauged theory whose associated anomaly theory is
\be\label{miano}
\cI_{d+1}=\exp\left(-2\pi is(d-p)\int B_{d-p-1}\cup_\eta\Bock(B_{p+1}^K)\right),\qquad s(d-p)=(-1)^{d-p}\,.
\ee
Note that this anomaly vanishes if the short exact sequence
\be\label{ext}
0\to\H p\to\G p\to\K p\to 0\,,
\ee
splits, as explained right before example (\ref{SO6}). We are thus led to the following statement.
\begin{state}[Gauging Converts Group Extension to Mixed Anomaly]{state1}
Gauging a $p$-form symmetry group $\H p$ lying in an extension of $p$-form symmetry groups (\ref{ext}) leads to a mixed 't Hooft anomaly of the form (\ref{miano}).
\end{state}

The converse of the above statement is true, as can be seen by using the fact that gauging dual symmetry is equivalent to ungauging.

\begin{state}[Gauging Converts Mixed Anomaly to Group Extension]{}
Gauging a $(d-p-2)$-form symmetry group having a mixed 't Hooft anomaly with a $p$-form symmetry $\K p$ of the form (\ref{miano}) leads to a dual $p$-form symmetry $\H p$ which combines with the residual $\K p$ symmetry to form an extended $\G p$ $p$-form symmetry, such that the three $p$-form symmetries are related via (\ref{ext}).
\end{state}

\begin{example}[$\Spin(6)$ vs. $SO(6)$ pure Yang-Mills]{}
The mixed 't Hooft anomaly (\ref{SO6A}) in pure $SO(6)$ Yang-Mills theory can be seen as an example of statement (\ref{state1}) as the $SO(6)$ theory is obtained by gauging $\Z_2$ subgroup of the $\Z_4$ electric 1-form symmetry of the pure $\Spin(6)$ Yang-Mills theory. The non-triviality of the anomaly is related to the fact that the short exact sequence
\be
0\to\Z_2\to\Z_4\to\Z_2\to0\,,
\ee
does not split.
\end{example}

\subsection{Spontaneous Symmetry Breaking and Confinement}
\subsubsection{General Concepts}
Just like ordinary 0-form global symmetries, higher $p$-form global symmetries can also exhibit spontaneous symmetry breaking (SSB). We will see that spontaneous breaking is closely related to \textbf{confinement}. 

\paragraph{SSB of 0-Form Symmetries.}
Let us begin by rephrasing spontaneous symmetry breaking of 0-form symmetries in a form that generalizes to $p$-form symmetries. The traditional definition is as follows.

\begin{note}[Spontaneous Breaking of 0-Form Symmetry]{}
A 0-form symmetry group $\G0$ is spontaneously broken if there exists a local operator $\cO$ satisfying the following two conditions:
\ben
\item $\cO$ is charged non-trivially under $\G0$,
\item The vev of $\cO$ is non-zero
\be\label{vev}
\langle\cO\rangle\neq0\,.
\ee
\een
\end{note}

The main requirement (\ref{vev}) can be equivalently stated in terms of the non-vanishing of an asymptotic two point correlation function of $\cO$
\be\label{vev2}
\lim_{|x-y|\to\infty}\langle\cO(x)\cO(y)\rangle=\langle\cO(x)\rangle\langle\cO(y)\rangle=\langle\cO\rangle\langle\cO\rangle\neq0\,,
\ee
where we have used cluster decomposition to convert a two point correlator at large distances into a product of one point correlators.

\paragraph{SSB of $p$-Form Symmetries.}
Generalizing (\ref{vev2}), the spontaneous breaking of a $p$-form symmetry is defined as follows.

\begin{note}[Spontaneous Breaking of $p$-Form Symmetry]{}
A $p$-form symmetry group $\G p$ is spontaneously broken if there is a $p$-dimensional operator $\cO_p$ charged non-trivially under $\G p$ having asymptotic correlation function
\be\label{vevp}
\lim_{R\to\infty}\left\langle\cO_p(S^p_R)\right\rangle\neq 0\,,
\ee
where $\cO_p$ is inserted along $S^p_R$, which is a $p$-dimensional sphere of radius $R$.
\end{note}
Note that for $p=0$, this matches (\ref{vev2}), as a 0-dimensional sphere consists of two points. 

\paragraph{Relationship to Confinement.}
Such an operator $\cO_p$ satisfying (\ref{vevp}) survives in the deep IR. On the other hand, if an operator $\cO'_p$ satisfies
\be\label{vevnp}
\lim_{R\to\infty}\left\langle\cO'_p(S^p_R)\right\rangle= 0\,,
\ee
then the operator $\cO_p$ does not exist in the IR.

Consequently, one says that an operator $\cO'_p$ satisfying (\ref{vevnp}) \textbf{confines} in the IR, whereas operator $\cO_p$ satisfying (\ref{vevp}) \textbf{deconfines} in the IR.

Thus, we have the following statement
\begin{state}[(De)Confinement and SSB]{}
If a $p$-dimensional operator charged under a $p$-form symmetry $\G p$ deconfines, then $\G p$ is spontaneously broken.

\vspace{3mm}

\ni On the other hand, if all charged $p$-dimensional operators confine, then $\G p$ is not broken spontaneously.
\end{state}

\paragraph{Relationship to Perimeter and Area Laws.}
Another well-known definition of confinement and deconfinement, in the context of line operators, is as follows:
\bit
\item If the asymptotic vev of a line operator $L$ behaves as
\be\label{perimeter}
\lim_{R\to\infty}\langle L(S^1_R)\rangle\sim\exp\left(-\alpha R\right)\,,
\ee
the line $L$ is said to exhibit \textbf{perimeter law}. In such a situation, $L$ \textbf{deconfines}.
\item On the other hand, if the asymptotic vev of a line operator $L$ behaves as
\be\label{area}
\lim_{R\to\infty}\langle L(S^1_R)\rangle\sim\exp\left(-\alpha R^2\right)\,,
\ee
the line $L$ is said to exhibit \textbf{area law}. In such a situation, $L$ \textbf{confines}.
\eit
Naively, both perimeter and area laws seem to satisfy the condition (\ref{vevnp}) for confinement. However, the crucial point is that we can modify a line operator $L$ satisfying perimeter law (\ref{perimeter}) by a counterterm localized along the line to obtain a new line operator $L'$
\be
L'(S^1_R):=L(S^1_R)\times\exp\left(\int_{S^1_R}\text{vol}\right)\,,
\ee
where \text{vol} is the volume form on $S^1_R$. The operator $L'$ now exhibits (\ref{vevp}) and hence deconfines. One then says that the original line $L$ deconfines, keeping the above understanding in mind.

On the other hand, one cannot add a similar \textit{local} counterterm to a line operator $L$ satisfying area law (\ref{area}):
\bit
\item If one adds a counterterm localized along the line, then the resulting line operator $L'$ still confines.
\item If one wants a deconfining line $L'$, the counterterm needs to be defined along a surface $\Sigma_2$ with boundary of $\Sigma_2$ being the locus of the line operator $L$, e.g.\
\be
L'(S^1_R,\Sigma_2):=L(S^1_R)\times\exp\left(\int_{\Sigma_2}\text{vol}\right)\,,
\ee
where $\partial\Sigma_2=S^1_R$ and $\text{vol}$ is the volume form on $\Sigma_2$. However, in such a situation the resulting line operator $L'$ depends on $\Sigma_2$, and hence is a non-genuine line operator attached to a surface operator.
\eit

\begin{figure}
\centering
\scalebox{1.1}{
\begin{tikzpicture}
\draw [thick] (-1,0.5) rectangle (3,-2.5);
\node at (1,-3) {$R$};
\draw [thick,-stealth](0.75,-3) -- (-1,-3);
\draw [thick,-stealth](1.25,-3) -- (3,-3);
\node at (-1.5,-1) {$T$};
\draw [thick,-stealth](-1.5,-0.75) -- (-1.5,0.5);
\draw [thick,-stealth](-1.5,-1.25) -- (-1.5,-2.5);
\node at (3.5,-1) {$L$};
\end{tikzpicture}
}
\caption{A line operator $L$ inserted along a rectangle in spacetime, with length $R$ along a spatial direction and length $T$ along time direction.}
\label{rect}
\end{figure}
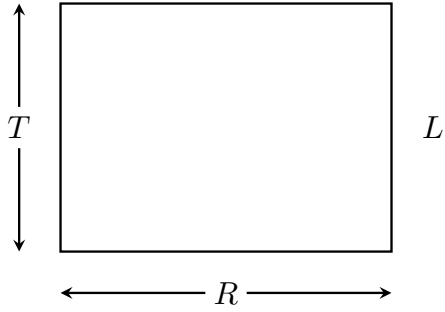

\paragraph{Potential Between Probe Particles.}
If we regard line operators as probe particles, the asymptotic vevs can be characterized in terms of potential energy between these probe particles. For this, consider inserting a line operator $L$ along a rectangle as shown in figure \ref{rect}. This configuration can now be regarded as an amplitude in which two probe particles sit separated by a distance $R$ for time interval $T$. The vev can then be evaluated as
\be
\langle L\rangle\sim\exp\left(-V(R)T\right)\,,
\ee
where $V(R)$ is the potential energy between the probe particles.

In such a characterization, the perimeter law corresponds to
\be\label{Ppot}
V(R)\sim \text{constant}\,,
\ee
and area law corresponds to
\be\label{Apot}
V(R)\sim R\,.
\ee
Thus, for probe particles showing area law, it costs infinite energy to separate them infinitely far away from each other. In other words, the probe particles remain confined.

Note that the potential (\ref{Apot}) for area law is as if the two particles are attached by a string for constant tension. This leads to the famous idea that confinement should be related to the existence of a dynamical string-like excitation, known as the \textbf{confining string}, such that the two probe particles live at its ends. 

\begin{example}[Meissner Effect]{}
The confining string is realized concretely in the context of Meissner effect in superconductivity. The setup is that of a 4d $U(1)$ gauge theory with a charged scalar $\phi$. As we provide a non-zero vev to the scalar
\be
\langle\phi\rangle\neq0\,,
\ee
the Dirac string associated to a 't Hooft line operator of magnetic charge $m$, which is a probe Dirac monopole, becomes a vortex string such that on a circle
\be
S^1=\{0\le\theta<2\pi\}\,,
\ee
around it we have
\be
\langle\phi\rangle=ve^{im\theta}\,.
\ee
This is due to the fact that we have a holonomy
\be
\oint_{S^1} A=m\,,
\ee
around the Dirac string. This vortex string acts as a confining string which confines the 't Hooft operator.
\end{example}

\paragraph{Coulomb Behavior.}
Consider a situation in which the IR effective theory is either a pure Maxwell theory or a pure non-abelian gauge theory. As is well-known, the potential between electric probe particles in such a theory falls with distance as
\be
V(R)\sim\frac1R\,.
\ee
Consequently a Wilson line $\cW$ in such a theory \textbf{deconfines}.

Magnetic probe particles also show Coulomb behavior, and hence we conclude that all dyonic line operators in such a theory deconfine. We are thus led to the following statements:
\begin{state}[SSB in Pure Maxwell Theory]{SSBMax}
The electric 1-form and magnetic $(d-3)$ form symmetries in pure Maxwell theory in $d\ge4$ are spontaneously broken.
\end{state}

\begin{state}[SSB in Pure Yang-Mills Theory]{}
The electric 1-form and magnetic $(d-3)$ form symmetries in a pure non-abelian gauge theory in $d\ge5$ are spontaneously broken.
\end{state}
The restriction of spacetime dimension $d$ in the above two statements arises because otherwise the beta function for the gauge coupling is negative and thus the gauge theory description is not a good effective description in the IR.

See \cite{Intriligator:1995au} for discussion of other interesting behaviors exhibited by line operators.

\subsubsection{Example: 4d $\cN=1$ SYM Theories with $\su(2)$ Gauge Algbebra}
Let us use the above concepts to discuss confinement and deconfinement in 4d $\cN=1$ $SU(2)$ and $SO(3)_\pm$ Super Yang-Mills (SYM) theories. 

\paragraph{Relationship to $\cN=2$ SYM with Soft SUSY Breaking.}
These $\cN=1$ SYM theories can be obtained as a limit
\be
\mu\to\infty\,,
\ee
of the $\cN=1$ gauge theories obtained by deforming pure $\cN=2$ $SU(2), SO(3)_\pm$ SYM by an $\cN=1$ preserving mass term
\be
\mu\,\Tr\,\Phi^2\,,
\ee
where $\Phi$ is the $\cN=1$ chiral multiplet sitting inside $\cN=2$ vector multiplet. Conjecturally, the phase structure of $\cN=1$ SYM theories is the same as the phase structure of $\cN=1$ gauge theories at small values of $\mu$.

\paragraph{Dual Meissner Effect in Seiberg-Witten Theory.}
The phase structure of the softly deformed theory for small values of $\mu$ was determined by Seiberg and Witten in their landmark paper \cite{Seiberg:1994rs}, where they exhibited an electric-magnetic dual version of Meissner effect leading to confinement of Wilson lines. Let us review this dual Meissner effect in more detail below.

\paragraph{Vacua After $\cN=1$ Deformation.}
The $\cN=2$ SYM theory has a moduli space of vacua, known as the Coulomb branch. The $\cN=1$ deformation $\mu$ lifts all of these vacua except for two special points on the Coulomb branch, namely the monopole and dyon points, which survive as vacua of the $\cN=1$ theory obtained after such a soft deformation.

\paragraph{IR Effective Theory.}
In both vacua, the IR theory is similar to the one discussed in the superconductivity example above, namely a $U(1)$ gauge theory with a scalar $\phi$ of gauge charge 1 that condenses
\be
\langle\phi\rangle\neq0\,.
\ee
Additionally, we have massive particle excitations of even magnetic charge. 

\paragraph{Confinement/Deconfinement of IR Line Operators.}
Now let us study the behavior exhibited by various line defects of the IR theory. First, note that the electrically charged condensate $\langle\phi\rangle$ completely screens the electric charge of probe particles. Hence the electric charge is irrelevant in the computation of asymptotic potential energy between probe particles. If the probe particles do not carry any magnetic charge, then one can separate them without any energy cost with potential (\ref{Ppot}), and hence all Wilson lines deconfine
\be
\cW^{IR}_n\longrightarrow\text{deconfine},\qquad n\in\Z\,.
\ee
For probe particles with magnetic charges, we need to separate the analysis of even and odd magnetic charges. The presence of dynamical particles with even magnetic charge means that the vortex strings attached to probe Dirac monopoles of even magnetic charge can break with these dynamical particles providing new ends. 

Consequently such vortex strings are unable to generate an area law potential, and we learn that 't Hooft lines of even magnetic charges deconfine
\be
\cH^{IR}_{2m}\longrightarrow\text{deconfine},\qquad m\in\Z\,.
\ee
On the other hand, the 't Hooft lines of odd magnetic charges confine as in the superconductivity example
\be
\cH^{IR}_{2m+1}\longrightarrow\text{confine},\qquad m\in\Z\,.
\ee
In general, for dyonic lines we have
\be\label{bIR}
\ba
\cD^{IR}_{n,2m}&=\cW^{IR}_n\cH^{IR}_{2m}\longrightarrow\text{deconfine},\qquad n,m\in\Z\,,\\
\cD^{IR}_{n,2m+1}&=\cW^{IR}_n\cH^{IR}_{2m+1}\longrightarrow\text{confine},\qquad n,m\in\Z\,.
\ea
\ee

\paragraph{Identification Between UV and IR Line Operators.}
In order to apply the above confinement and deconfinement results to the original UV theory, we need to know the identification between the UV and IR line operators, which was described in detail by Seiberg and Witten. The identification in both vacua involves an electric-magnetic duality, but the duality transformation is different in both vacua:
\bit
\item \textbf{Monopole Vacuum}: In the monopole vacuum, the IR scalar $\phi$ carries a magnetic charge from the UV point of view, and thus one says that the $\cN=1$ $SU(2)$ theory exhibits monopole condensation in this vacuum. The map of line operators is
\be\label{m1}
\ba
\cW^{UV}_j&\longrightarrow\cH^{IR}_{2j},\qquad j\in\Z/2\,,\\
\cH^{UV}_n&\longrightarrow\cW^{IR}_{n},\qquad n\in\Z\,,\\
\cD^{UV}_{j,n}\equiv\cW^{UV}_j\cH^{UV}_n&\longrightarrow\cD^{IR}_{n,2j},\qquad j\in\Z/2,~n\in\Z\,,
\ea
\ee
where $\cH^{UV}_n$ is a 't Hooft operator whose associated cocharacter
\be
\phi:~U(1)\to SO(3)\,,
\ee
has winding number $n$ around a maximal torus
\be
U(1)\subset SO(3)\,.
\ee
\item \textbf{Dyon Vacuum}: In the dyon vacuum, the IR scalar $\phi$ carries a dyonic charge from the UV point of view, and thus one says that the $\cN=1$ $SU(2)$ theory exhibits dyon condensation in this vacuum. The map of line operators is
\be\label{m2}
\ba
\cW^{UV}_j&\longrightarrow\cH^{IR}_{2j},\qquad j\in\Z/2\,,\\
\cH^{UV}_n&\longrightarrow\cD^{IR}_{n,-n},\qquad n\in\Z\,,\\
\cD^{UV}_{j,n}\equiv\cW^{UV}_j\cH^{UV}_n&\longrightarrow\cD^{IR}_{n,2j-n},\qquad j\in\Z/2,~n\in\Z\,.
\ea
\ee
\eit

\paragraph{Confinement/Deconfinement of UV Line Operators.}
Using the above maps (\ref{m1}), (\ref{m2}) and the behavior (\ref{bIR}) of IR lines, we can now deduce the behavior exhibited by UV lines:
\bit
\item \textbf{Monopole Vacuum}: In the monopole vacuum, the UV lines exhibit the following behaviors:
\be
\ba
\cW_j^{UV}&\longrightarrow\text{confine},\qquad j\not\in\Z\,,\\
\cW_j^{UV}&\longrightarrow\text{deconfine},\qquad j\in\Z\,,\\
\cH_n^{UV}&\longrightarrow\text{deconfine},\qquad n\in\Z\,,\\
\cD^{UV}_{j,n}&\longrightarrow\text{confine},\qquad j\not\in\Z,~n\in\Z\,,\\
\cD^{UV}_{j,n}&\longrightarrow\text{deconfine},\qquad j\in\Z,~n\in\Z\,.
\ea
\ee
\item \textbf{Dyon Vacuum}: In the dyon vacuum, the UV lines exhibit the following behaviors:
\be
\ba
\cW_j^{UV}&\longrightarrow\text{confine},\qquad j\not\in\Z\,,\\
\cW_j^{UV}&\longrightarrow\text{deconfine},\qquad j\in\Z\,,\\
\cH_n^{UV}&\longrightarrow\text{deconfine},\qquad n\in2\Z\,,\\
\cH_n^{UV}&\longrightarrow\text{confine},\qquad n\not\in2\Z\,,\\
\cD^{UV}_{j,n}&\longrightarrow\text{deconfine},\qquad 2j-n\in2\Z\,,\\
\cD^{UV}_{j,n}&\longrightarrow\text{confine},\qquad 2j-n\not\in2\Z\,.
\ea
\ee
\eit

\paragraph{$SU(2)$ Theory.}
Let us now pick gauge group
\be
\cG=SU(2)\,,
\ee
for the UV gauge theory, and study what happens in the IR. The charged genuine line operators are
\be
\cD_{j,n}^{UV},\qquad j\not\in\Z,~n\in2\Z\,,
\ee
which confine in both vacua. Thus, we learn the following.
\begin{state}[1-Form SSB in 4d $\cN=1$ $SU(2)$ SYM]{}
The $\Z_2$ 1-form symmetry of 4d $\cN=1$ $SU(2)$ SYM theory is spontaneously \textbf{unbroken} in both vacua.
\end{state}

The IR theory in both vacua should be a 4d TQFT with $\Z_2$ 1-form symmetry. To determine this TQFT, we need to study the operators surviving in the IR:
\bit
\item \textbf{Monopole Vacuum}: We need to take into account line operators
\be\label{LSU2}
\cD^{UV}_{j,n},\qquad j\in\Z,~n\in\Z\,,
\ee
and the surface operator
\be
U_2\,,
\ee
generating the $\Z_2$ 1-form symmetry. The genuine line operators are for $n\in2\Z$, which are all uncharged under the 1-form symmetry. This implies that the 1-form symmetry must be non-faithful in the IR, i.e.\ $U_2$ must be isomorphic to the identity surface operator in the IR. This follows from the fact that any non-identity genuine topological operator in a TQFT has to braid non-trivially with some other non-identity genuine topological operator of the TQFT. Since $U_2$ does not braid non-trivially with any of the lines, it must be the identity operator. This implies that all line operators in (\ref{LSU2}) are genuine lines, but there is no non-identity surface operator available for them to braid with. Consequently, all the IR line operators must be identity as well. Thus, we obtain a completely trivial theory in the IR which carries only identity operators.

So far we have not included the coupling of $\Z_2$ 1-form symmetry in the IR theory. There are two possibilities because of the existence of a non-trivial SPT phase (\ref{4dSPT}) for $\Z_2$ 1-form symmetry in 4d. In order to determine this, we need to choose a coupling for the $\Z_2$ 1-form symmetry in the UV, which is a choice of whether the non-trivially charged lines in the twisted sector are
\be\label{c1}
\cD^{UV}_{j,n},\qquad j\not\in\Z,~n\not\in2\Z\,,
\ee
or
\be\label{c2}
\cD^{UV}_{j,n},\qquad j\in\Z,~n\not\in2\Z\,.
\ee
Conventionally, the choice (\ref{c1}) is known simply as the $SU(2)$ theory, while the choice (\ref{c2}) is known as the $TSU(2)$ theory. For the $SU(2)$ theory, none of the charged lines survive in the IR and hence we obtain trivial SPT phase. On the other hand, for $TSU(2)$ theory, we do have charged lines in the twisted sector in IR, and hence we obtain the non-trivial SPT phase. Thus the IR 4d TQFTs with $\Z_2$ 1-form symmetry are
\be
\ba
\fT^{M}_{SU(2)}&=\text{Trivial}\,,\\
\fT^{M}_{TSU(2)}&=\text{SPT}\,,
\ea
\ee
where SPT denotes the non-trivial SPT phase and the superscript $M$ denotes the monopole vacuum.
\item \textbf{Dyon Vacuum}: The line operators that need to be taken into account are
\be
\cD^{UV}_{j,n}\longrightarrow\text{deconfine},\qquad 2j-n\in2\Z\,.
\ee
The genuine line operators have $j\in\Z$, $n\in2\Z$ which are uncharged under 1-form symmetry. Thus $U_2$ must again be identity operator and similar to the case for monopole vacuum all line operators must also be identity. The IR theory is thus trivial if one does not take into account the $\Z_2$ 1-form symmetry.

Taking into account the $\Z_2$ 1-form symmetry, the 4d TQFTs are
\be
\ba
\fT^{D}_{SU(2)}&=\text{SPT}\,,\\
\fT^{D}_{TSU(2)}&=\text{Trivial}\,,
\ea
\ee
where the superscript $D$ denotes the dyon vacuum.
\eit
Traditionally, the vacuum carrying trivial SPT is referred to as \textbf{confining vacuum}, while the non-trivial SPT is referred to as \textbf{oblique confining vacuum}.

Note that the definition of $SO(3)_+$ vs. $TSO(3)_+$ theories is chosen such that gauging the $\Z_2$ 1-form symmetry of the $SU(2)$ theory leads to the $SO(3)_+$ theory (as opposed to the $TSO(3)_+$ theory).

\paragraph{$SO(3)_+$ Theory.}
Now pick the gauge group
\be
\cG=SO(3)\,,
\ee
with trivial discrete theta angle. The charged genuine line operators are
\be\label{cgl}
\cD_{j,n}^{UV},\qquad j\in\Z,~n\not\in2\Z\,,
\ee
which deconfine in the monopole vacuum, but confine in the dyon vacuum. Thus, we learn the following.
\begin{state}[1-Form SSB in 4d $\cN=1$ $SO(3)_+$ SYM]{}
The $\Z_2$ 1-form symmetry of 4d $\cN=1$ $SO(3)_+$ SYM theory is spontaneously \textbf{broken} in one vacuum and spontaneously \textbf{unbroken} in the other vacuum.
\end{state}

Let us study the IR TQFT in the two vacua.
\bit
\item \textbf{Monopole Vacuum}: Since we have charged genuine lines (\ref{cgl}) that deconfine, $U_2$ is a non-identity surface operator. Similarly all the lines (\ref{cgl}) are non-identity topological operators. However, all these lines are the same in the IR. This is because if we had two different lines $L_1$ and $L_2$, then $L_1^{-1}L_2$ would be a non-identity line that does not braid non-trivially with any surface, leading to a contradiction. Let us denote this IR line operator as $U_1$. Thus we have an \textbf{emergent 2-form symmetry} in the IR as a consequence of deconfinement! We can recognize the properties of $U_2$ and $U_1$ as that of surface and line operators in a 4d $\Z_2$ gauge theory
\be
\pi\int a_1\cup\delta b_2\,,
\ee
which is the IR effective theory
\be
\fT^M_{SO(3)_+}=\text{$\Z_2$ Gauge Theory}\,.
\ee
We can modify the 1-form coupling in the UV and begin with $TSO(3)_+$ theory, but this only impacts twisted sector lines, which all confine, and hence the coupling does not impact the IR physics
\be
\fT^M_{TSO(3)_+}=\text{$\Z_2$ Gauge Theory}\,.
\ee
\item \textbf{Dyon Vacuum}: The story here is similar to the $SU(2)$ cases discussed above, and one finds that
\be
\ba
\fT^D_{SO(3)_+}&=\text{SPT}\,,\\
\fT^D_{TSO(3)_+}&=\text{Trivial}\,,
\ea
\ee
where the 1-form coupling for $SO(3)_+$ is chosen such that the charged twisted sector lines are
\be
\cD_{j,n},\qquad j\not\in\Z,~n\not\in2\Z\,,
\ee
and for $TSO(3)_+$ the charged twisted sector lines are
\be
\cD_{j,n},\qquad j\not\in\Z,~n\in2\Z\,.
\ee
\eit

\paragraph{$SO(3)_-$ Theory.}
The analysis is similar to the $SO(3)_+$ case and is left to the interested reader. The final results are
\be
\ba
\fT_{SO(3)_-}^{M}&=\text{SPT}\,,\\
\fT_{TSO(3)_-}^{M}&=\text{Trivial}\,,\\
\fT_{SO(3)_-}^{D}&=\text{$\Z_2$ Gauge Theory}\,,\\
\fT_{TSO(3)_-}^{D}&=\text{$\Z_2$ Gauge Theory}\,.
\ea
\ee
Note that the definition of $SO(3)_-$ vs. $TSO(3)_-$ theories is chosen such that gauging the $\Z_2$ 1-form symmetry of the $TSU(2)$ theory leads to the $SO(3)_-$ theory (as opposed to the $TSO(3)_-$ theory).

\begin{tech}[Confining Non-Lagrangian Theories]{}
In the above study of confinement/deconfinement in 4d $\cN=1$ SYM, the Lagrangian descriptions were crucial for obtaining the presented results. However, the definitions of confinement and deconfinement do not require any Lagrangian description. This raises the question of how to study confinement/deconfinement in non-Lagrangian or strongly coupled 4d theories. An answer to this question was provided recently in \cite{Bhardwaj:2021zrt} for theories that admit a construction in terms of M5 branes in M-theory, or as compactifications of 6d $\cN=(2,0)$ SCFTs, where examples of confining non-Lagrangian theories are also provided.
\end{tech}

\paragraph{$S$ and $T$ Operations.}
All the UV theories considered above are related by gaugings. We have already discussed two such gaugings
\be\label{SLp}
\ba
SU(2)/\Z_2^{(1)}&=SO(3)_+\,,\\
TSU(2)/\Z_2^{(1)}&=SO(3)_-\,.
\ea
\ee
It is customary to denote the gauging by $S$. In terms of this, the above gaugings are represented as relations
\be
\ba
S(SU(2))&=SO(3)_+\,,\\
ST(SU(2))&=SO(3)_-\,.
\ea
\ee
Thus, $S$ is the operation of gauging, and $T$ is the operation of stacking an SPT phase. It can be shown \cite{Gaiotto:2014kfa,Bhardwaj:2020ymp} that these operations obey
\be\label{SL}
(ST)^3=1\,,
\ee
which is an identity also satisfied by the $S$ and $T$ matrices generating the $SL(2,\Z)$ group. For this reason, the above and $S$ and $T$ operations are referred to as ``$SL(2,\Z)$ operations on 4d QFTs with $\Z_2$ 1-form symmetry''. It should be noted that $T$ is an order two operation in our context, rather than the usual infinite order operation. Combining it with the fact that gauging twice is the same as ungauging, i.e.
\be
S^2=1\,.
\ee
we find that the group of these $S$ and $T$ transformations is $PSL(2,\Z_2)$.

The full web of gaugings can be deduced by using the relationship (\ref{SL}) along with (\ref{SLp}). For example, if one wants to know the result of gauging the $\Z_2$ 1-form symmetry of  the $TSO(3)_-$ theory, one is computing
\be
S(TSO(3)_-)=STST(SU(2))=TS(SU(2))=TSO(3)_+\,.
\ee
Alternatively, one can use the exchange of charged and twisted sectors of line operators under 1-form gauging to deduce the theory after gauging.

\paragraph{Compatibility Between UV and IR Gaugings.}
One can also consider gauging 1-form symmetries of the IR TQFTs discussed above. We have
\be
\ba
S(\text{Trivial})&=\text{$\Z_2$ Gauge Theory}\,,\\
S(\text{SPT})&=\text{SPT}\,,\\
S(\text{$\Z_2$ Gauge Theory})&=\text{Trivial}\,.
\ea
\ee
This can be easily seen by exchanging the twisted and charged sectors of line operators: 
\bit
\item In the trivial theory, we have a twisted sector line that is uncharged. This is simply the identity line living at the end of $U_2$, which as we discussed is simply the identity surface operator. After gauging, this line gives rise to a charged line in the untwisted sector, which means that the gauged theory is the $\Z_2$ gauge theory.
\item In the SPT theory, we have a twisted sector line that is charged. After gauging, it remains in the twisted sector and also remains charged. Thus, the gauged theory is SPT again.
\item In the $\Z_2$ gauge theory, we have an untwisted sector line that is charged. After gauging, it becomes a line in the twisted sector that is uncharged. Thus, the gauged theory is the trivial theory.
\eit
The $T$ operations are
\be
\ba
T(\text{Trivial})&=\text{SPT}\,,\\
T(\text{SPT})&=\text{Trivial}\,,\\
T(\text{$\Z_2$ Gauge Theory})&=\text{$\Z_2$ Gauge Theory}\,.
\ea
\ee

Now $S$ and $T$ transformations should commute with the RG flow. Indeed, this can be explicitly checked. For example, we should have
\be
S\left(\fT_{SU(2)}^M\right)=\fT_{S(SU(2))}^M=\fT_{SO(3)_+}^M\,,
\ee
which is indeed the case. As another example, we should have
\be
S\left(\fT_{TSO(3)_-}^D\right)=\fT^D_{TSO(3)_+}\,,
\ee
which the reader can easily see is also true.

\subsection{Symmetry TFTs}
\label{sec:SymTFTsection}
We have seen that any gauging of discrete symmetries can be ungauged, simply by a subsequent gauging of the dual symmetry. This means that discrete gaugings do not change the full information of observables in a theory. However, as we have seen, discrete gaugings do mix different types of observables into each other, e.g.\ the exchange of twisted and charged sectors discussed in section \ref{tcex}. As a consequence, a physical observable may appear in different contexts in different \textit{gauge frames} associated to a discrete symmetry. Symmetry topological field theories (TFTs), abbreviated \textbf{SymTFTs} or \textbf{Symmetry TFTs}, are powerful tools that detangle the physical observables from their behavior in different gauge frames. They also encode the 't Hooft anomalies of the symmetry and possible higher-group structures. We refer the reader to \cite{Freed:2022qnc} for general logic of SymTFTs, and to \cite{Gaiotto:2014kfa,Gaiotto:2020iye,Apruzzi:2021nmk} for detailed discussions of SymTFTs in the context of invertible symmetries. Recently, detailed discussions of SymTFTs for non-invertible symmetries appeared in \cite{Bhardwaj:2023ayw}, which also elucidated the general concepts of \cite{Freed:2022qnc}.

\paragraph{Sandwich Construction.}
The basic idea of SymTFT is encoded in the sandwich construction. Let us first present the idea abstractly, and then we will concretely illustrate it.

\begin{figure}
\centering
\scalebox{1}{
\begin{tikzpicture}
\draw [yellow, fill= yellow, opacity =0.1]
(0,0) -- (0,4) -- (2,5) -- (7,5) -- (7,1) -- (5,0)--(0,0);
\begin{scope}[shift={(0,0)}]
\draw [white, thick, fill=white,opacity=1]
(0,0) -- (0,4) -- (2,5) -- (2,1) -- (0,0);
\end{scope}
\begin{scope}[shift={(5,0)}]
\draw [white, thick, fill=white,opacity=1]
(0,0) -- (0,4) -- (2,5) -- (2,1) -- (0,0);
\end{scope}
\begin{scope}[shift={(5,0)}]
\draw [red, thick, fill=red,opacity=0.2] 
(0,0) -- (0, 4) -- (2, 5) -- (2,1) -- (0,0);
\draw [red, thick]
(0,0) -- (0, 4) -- (2, 5) -- (2,1) -- (0,0);
\node at (1,3.5) {$\Bphys_{\mathfrak{T}}$};
\end{scope}
\begin{scope}[shift={(0,0)}]
\draw [blue, thick, fill=blue,opacity=0.2]
(0,0) -- (0, 4) -- (2, 5) -- (2,1) -- (0,0);
\draw [blue, thick]
(0,0) -- (0, 4) -- (2, 5) -- (2,1) -- (0,0);
\node at (1,3.5) {$\Bsym_\cS$};
\end{scope}
\node at (3.3, 3.5) {$\mathfrak{Z}({\mathcal{S}})$};
\draw[dashed] (0,0) -- (5,0);
\draw[dashed] (0,4) -- (5,4);
\draw[dashed] (2,5) -- (7,5);
\draw[dashed] (2,1) -- (7,1);
    \draw [blue, thick] (1, 1) -- (1, 2) ; 
    \draw [blue, thick] (0.5, 3) -- (1, 2) -- (1.5, 3.2) ; 
    \node[blue] at (0.5, 1.5) {$\cS$};
\begin{scope}[shift={(9,0)}]
\node at (-1,2.3) {$=$} ;
\draw [purple, thick, fill=purple,opacity=0.2] 
(0,0) -- (0, 4) -- (2, 5) -- (2,1) -- (0,0);
\draw [purple, thick]
(0,0) -- (0, 4) -- (2, 5) -- (2,1) -- (0,0);
\node at (1,3.5) {$\fT$};    
    \draw [blue, thick] (1, 1) -- (1, 2) ; 
    \draw [blue, thick] (0.5, 3) -- (1, 2) -- (1.5, 3.2) ; 
    \node[blue] at (0.5, 1.5) {$\cS$};
\end{scope}
\end{tikzpicture}
}
\caption{The sandwich construction as explained in the text. We have also displayed topological operators and their junctions generating symmetry $\cS$ as a Y-shape. These operators are localized on the symmetry boundary $\Bsym_\cS$.}
\label{SymTFT1}
\end{figure}
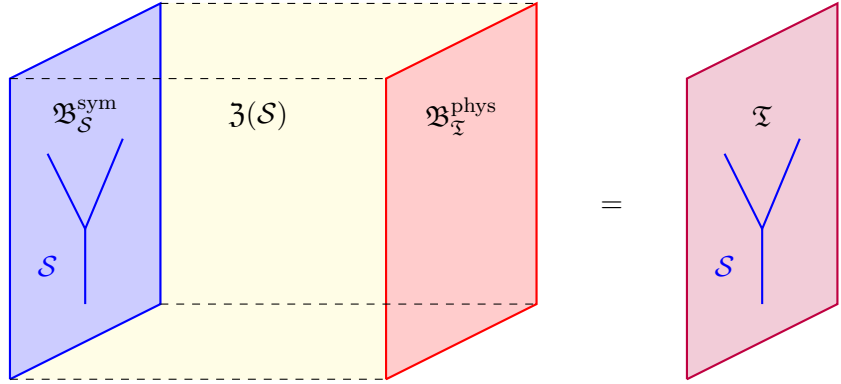

Consider a $d$-dimensional theory $\fT$ with some discrete symmetry $\cS$. This could be a higher-form or a higher-group (or even a non-invertible) symmetry. According to the sandwich construction, we can realize $\fT$ as an interval compactification involving the following ingredients:
\bit
\item \textbf{Symmetry TFT}: This is a $(d+1)$-dimensional TQFT $\fZ(\cS)$ associated only to the symmetry $\cS$ and independent of the theory $\fT$.
\item \textbf{Symmetry Boundary Condition}: This is a topological boundary condition $\fB^{\text{sym}}_\cS$ of the SymTFT $\fZ(\cS)$ which also only depends on the symmetry $\cS$ and independent of the theory $\fT$.
\item \textbf{Physical Boundary Condition}: This is a possibly non-topological boundary condition $\fB^{\text{phys}}_\fT$ of the SymTFT $\fZ(\cS)$ which is different for different theories $\fT$.
\eit
See figure \ref{SymTFT1}.

The symmetry $\cS$ is separated onto the symmetry boundary $\fB^{\text{sym}}_\cS$, while the dynamics of the theory $\fT$ lives on the boundary $\fB^{\text{phys}}_\fT$.

\paragraph{SymTFTs for Higher-Form Symmetries.}
Let the symmetry $\cS$ be a product of discrete higher-form symmetries
\be
\cS=\prod_{i=1}^n\G{p_i}\,,
\ee
with a mixed 't Hooft anomaly captured by an anomaly theory
\be\label{invSPT}
\cI_{d+1}[A_{p_1+1},A_{p_2+1},\cdots, A_{p_n+1}]=\exp\left(i\int \cL^{\text{eff}}_{d+1}[A_{p_1+1},A_{p_2+1},\cdots, A_{p_n+1}]\right)\,,
\ee
where $A_{p_i+1}$ are background fields for $\G{p_i}$ symmetries and $\cL^{\text{eff}}_{d+1}$ is the effective Lagrangian for the anomaly theory. We assume that any 0-form symmetries appearing in $\cS$ are abelian. For non-abelian 0-form symmetries, we need a generalization of what is discussed here. See \cite{Bhardwaj:2023ayw} for more details.

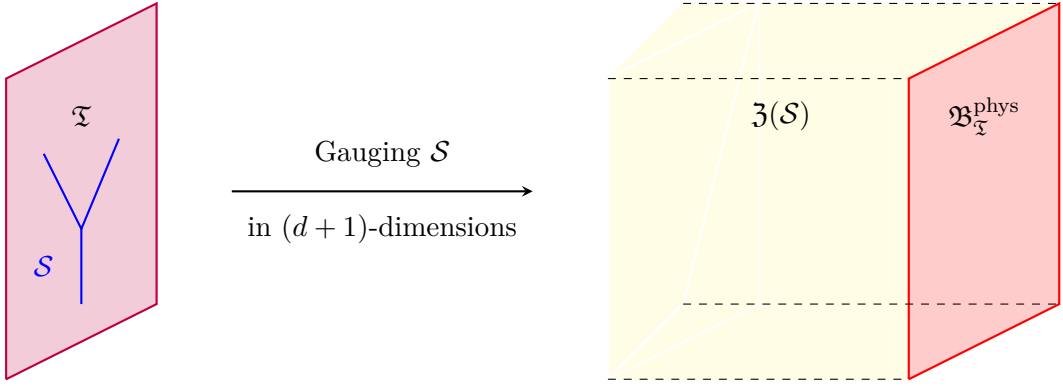
\begin{figure}
\centering
\scalebox{1}{
\begin{tikzpicture}
\draw [yellow, fill= yellow, opacity =0.1]
(1,0) -- (1,4) -- (2,5) -- (7,5) -- (7,1) -- (5,0)--(1,0);
\begin{scope}[shift={(1,0)}]
\draw [yellow!0.1, thick, fill=white,opacity=1]
(0,0) -- (0,4) -- (2,5) -- (2,1) -- (0,0);
\end{scope}
\begin{scope}[shift={(5,0)}]
\draw [yellow!0.1, thick, fill=white,opacity=1]
(0,0) -- (0,4) -- (2,5) -- (2,1) -- (0,0);
\end{scope}
\begin{scope}[shift={(5,0)}]
\draw [red, thick, fill=red,opacity=0.2] 
(0,0) -- (0, 4) -- (2, 5) -- (2,1) -- (0,0);
\draw [red, thick]
(0,0) -- (0, 4) -- (2, 5) -- (2,1) -- (0,0);
\node at (1,3.5) {$\Bphys_{\mathfrak{T}}$};
\end{scope}
\begin{scope}[shift={(1,0)}]
\draw [yellow!0.1,fill=yellow,opacity=0.1]
(0,0) -- (0,4) -- (2,5) -- (2,1) -- (0,0);
\draw [yellow!0.1, thick]
(0,0) -- (0,4) -- (2,5) -- (1,1) -- (0,0);
\end{scope}
\node at (3.3, 3.5) {$\mathfrak{Z}({\mathcal{S}})$};
\draw[dashed] (1,0) -- (5,0);
\draw[dashed] (1,4) -- (5,4);
\draw[dashed] (2,5) -- (7,5);
\draw[dashed] (2,1) -- (7,1);
\begin{scope}[shift={(-7,0)}]
\draw [purple, thick, fill=purple,opacity=0.2] 
(0,0) -- (0,4) -- (2,5) -- (2,1) -- (0,0);
\draw [purple, thick]
(0,0) -- (0,4) -- (2,5) -- (2,1) -- (0,0);
\node at (1,3.5) {$\fT$};    
    \draw [blue, thick] (1,1) -- (1,2) ; 
    \draw [blue, thick] (0.5,3) -- (1,2) -- (1.5,3.2) ; 
    \node[blue] at (0.5,1.5) {$\cS$};
\end{scope}
\draw [thick,-stealth](-4,2.5) -- (0,2.5);
\node at (-2,3) {Gauging $\cS$};
\node at (-2,2) {in $(d+1)$-dimensions};
\end{tikzpicture}
}
\caption{The SymTFT $\fZ(\cS)$ and physical boundary $\Bphys_\fT$ for a $d$-dimensional theory $\fT$ having invertible symmetry $\cS$ are constructed by gauging $\cS$ in $(d+1)$-dimensions.}
\label{SymTFT2}
\end{figure}

Then the right half of the sandwich construction involving the SymTFT and the physical boundary is obtained by gauging all the $\G{p_i}$ symmetries. Note that this is impossible in $d$-dimensions unless the 't Hooft anomaly vanishes, but here we are performing the gauging in $(d+1)$-dimensions, which is possible after attaching SPT phase $\cI_{d+1}$ to the theory $\fT$. See figure \ref{SymTFT2}. 

As an SPT phase is a trivial theory once we forget about the symmetry, the gauging procedure leads to a discrete higher-form gauge theory, but this theory generally is not of the type discussed in section \ref{hfgt}. The presence of the SPT phase modifies the action of the gauge theory to
\be\label{SymAct}
\sum_{i=1}^n 2\pi\int a_{p_i+1}\cup_\eta \delta b_{d-p_i-1}+\int\cL^{\text{eff}}_{d+1}[a_{p_1+1},a_{p_2+1},\cdots, a_{p_n+1}]\,.
\ee
Now, to complete the sandwich construction, we need to provide the symmetry boundary condition. This is comprised of
\bit
\item \textbf{Dirichlet} boundary condition (b.c.) for all gauge fields $a_{p_i+1}$.
\item \textbf{Neumann} b.c. for all gauge fields $b_{d-p_i-1}$.
\eit
Indeed, now the interval compactification simply undoes the gauging of $\G{p_i}$ symmetries, because $\fB^{\text{sym}}_\cS$ carries Dirichlet b.c. for $a_{p_i+1}$. As a result of it, the interval compactification performs the replacement
\be\label{eq:dirichlet}
a_{p_i+1}\to A_{p_i+1}\,,
\ee
leaving behind the SPT phase (\ref{invSPT}).

\begin{figure}
\centering
\scalebox{1}{
\begin{tikzpicture}
\draw [yellow, fill= yellow, opacity =0.1]
(0,0) -- (0,4) -- (2,5) -- (6.5,5) -- (6.5,1) -- (4.5,0)--(0,0);
\begin{scope}[shift={(0,0)}]
\draw [white, thick, fill=white,opacity=1]
(0,0) -- (0,4) -- (2,5) -- (2,1) -- (0,0);
\end{scope}
\begin{scope}[shift={(0,0)}]
\draw [blue, thick, fill=blue,opacity=0.2]
(0,0) -- (0, 4) -- (2, 5) -- (2,1) -- (0,0);
\draw [blue, thick]
(0,0) -- (0, 4) -- (2, 5) -- (2,1) -- (0,0);
\node at (1,3.5) {$\Bsym_\cS$};
\end{scope}
\node at (5.5,3.5) {$\mathfrak{Z}({\mathcal{S}})$};
\draw[dashed] (0,0) -- (4.5,0);
\draw[dashed] (0,4) -- (4.5,4);
\draw[dashed] (2,5) -- (6.5,5);
\draw[dashed] (2,1) -- (6.5,1);
\draw [thick,blue](3.5,3.5) -- (3.5,1);
\draw [-stealth](3,2.5) -- (2.5,2.5);
\node[blue] at (4,2) {$U_{g_i}$};
\node at (7.5,2.5) {=};
\begin{scope}[shift={(8.5,0)}]
\draw [yellow, fill=yellow, opacity=0.1]
(0,0) -- (0,4) -- (2,5) -- (6.5,5) -- (6.5,1) -- (4.5,0)--(0,0);
\begin{scope}[shift={(0,0)}]
\draw [white, thick, fill=white,opacity=1]
(0,0) -- (0,4) -- (2,5) -- (2,1) -- (0,0);
\end{scope}
\begin{scope}[shift={(0,0)}]
\draw [blue, thick, fill=blue,opacity=0.2]
(0,0) -- (0,4) -- (2,5) -- (2,1) -- (0,0);
\draw [blue, thick]
(0,0) -- (0,4) -- (2,5) -- (2,1) -- (0,0);
\node at (1,3.5) {$\Bsym_\cS$};
\end{scope}
\node at (5.5,3.5) {$\mathfrak{Z}({\mathcal{S}})$};
\draw[dashed] (0,0) -- (4.5,0);
\draw[dashed] (0,4) -- (4.5,4);
\draw[dashed] (2,5) -- (6.5,5);
\draw[dashed] (2,1) -- (6.5,1);
\draw [thick,blue](1.5,3.5) -- (1.5,1);
\node[blue] at (1,2) {$U_{g_i}$};
\end{scope}
\begin{scope}[shift={(0,-7)}]
\draw [yellow, fill=yellow, opacity=0.1]
(0,0) -- (0,4) -- (2,5) -- (6.5,5) -- (6.5,1) -- (4.5,0)--(0,0);
\begin{scope}[shift={(0,0)}]
\draw [white, thick, fill=white,opacity=1]
(0,0) -- (0,4) -- (2,5) -- (2,1) -- (0,0);
\end{scope}
\begin{scope}[shift={(0,0)}]
\draw [blue, thick, fill=blue,opacity=0.2]
(0,0) -- (0,4) -- (2,5) -- (2,1) -- (0,0);
\draw [blue, thick]
(0,0) -- (0,4) -- (2,5) -- (2,1) -- (0,0);
\node at (1,3.5) {$\Bsym_\cS$};
\end{scope}
\node at (5.5,3.5) {$\mathfrak{Z}({\mathcal{S}})$};
\draw[dashed] (0,0) -- (4.5,0);
\draw[dashed] (0,4) -- (4.5,4);
\draw[dashed] (2,5) -- (6.5,5);
\draw[dashed] (2,1) -- (6.5,1);
\draw [thick,red](3.5,3.5) -- (3.5,1);
\draw [-stealth](3,2.5) -- (2.5,2.5);
\node[red] at (4,2) {$U_{\wh g_i}$};
\end{scope}
\node at (7.5,-4.5) {=};
\begin{scope}[shift={(8.5,-7)}]
\draw [yellow, fill=yellow, opacity=0.1]
(0,0) -- (0,4) -- (2,5) -- (6.5,5) -- (6.5,1) -- (4.5,0)--(0,0);
\begin{scope}[shift={(0,0)}]
\draw [white, thick, fill=white,opacity=1]
(0,0) -- (0,4) -- (2,5) -- (2,1) -- (0,0);
\end{scope}
\begin{scope}[shift={(0,0)}]
\draw [blue, thick, fill=blue,opacity=0.2]
(0,0) -- (0,4) -- (2,5) -- (2,1) -- (0,0);
\draw [blue, thick]
(0,0) -- (0,4) -- (2,5) -- (2,1) -- (0,0);
\node at (1,3.5) {$\Bsym_\cS$};
\end{scope}
\node at (5.5,3.5) {$\mathfrak{Z}({\mathcal{S}})$};
\draw[dashed] (0,0) -- (4.5,0);
\draw[dashed] (0,4) -- (4.5,4);
\draw[dashed] (2,5) -- (6.5,5);
\draw[dashed] (2,1) -- (6.5,1);
\end{scope}
\end{tikzpicture}
}
\caption{Fate of bulk topological operators of the SymTFT as they are pushed to the symmetry boundary.}
\label{SymTFT3}
\end{figure}
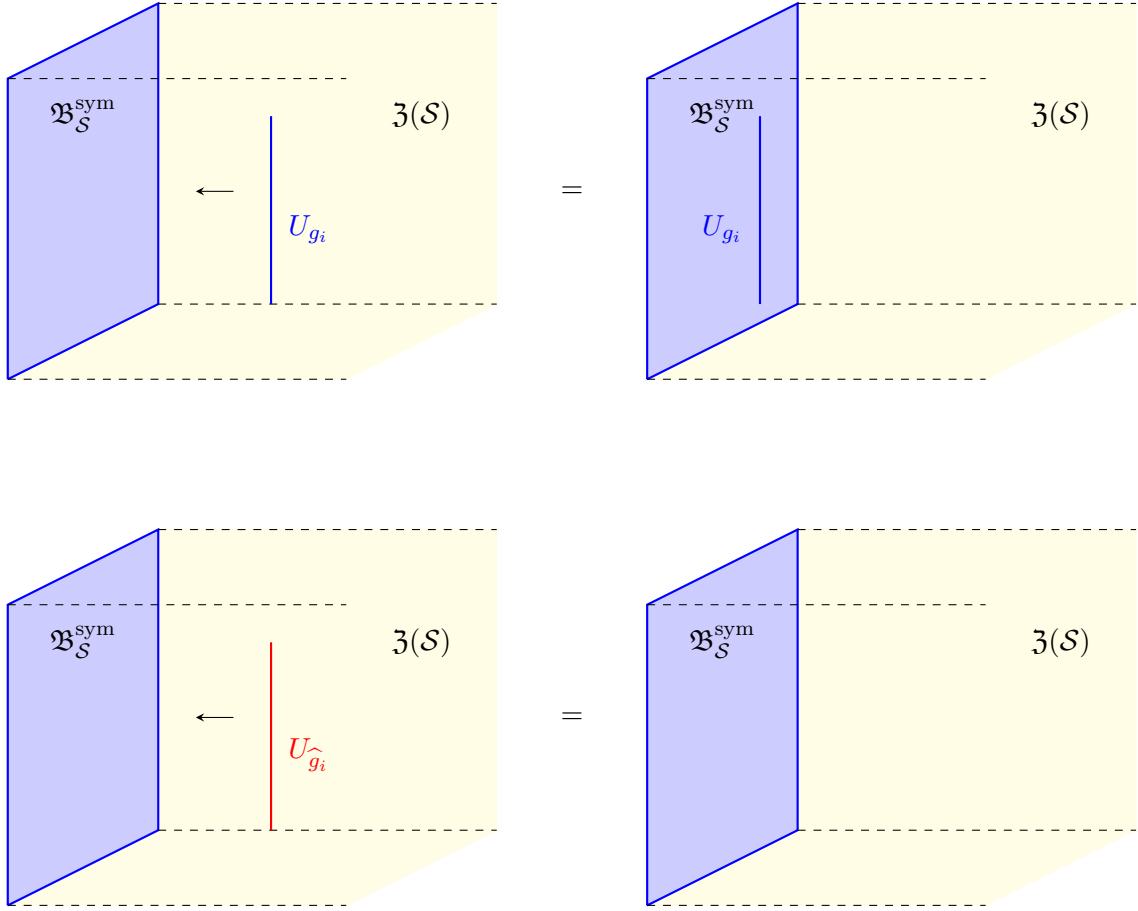

Both the symmetries and charged operators of $\fT$ are encoded in topological operators of the SymTFT. Recall from the discussion in section \ref{hfgt} that there are two types of topological operators
\be
\ba
U_{\wh g_i}&=\int \wh g_i(a_{p_i+1}),\qquad \wh g_i\in\wh G^{(p_i)}\,,\\
U_{g_i}&=\int b_{d-p_i-1}(g_i),\qquad g_i\in G^{(p_i)}\,,
\ea
\ee
in the bulk. When pushed to the symmetry boundary, the Dirichlet b.c. forces $U_{\wh g_i}$ to become invisible, while due to the Neumann b.c. $U_{g_i}$ survive and generate the symmetry $\cS$ of $\fB^{\text{sym}}_\cS$. See figure \ref{SymTFT3}.

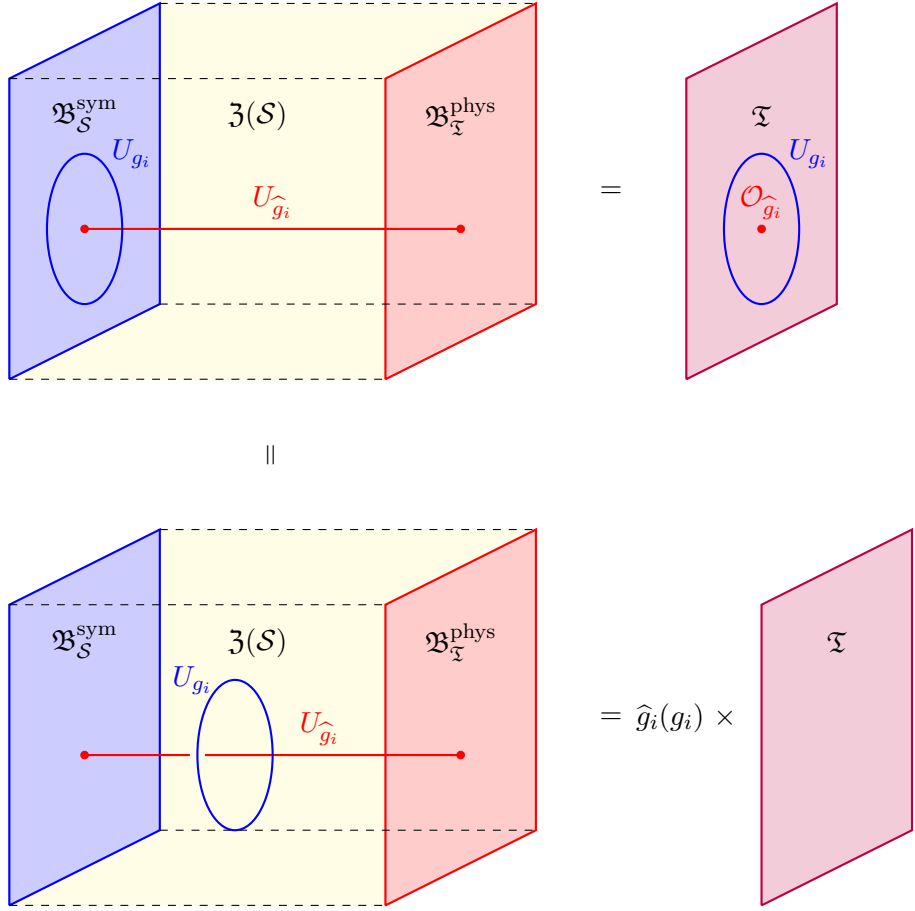
\begin{figure}
\centering
\scalebox{1}{
\begin{tikzpicture}
\draw [yellow, fill= yellow, opacity =0.1]
(0,0) -- (0,4) -- (2,5) -- (7,5) -- (7,1) -- (5,0)--(0,0);
\begin{scope}[shift={(0,0)}]
\draw [white, thick, fill=white,opacity=1]
(0,0) -- (0,4) -- (2,5) -- (2,1) -- (0,0);
\end{scope}
\begin{scope}[shift={(5,0)}]
\draw [white, thick, fill=white,opacity=1]
(0,0) -- (0,4) -- (2,5) -- (2,1) -- (0,0);
\end{scope}
\begin{scope}[shift={(5,0)}]
\draw [red, thick, fill=red,opacity=0.2] 
(0,0) -- (0, 4) -- (2, 5) -- (2,1) -- (0,0);
\draw [red, thick]
(0,0) -- (0, 4) -- (2, 5) -- (2,1) -- (0,0);
\node at (1,3.5) {$\Bphys_{\mathfrak{T}}$};
\end{scope}
\begin{scope}[shift={(0,0)}]
\draw [blue, thick, fill=blue,opacity=0.2]
(0,0) -- (0, 4) -- (2, 5) -- (2,1) -- (0,0);
\draw [blue, thick]
(0,0) -- (0, 4) -- (2, 5) -- (2,1) -- (0,0);
\node at (1,3.5) {$\Bsym_\cS$};
\end{scope}
\node at (3.3, 3.5) {$\mathfrak{Z}({\mathcal{S}})$};
\draw[dashed] (0,0) -- (5,0);
\draw[dashed] (0,4) -- (5,4);
\draw[dashed] (2,5) -- (7,5);
\draw[dashed] (2,1) -- (7,1);
\begin{scope}[shift={(9,0)}]
\node at (-1,2.5) {$=$} ;
\draw [purple, thick, fill=purple,opacity=0.2] 
(0,0) -- (0, 4) -- (2, 5) -- (2,1) -- (0,0);
\draw [purple, thick]
(0,0) -- (0, 4) -- (2, 5) -- (2,1) -- (0,0);
\node at (1,3.5) {$\fT$};
\end{scope}
\draw [thick,red](1,2) node (v1) {} -- (6,2);
\draw [thick,blue] (v1) ellipse (0.5 and 1);
\node[blue] at (1.5,3) {~~$U_{g_i}$};
\node[red] at (3.5,2.35) {$U_{\wh g_i}$};
\draw [red,fill=red] (10,2) node (v2) {} ellipse (0.05 and 0.05);
\draw [thick,blue] (v2) ellipse (0.5 and 1);
\node[blue] at (10.5,3) {~~$U_{g_i}$};
\node[red] at (10,2.35) {$\cO_{\wh g_i}$};
\draw [red,fill=red] (6,2) node (v2) {} ellipse (0.05 and 0.05);
\draw [red,fill=red] (1,2) node (v2) {} ellipse (0.05 and 0.05);
\node[rotate=90] at (3.5,-1) {=};
\begin{scope}[shift={(0,-7)}]
\draw [yellow, fill=yellow, opacity=0.1]
(0,0) -- (0,4) -- (2,5) -- (7,5) -- (7,1) -- (5,0)--(0,0);
\begin{scope}[shift={(0,0)}]
\draw [white, thick, fill=white,opacity=1]
(0,0) -- (0,4) -- (2,5) -- (2,1) -- (0,0);
\end{scope}
\begin{scope}[shift={(5,0)}]
\draw [white, thick, fill=white,opacity=1]
(0,0) -- (0,4) -- (2,5) -- (2,1) -- (0,0);
\end{scope}
\begin{scope}[shift={(5,0)}]
\draw [red, thick, fill=red,opacity=0.2] 
(0,0) -- (0,4) -- (2,5) -- (2,1) -- (0,0);
\draw [red, thick]
(0,0) -- (0,4) -- (2,5) -- (2,1) -- (0,0);
\node at (1,3.5) {$\Bphys_{\mathfrak{T}}$};
\end{scope}
\begin{scope}[shift={(0,0)}]
\draw [blue, thick, fill=blue,opacity=0.2]
(0,0) -- (0,4) -- (2,5) -- (2,1) -- (0,0);
\draw [blue, thick]
(0,0) -- (0,4) -- (2,5) -- (2,1) -- (0,0);
\node at (1,3.5) {$\Bsym_\cS$};
\end{scope}
\node at (3.3,3.5) {$\mathfrak{Z}({\mathcal{S}})$};
\draw[dashed] (0,0) -- (5,0);
\draw[dashed] (0,4) -- (5,4);
\draw[dashed] (2,5) -- (7,5);
\draw[dashed] (2,1) -- (7,1);
\draw [thick,red](2.6,2)  -- (6,2);
\draw [thick,blue] (3,2) ellipse (0.5 and 1);
\node[blue] at (2.5,3) {$U_{g_i}$~~};
\node[red] at (4,2.35) {~~$U_{\wh g_i}$};
\draw [red,fill=red] (6,2) node (v2) {} ellipse (0.05 and 0.05);
\draw [red,fill=red] (1,2) node (v2) {} ellipse (0.05 and 0.05);
\end{scope}
\draw [thick,red](2.4,-5) -- (1,-5);
\begin{scope}[shift={(9,-7)}]
\node at (-1,2.5) {$=$} ;
\draw [purple, thick, fill=purple,opacity=0.2] 
(1,0) -- (1,4) -- (3,5) -- (3,1) -- (1,0);
\draw [purple, thick]
(1,0) -- (1,4) -- (3,5) -- (3,1) -- (1,0);
\node at (2,3.5) {$\fT$};
\end{scope}
\node at (9,-4.5) {$\wh g_i(g_i)~\times$};
\end{tikzpicture}
}
\caption{Compactifying the bulk topological operator $U_{\wh g_i}$ along the interval produces an operator $\cO_{\wh g_i}$ in the theory $\fT$ having charge $\wh g_i\in\whG{p_i}$ under the $\G{p_i}$ $p_i$-form symmetry.}
\label{SymTFT4}
\end{figure}

The fact that $U_{\wh g_i}$ become invisible on $\fB^{\text{sym}}_\cS$ means that they can end along $\fB^{\text{sym}}_\cS$. We can thus sandwich the $(p_i+1)$-dimensional $U_{\wh g_i}$ operators between the two boundaries to construct $p_i$-dimensional operators in the theory $\fT$, which are generally non-topological as an end of $U_{\wh g_i}$ along $\fB^{\text{phys}}_\fT$ is generally non-topological. See figure \ref{SymTFT4}. These are precisely the operators of $\fT$ charged under $p_i$-form symmetries! The fact that these operators are charged follows from the fact that $U_{\wh g_i}$ are charged under $U_{g_i}$ in the bulk SymTFT. See figure \ref{SymTFT4}.

\paragraph{Gauging as a Change of b.c.}
Let us return to our general setup and replace the topological boundary condition $\Bsym_\cS$ by another topological boundary condition $\Bsym_{\cS'}$ of the SymTFT $\fZ(\cS)$. The b.c. $\Bsym_{\cS'}$ has topological operators living on it which constitute a symmetry $\cS'$ which in general is different from $\cS$.

\begin{figure}
\centering
\scalebox{1}{
\begin{tikzpicture}
\draw [yellow, fill= yellow, opacity =0.1]
(0,0) -- (0,4) -- (2,5) -- (7,5) -- (7,1) -- (5,0)--(0,0);
\begin{scope}[shift={(0,0)}]
\draw [white, thick, fill=white,opacity=1]
(0,0) -- (0,4) -- (2,5) -- (2,1) -- (0,0);
\end{scope}
\begin{scope}[shift={(5,0)}]
\draw [white, thick, fill=white,opacity=1]
(0,0) -- (0,4) -- (2,5) -- (2,1) -- (0,0);
\end{scope}
\begin{scope}[shift={(5,0)}]
\draw [red, thick, fill=red,opacity=0.2] 
(0,0) -- (0, 4) -- (2, 5) -- (2,1) -- (0,0);
\draw [red, thick]
(0,0) -- (0, 4) -- (2, 5) -- (2,1) -- (0,0);
\node at (1,3.5) {$\Bphys_{\mathfrak{T}}$};
\end{scope}
\begin{scope}[shift={(0,0)}]
\draw [blue, thick, fill=blue,opacity=0.2]
(0,0) -- (0, 4) -- (2, 5) -- (2,1) -- (0,0);
\draw [blue, thick]
(0,0) -- (0, 4) -- (2, 5) -- (2,1) -- (0,0);
\node at (1,3.5) {$\Bsym_{\cS'}$};
\end{scope}
\node at (3.3, 3.5) {$\mathfrak{Z}({\mathcal{S}})$};
\draw[dashed] (0,0) -- (5,0);
\draw[dashed] (0,4) -- (5,4);
\draw[dashed] (2,5) -- (7,5);
\draw[dashed] (2,1) -- (7,1);
    \draw [red, thick] (1, 1) -- (1, 2) ; 
    \draw [red, thick] (0.5, 3) -- (1, 2) -- (1.5, 3.2) ; 
    \node[red] at (0.5, 1.5) {$\cS'$};
\begin{scope}[shift={(9,0)}]
\node at (-1,2.3) {$=$} ;
\draw [purple, thick, fill=purple,opacity=0.2] 
(0,0) -- (0, 4) -- (2, 5) -- (2,1) -- (0,0);
\draw [purple, thick]
(0,0) -- (0, 4) -- (2, 5) -- (2,1) -- (0,0);
\node at (1,3.5) {$\fT'=\fT/\cS$};    
    \draw [red, thick] (1, 1) -- (1, 2) ; 
    \draw [red, thick] (0.5, 3) -- (1, 2) -- (1.5, 3.2) ; 
    \node[red] at (0.5, 1.5) {$\cS'$};
\end{scope}
\end{tikzpicture}
}
\caption{Performing the sandwich construction with another topological boundary condition $\Bsym_{\cS'}$ produces a theory $\fT'$ obtained by gauging (part of) the symmetry $\cS$, possibly with discrete torsion. The topological operators of $\Bsym_{\cS'}$ generate a symmetry $\cS'$ of $\fT'$ which is a combination of dual and residual symmetries under gauging.}
\label{SymTFT5}
\end{figure}
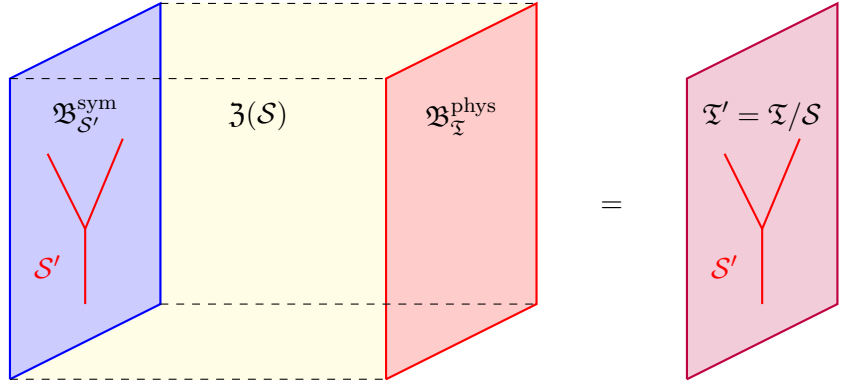

Performing now the sandwich construction with b.c. $\Bsym_{\cS'}$ leads to a $d$-dimensional theory $\fT'$ that has symmetry $\cS'$. The two $d$-dimensional theories are related by a discrete gauging. That is, we can obtain $\fT'$ by gauging a part of the symmetry $\cS$ of $\fT$, possibly with discrete torsion. See figure \ref{SymTFT5}. $\cS'$ is the symmetry obtained by combining the dual symmetries arising from gauging along with residual symmetries that are left ungauged.

\begin{example}[Gauging a Higher-Form Symmetry]{}
Consider a theory $\fT$ carrying a non-anomalous discrete $p$-form symmetry $\G p$ (for $p=0$ we only consider abelian $\G0$). As we discussed above, the action for the SymTFT is
\be
2\pi\int a_{p+1}\cup_\eta \delta b_{d-p-1}\,.
\ee
Let us choose a topological boundary condition $\Bsym_{\cS'}$
\bit
\item \textbf{Neumann} boundary condition (b.c.) for gauge field $a_{p+1}$.
\item \textbf{Dirichlet} b.c. for gauge field $b_{d-p-1}$.
\eit
The sandwich construction now constructs the theory
\be
\fT'=\fT/\G p\,.
\ee
Indeed, the topological operators on $\Bsym_{\cS'}$ are $U_{\wh g}$ which generate
\be
\G{d-p-2}=\whG p\,,
\ee
symmetry of $\fT/\G p$. The topological operators $U_g$ now end on $\Bsym_{\cS'}$ and thus give rise to charged operators of $\fT/\G p$.
\end{example}

\paragraph{Obstructions to Gauging.}
Obstructions to gauging, i.e.\ 't Hooft anomalies of $\cS$, now manifest as obstructions for the existence of topological boundary conditions of $\fZ(\cS)$. Consider again higher-form symmetries, in which case the SymTFT has action (\ref{SymAct}). The presence of $\cL^{\text{eff}}_{d+1}$ modifies the equations of motion for $b_{d-p_i-1}$ fields as
\be
\delta b_{d-p_i-1}=f_i[a_{p_1+1},a_{p_2+1},\cdots, a_{p_n+1}]\,,
\ee
where $f_i$ is some $(d-p_i)$-cochain. Now we can impose Dirichlet b.c. on $b_{d-p_i-1}$ for some fixed $i$ only if we are imposing Dirichlet b.c.\ on enough $a_{p_j+1}$ fields such that $f_i$ becomes purely a background field. This condition might not be solvable as it might require imposing Dirichlet b.c.\ on both $b_{d-p_i-1}$ and $a_{p_i+1}$ for same $i$, which is a contradiction. Let us see this explicitly in an example:

\begin{example}[Pure $SO(6)$ Yang-Mills theory]{}
Recall from example \ref{SO6} that pure $SO(6)$ gauge theory has a mixed 't Hooft anomaly between $\G1=\Z^e_2$ electric 1-form symmetry and $\G{d-3}=\Z^m_2$ $(d-3)$-form symmetry given by
\be\label{thoSO6}
\cI_{d+1}[A_2^e,A_{d-2}^m]=\exp\left(\pi i\int A_{d-2}^m\cup\Bock(A_2^e)\right)\,.
\ee
The SymTFT has action
\be
\pi \int \left(a_2^e\cup \delta b_{d-2}^e+a_{d-2}^m\cup \delta b_{2}^m+a_{d-2}^m\cup\Bock(a_2^e)\right)\,.
\ee
The equation of motion descending from $a_{d-2}^m$ is
\be
\delta b_2^m=\Bock(a_2^e)\,,
\ee
which means that imposing Dirichlet b.c.\ on $b_2^m$ requires imposing Dirichlet b.c.\ on $a_2^e$. In other words, we cannot impose Dirichlet b.c.\ on both $b_2^m$ and $b_{d-2}^e$ simultaneously, since the latter Dirichlet b.c.\ implies Neumann b.c. for $a_2^e$. Said differently, we cannot gauge both the electric and magnetic higher-form symmetries simultaneously, which is also the obstruction imposed by the 't Hooft anomaly (\ref{thoSO6}).
\end{example}

\section{Higher-Group Symmetries}\label{hgs}
Higher group symmetries arise when higher-form symmetries of different degrees mix with each other. This means that performing a gauge transformation of background field $B_{p+1}$ for a $p$-form symmetry shifts also the background field $B_{q+1}$ for a $q$-form symmetry. If $r$ is the largest integer for which an $r$-form symmetry participates in the higher-group, then the higher-group is an $\bm(r+1)$\textbf{-group}. We will focus mainly on simplest types of higher-groups, namely 2-group symmetries, which involve only 0-form and 1-form symmetries. 

A lot of work has been performed on this subject. The foundational papers discussing higher-group symmetries in QFT are \cite{Sharpe:2015mja,Tachikawa:2017gyf,Cordova:2018cvg,Benini:2018reh}. The discussion of higher-group symmetries in strongly coupled QFTs via string theory and compactifications of higher-dimensional QFTs began in \cite{Apruzzi:2021vcu,Bhardwaj:2021wif,Apruzzi:2021mlh}. See \cite{Hsin:2020nts,Perez-Lona:2023llv,Pantev:2022kpl,Lee:2021crt,Brennan:2020ehu,Cordova:2020tij,Iqbal:2020lrt,Barkeshli:2022edm,Carta:2022fxc,Nawata:2023rdx,Bhardwaj:2022scy,DelZotto:2022joo,DelZotto:2020sop,DelZotto:2015isa,Yu:2020twi,DeWolfe:2020uzb,Hidaka:2020izy,Hidaka:2021mml,Hidaka:2021kkf} for a sample of other works (in no particular order) on higher-group symmetries.

\subsection{Continuous Higher-Group Symmetries}
\paragraph{Example: Gauging Mixed 't Hooft Anomaly.}
Let us begin with an example of a higher-group symmetry in which the constituent $p$-form symmetry groups are all continuous \cite{Cordova:2018cvg}. Gauged versions of such continuous higher-groups are famously encountered in the Green-Schwarz mechanism of anomaly cancellation.

Let us begin with a $d$-dimensional theory having a $p$-form symmetry $U(1)^{(p)}$ and a $(d-2p-4)$-form symmetry $U(1)^{(d-2p-4)}$ with a mixed 't Hooft anomaly
\begin{equation}
    W[B_{p+1}+ d\Lambda_p,B_{d-2p-3} + d\Lambda_{d-2p-4}] = W[B_{p+1},B_{d-2p-3}]  + 2\pi \kappa \int \Lambda_{p}\wedge H_{p+2} \wedge H_{d-2p-2} \,.
    \label{anomalycontin}
\end{equation}
Here $\kappa \in \mathbb{Z}$ characterizes the anomaly, $W$ is the effective action and $H_{p+2}$ and $H_{d-2p-2}$ are the field strengths of background fields $B_{p+1}$ and $B_{d-2p-3}$ respectively. 

Since there is no anomaly term dependent upon $\Lambda_{d-2p-4}$, we can gauge the $U(1)^{(d-2p-4)}$ symmetry. Naively, the gauging would convert the mixed 't Hooft anomaly into an ABJ anomaly, which would indicate that the gauged theory does not carry the $U(1)^{(p)}$ symmetry. However, this ABJ anomaly can be cancelled in a fashion similar to the Green-Schwarz mechanism. Notice that the gauging procedure introduces a new $(2p+1)$-form symmetry $U(1)^{(2p+1)}$ generated by Noether current
\be
j^{d-2p-2}=H_{d-2p-2}\,.
\ee
As we know, the background field $B_{2p+2}$ of this symmetry couples as
\begin{equation}
    2\pi \int B_{2p+2} \wedge H_{d-2p-2}\,.
    \label{newaction}
\end{equation}
Naively, under $U(1)^{(2p+1)}$ gauge transformations, this background field transforms as $B_{2p+2} \rightarrow B_{2p+2} + d\Lambda_{2p+1}$, but if we were to modify the gauge transformations of $B_{2p+2}$ in such a way that $B_{2p+2}$ also transforms under $\Lambda_{p}$, we would be able to cancel the ABJ anomaly (\ref{anomalycontin}) by this new term in the action \eqref{newaction}. More explicitly, let $B_{2p+2}$ transform as:
\begin{equation}
    B_{2p+2} \rightarrow B_{2p+2} + d\Lambda_{2p+1} - \kappa \Lambda_{p}\wedge H_{p+2}\,.
    \label{gaugetrans}
\end{equation}
Then the anomaly \eqref{anomalycontin} is canceled and the effective action transforms as:
\begin{equation}
    W[B_{p+1}+ d\Lambda_{p}, B_{2p+2} + d\Lambda_{2p+1} - \kappa \Lambda_{p}\wedge H_{p+2} ] = W[B_{p+1},B_{2p+2}]\,.
\end{equation}
The transformation of $B_{2p+2}$ under gauge transformations for $B_{p+1}$ implies that  $U(1)^{(p)}$ and $U(1)^{(2p+1)}$ symmetries combine to form a $\bm(2p+2)$\textbf{-group symmetry}. The simplest case is $p=0$, in which case we have a \textbf{2-group symmetry}.

Note that the ordinary field strength $H_{2p+3} = dB_{2p+2}$ is no longer gauge invariant, but instead needs to be modified to
\begin{equation}
    H_{2p+3} = dB_{2p+2} + \kappa B_{p+1} \wedge H_{p+2}\,,
    \label{fieldstrength}
\end{equation}
which is now no longer closed. In fact, the term containing $\kappa$ characterizing the higher-group is captured precisely in the failure of closure of $H_{2p+3}$
\be
dH_{2p+3}= \kappa H_{p+2} \wedge H_{p+2}\,.
\ee

\paragraph{General Structure.}
In general, if we have relationships of the form
\be
H_{p+2}=dB_{p+1}+\Theta(B_{p_1+1},B_{p_2+1},\cdots, B_{p_n+1})\,,
\ee
for the field strength $H_{p+2}$ of a continuous $p$-form symmetry, where $\Theta$ is a function of background fields $B_{p_i+1}$ of other continuous $p_i$-form symmetries, such that $H_{p+2}$ is invariant under all gauge transformations, then we say that we have a continuous higher-group symmetry.

\subsection{Discrete Higher-Group Symmetries}
\subsubsection{Motivation: Group Extension}
Discrete higher-group symmetries involve extensions of $p$-form symmetries by other $p_i$-form symmetries. Thus, we begin by recalling the simplest case of extensions, namely extensions of abelian groups.

\paragraph{Extension Class of a Group Extension.}
Let $\K0$ and $\H0$ be two abelian groups that sit in a short exact sequence
\be\label{ses2}
0\to\H0\to\G0\to\K0\to0\,.
\ee
If the short exact sequence does not split, then we cannot express $\G0$ as a direct product of $\H0$ and $\K0$
\be
\G0\not\cong\H0\times\K0\,.
\ee
Physically, we should think of these groups as 0-form symmetries of a theory.

One way to understand the extension $\G0$ is to represent its elements as pairs
\be
g=(h,k),\qquad h\in\H0,~k\in\K0\,.
\ee
Then the group operation takes the form
\be
g_1+g_2=(h_1+h_2+e(k_1,k_2),k_1+k_2)\,,
\ee
described in terms of a $\H0$-valued 2-cochain $e$ on $\K0$. Imposing associativity of group operation implies that $e$ is a 2-cocycle
\be
\delta e=0\,.
\ee
Finally there is a freedom of shifting
\be
(h,k)\to (h+f(k),k)\,,
\ee
which does not change the extension, where $f$ is an $\H0$-valued 1-cochain on $\K0$. This shift modifies $e$ as
\be
e\to e+\delta f\,.
\ee
Thus, the short exact sequence (\ref{ses2}), and hence the form of the extended group $\G0$, is described a cohomology class
\be
[e]\in H^2(\K0,\H0)\,.
\ee

\begin{figure}
\centering
\scalebox{1.1}{
\begin{tikzpicture}
\draw [thick,blue](0,1.5) -- (0,-0.5) (0,-0.5) -- (-1.5,-2) (0,-0.5) -- (1.5,-2);
\draw [thick,red](2,1) -- (0,-0.5);
\draw [red,fill=red] (0,-0.5) ellipse (0.05 and 0.05);
\node[blue] at (-1.5,-2.5) {$k_1$};
\node[blue] at (1.5,-2.5) {$k_2$};
\node[blue] at (0,2) {$k_1k_2$};
\node[red] at (2.5,1.5) {$e(k_1,k_2)$};
\end{tikzpicture}
}
\caption{Group extension (\ref{ses2}) in terms of topological codimension-1 operators for $\K0$ (shown in blue) and $\H0$ (shown in red).}
\label{extf}
\end{figure}
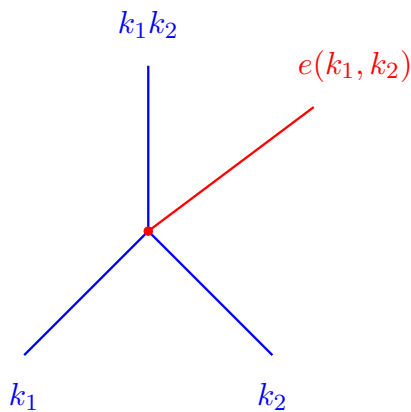

\paragraph{Interpretation in Terms of Topological Operators.}
The cocycle $e$ is physically interpreted as follows: as two codimension-1 topological operators associated to $k_1,k_2\in\K0$ come together to form the topological operator associated to $k_1k_2\in\K0$, the resulting codimension-2 junction acts as a source of a codimension-1 topological operator associated to
\be
e(k_1,k_2)\in\H0\,.
\ee
See figure \ref{extf}.

\paragraph{Implication for Background Fields.}
The fact that junctions of $\K0$ topological operators provide sources for $\H0$ topological operators means that the background field for $\H0$ is not closed when a $\K0$ background is turned on. It is easy to see that the two background fields are related as
\be\label{Be}
\delta B_1^H=\left(B_1^K\right)^*e\,,
\ee
where $B_1^H$ and $B_1^K$ are background fields for $\H0$ and $\K0$ respectively, and $\left(B_1^K\right)^*e$ is pullback of $e$ via $B_1^K$, where such pullbacks were discussed right after (\ref{2dSPT}).

\subsubsection{General Structure}
\paragraph{Higher-Groups.}
Generalizing (\ref{Be}), we say that we have a discrete higher-group symmetry if a background field of a $\G p$ $p$-form symmetry is not closed
\be\label{BT}
\delta B_{p+1}+\Theta(B_{p_1+1},B_{p_2+1},\cdots B_{p_n+1})=0\,,
\ee
where $\Theta$ is a $\G p$-valued $(p+2)$-cocycle on spacetime which is a function of other background fields $B_{p_i+1}$. Similarly, background fields for other higher-form symmetries in the higher-group may be non-closed as well.

There is a neat interpretation of (\ref{BT}) in terms of topological operators. The Poincaré dual of $\Theta(B_{p_1+1},B_{p_2+1},\cdots B_{p_n+1})$ is a $(d-p-2)$-cycle which acts as a source of the codimension-$(p+1)$ $\G p$ topological operators.

\paragraph{2-Groups.}
Let us specialize to a 2-group symmetry comprised of a $\G0$ 0-form symmetry and a $\G1$ 1-form symmetry. A 2-group symmetry is then specified by a \textbf{Postnikov class},
\be
[\Theta]\in H^3\left(\G0,\G1\right)\,,
\ee
capturing the non-closure of the background field $B_2$ of the 1-form symmetry
\be
\delta B_2+B_1^*\Theta=0\,,
\ee
where $B_1$ is the background field for 0-form symmetry.

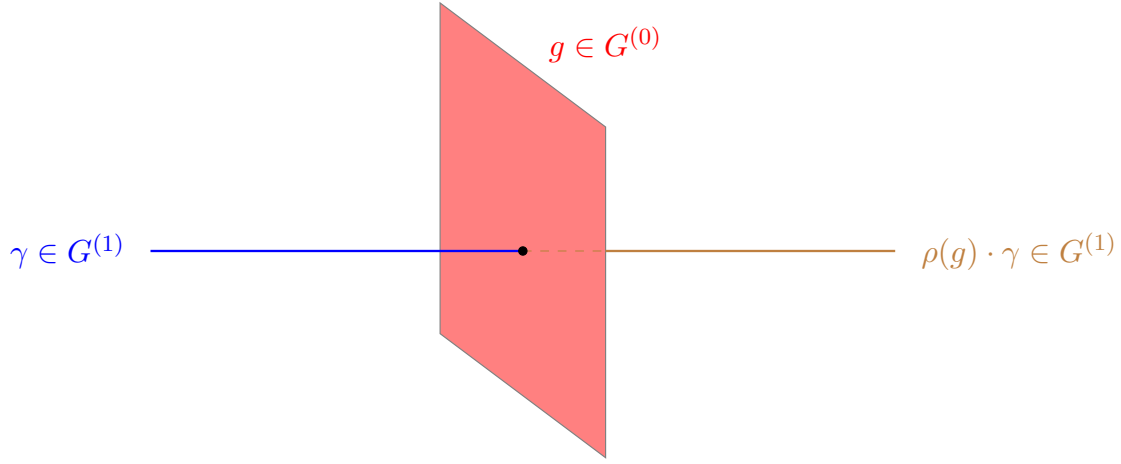
\begin{figure}
\centering
\scalebox{1.1}{
\begin{tikzpicture}
\draw[fill=red,opacity=0.5] (-1,1) -- (-1,2.5) -- (1,1) -- (1,-3) -- (-1,-1.5)--(-1,1);
\draw[thick,blue] (-4.5,-0.5) -- (0,-0.5);
\draw [thick,brown] (4.5,-0.5) -- (1,-0.5);
\draw [brown,dashed](0,-0.5) -- (1,-0.5);
\node[blue] at (-5.5,-0.5) {$\gamma\in G^{(1)}$};
\node[brown] at (6,-0.5) {$\rho(g)\cdot\gamma\in G^{(1)}$};
\node[red] at (1,2) {$g\in G^{(0)}$};
\draw [black,fill=black] (0,-0.5) ellipse (0.05 and 0.05);
\end{tikzpicture}
}
\caption{A codimension-1 topological operator generating a 0-form symmetry $g\in\G0$ acting on a codimension-2 topological operator generating a 1-form symmetry $\gamma\in\G1$ and converting it into a codimension-2 topological operator generating a 1-form symmetry $\rho(g)\cdot\gamma\in\G1$.}
\label{2gAct}
\end{figure}

\paragraph{More General 2-Groups.}
In the above discussion, we have not included non-trivial actions of 0-form symmetries on $p$-form symmetries. See figure \ref{2gAct}. Including such actions in the case of 2-groups, we find the most general discrete 2-group $\bG^{(2)}$ is specified as
\be
\bG^{(2)}=\left\{\G0,\G1,\rho,[\Theta]\right\}\,,
\ee
where $\G0$ and $\G1$ are discrete 0-form and 1-form symmetry groups, $\rho$ is the action of $\G0$ on $\G1$ which is a homomorphism
\be
\rho:~\G0\to\text{Aut}\left(\G1\right)\,,
\ee
and finally the \textbf{Postnikov class} is
\be
[\Theta]\in H^3_\rho\left(\G0,\G1\right)\,,
\ee
where the cohomology group is twisted by $\rho$.

\begin{tech}[Twisted Group Cohomology]{}
Given an action of the form
\be
\rho:~\G0\to\text{Aut}(\G1)\,,
\ee
we can modify the differential for group cohomology as
\be
\ba
\delta_\rho\alpha_p(g_1,g_2,\cdots,g_{p+1})=&\rho(g_1)\cdot\alpha_p(g_2,g_3,\cdots,g_{p+1})\alpha_p^{s(p+1)}(g_1,g_2,\cdots,g_p)\\
&\prod_{i=1}^p\alpha_p^{s(i)}(g_1,g_2,\cdots,g_{i-1},g_ig_{i+1},g_{i+2},g_{i+3}\cdots,g_{p+1})\,,
\ea
\ee
where $\alpha_p$ is an $\G1$-valued $p$-group-cochain on $\G0$ and
\be
\rho(g_1)\cdot\alpha_p(g_2,g_3,\cdots,g_{p+1})\,,
\ee
denotes the action of the automorphism $\rho(g_1)\in\text{Aut}(\G1)$ on $\alpha_p(g_2,g_3,\cdots,g_{p+1})\in \G1$.

The cohomology group obtained this way is denoted
\be
H^p_\rho\left(\G0,\G1 \right)\,,
\ee
and is referred to as the $p$-th \textbf{twisted group cohomology} of $\G0$-valued in $\G1$ with twist $\rho$.
\end{tech}

\paragraph{Split 2-Groups.}
When the action $\rho$ is non-trivial but the Postnikov class vanishes
\be
[\Theta]=0\in H^3_\rho\left(\G0,\G1\right)\,,
\ee
then the resulting 2-group is referred to as a \textbf{split 2-group}.

\begin{example}[Charge Conjugation in Pure $SU(N)$ Yang-Mills]{}
Consider $d$-dimensional pure $SU(N)$ Yang-Mills theory, which has an electric 1-form symmetry
\be
\G1=\Z_N\,.
\ee
The theory also contains a
\be
\G0=\Z_2\,,
\ee
charge conjugation 0-form symmetry associated to an outer-automorphism of $SU(N)$ which flips the Dynkin diagram.

The two symmetries combine to form a split 2-group as the outer-automorphism of $SU(N)$ acts as an outer-automorphism of its $\Z_N$ center such that
\be
p\longrightarrow -p,\qquad p\in\{0,1,2,\cdots,N-1\}\,.
\ee
\end{example}

\subsubsection{Relationship to Mixed 't Hooft Anomalies}
Consider a higher-group in which the only non-closure relationship is (\ref{BT}) and assume that the 't Hooft anomaly of the higher-group symmetry is such that it trivializes on the constituent $\G p$ $p$-form symmetry group. In such a situation, we can gauge $\G p$. In the gauged theory, we obtain a dual
\be
\G{d-p-2}=\whG p\,,
\ee
$(d-p-2)$-form symmetry. The fact that $\G p$ participated in a higher-group symmetry (\ref{BT}) is traded with the fact that the dual $\G{d-p-2}$ symmetry participates in a mixed 't Hooft anomaly with associated anomaly theory
\be
\cI_{d+1}=\exp\left(-2\pi is(d-p)\int B_{d-p-1}\cup_\eta\Theta(B_{p_1+1},B_{p_2+1},\cdots B_{p_n+1})\right),\qquad s(d-p)=(-1)^{d-p}\,.
\ee
The argument is the same the one taking (\ref{GHK}) to (\ref{miano}), already discussed in the context of group extensions.

\subsection{Mixed Continuous-Discrete 2-Group Symmetries}
Above we have discussed higher-groups in which all participating higher-form symmetries are continuous, and higher-groups in which all participating higher-form symmetries are discrete. One can also have higher-groups in which some higher-form symmetries are continuous and some higher-form symmetries are discrete. 

In this subsection, we discuss 2-group symmetries in which the 0-form symmetry group $\G0$ is continuous while the 1-form symmetry group $\G1$ is discrete. The defining relation for such a 2-group is
\be
\delta B_2+B_1^*\Theta=0\,,
\ee
where $\Theta$ is a representative of a class
\be
[\Theta]\in H^3(B\G0,\G1)\,,
\ee
referred to again as a \textbf{Postnikov class}, where $H^3(B\G0,\G1)$ is the third cohomology group of the classifying space $B\G0$ of $\G0$ with $\G1$ coefficients and $B_1^*\Theta$ is the pullback of $\Theta$ under a map $B_1$ from spacetime to $B\G0$ describing a principal bundle for 0-form background. 

\begin{tech}[Classifying Spaces]{}
A classifying space $BG$ for some group $G$ is obtained by beginning with a space $EG$ having two properties: 
\bit
\item All homotopy groups of $EG$ are trivial,
\item There is a free action (i.e.\ without fixed points) of $G$ on $EG$.
\eit
We then let
\be
BG:= EG/G\,,
\ee
where the quotient of $EG$ is taken by the free $G$ action.

Note that there is no unique classifying space $BG$. Given a fixed group $G$, there can be multiple spaces $EG$ and $BG$ satisfying the above criteria.

Note that $EG$ is naturally a principal $G$-bundle over $BG$, and is known as the universal $G$-bundle as one can obtain any principal bundle $P$ over any manifold $M$ as a pullback
\be
P=f^*(EG)\,,
\ee
of this universal bundle $EG$ over $BG$, under a map
\be
f:~M\to BG\,.
\ee
\end{tech}

We will focus on the study of \textbf{Bockstein 2-groups} which arise very naturally in gauge theories in various dimensions \cite{Hsin:2020nts,Bhardwaj:2021wif}.

\subsubsection{Global Form of 0-Form Symmetries}\label{GF}
Before describing Bockstein 2-groups, we have to describe some preliminaries regarding the precise global form of continuous Lie groups of 0-form symmetries. We will adopt the approach discussed in section 2 of \cite{Bhardwaj:2021ojs}.

\paragraph{0-Form Symmetry Algebra.}
Consider a continuous 0-form symmetry based on a compact semi-simple non-abelian Lie algebra
\be
\ff=\text{Lie algebra of 0-form symmetry}\,.
\ee
This 0-form algebra is simply the algebra of Noether current local operators
\be
j^\mu_a(x)\,,
\ee
where $a$ is an index in $\ff$, while $\mu$ is spacetime index.

\paragraph{0-Form Symmetry Group.}
Let 
\be
F=\text{Simply connected Lie group associated to $\ff$}\,.
\ee
It may happen that not all elements of $F$ act non-trivially on the theory, meaning that $F$ is not completely faithful in the sense discussed in section \ref{SPT}.

In general, a subgroup
\be
\cZ\subseteq Z(F)\,,
\ee
may be non-faithful, where $Z(F)$ is the center of $F$. Then, we have the following definition
\begin{note}[0-Form Symmetry Group]{}
The completely faithfully acting 0-form symmetry group is
\be
\G0=F/\cZ\,,
\ee
which is referred to simply as the \textbf{0-form symmetry group} of the theory, or sometimes as the global form of continuous 0-form symmetry.
\end{note}

How do we determine the group $\cZ$? This can be done by studying the action of $\ff$ on genuine local operators, which form representations $R_i$ of $\ff$. Then $\cZ$ is
\be\label{cZ}
\cZ=\text{Subgroup of $Z(F)$ leaving all local operator representations $R_i$ invariant}\,.
\ee
Including discrete 0-form symmetries along with continuous 0-form symmetries leads to disconnected 0-form symmetry groups. See \cite{Bhardwaj:2022scy} for more details.

\paragraph{0-Form Symmetry Bundles.}
A background field for a continuous 0-form symmetry is a connection on a principal bundle for the 0-form symmetry group $\G0$. The possible bundles are sensitive to the global form. In particular a $\G0$ bundle may not be liftable to an $F$ bundle, with an obstruction class $\left[w_2^f\right]$ which is $\cZ$-valued. 

\paragraph{Flavor Symmetry Groups in Gauge Theories.}
In a gauge theory, local operators arise from combinations of field strength and matter fields. As we have discussed earlier, if such a combination transforms under the gauge group $\cG$, then it gives rise to a non-genuine local operator living at the end of a Wilson line operator. Thus, genuine local operators correspond to gauge invariant combinations of field strength and matter fields. This leads to the following computation.

Let $\phi_i$ be matter fields transforming in irreducible representations $R^g_i$ of $\cG$ and irreducible representations of $R^f_i$ of $F$. This information provides us with charges
\be
\left(q^g_i,q^f_i\right)\in\wh Z(\cG)\times\wh Z(F)\,.
\ee
These charges generate a subgroup
\be\label{cM}
Q\subseteq\wh Z(\cG)\times\wh Z(F)\,.
\ee
The 0-form center charges of gauge invariant operators are
\be
Q_{\text{genuine}}=Q\cap\wh Z(F)\subseteq\wh Z(F)\,.
\ee
The subgroup $\cZ\subseteq Z(F)$ we are after is then the subgroup that pairs trivially with $Q_{\text{genuine}}$, whose Pontryagin dual can be computed as
\be\label{Qgc}
\wh\cZ=\frac{\wh Z(F)}{Q_{\text{genuine}}}\,.
\ee

\begin{example}[Abelian Gauge Theory]{AGT}
Consider a $d$-dimensional 
\be
\cG=U(1)\,,
\ee
gauge theory with 2 massless complex scalars of gauge charge
\be
q^g=2\,.
\ee
There is an
\be
\ff=\su(2)\,,
\ee
flavor symmetry algebra rotating the two complex scalars in fundamental representation of $\su(2)$. The associated simply connected group is
\be
F=SU(2)\,.
\ee
We have
\be
\wh Z(\cG)\times\wh Z(F)=\Z\times\Z_2\,,
\ee
with the matter fields generating the subgroup
\be
Q=\langle(2,1)\rangle\in\Z\times\Z_2\,,
\ee
from which we compute
\be
Q_{\text{genuine}}=Q\cap \wh Z(F)=0\,,
\ee
implying that
\be
\cZ=Z(F)=\Z_2\,,
\ee
and the 0-form flavor symmetry group is
\be
\G0=F/\cZ=SU(2)/\Z_2=SO(3)\,.
\ee
\end{example}

\begin{example}[Non-Abelian Gauge Theory]{NAGT}
Consider a $d$-dimensional non-abelian gauge theory with gauge group
\be
\cG=\Spin(4N+2)\,,
\ee
which has
\be
\wh Z(\cG)=\Z_4\,.
\ee
Additionally we have $2M$ complex scalars all transforming in vector representation of the gauge group for which we have
\be
q^g=2\in\Z_4\,,
\ee
for all $i$. We tune the theory such that we have a flavor algebra
\be
\ff=\sp(M) =\u\sp(2M)\,,
\ee
under which the scalars transform in $2M$-dimensional fundamental representation. The associated simply connected group is
\be
F=\text{Sp}(M)\,,
\ee
for which we have
\be
\wh Z(F)=\Z_2\,,
\ee
and the matter fields provide the charge
\be
q^f=1\in\Z_2\,,
\ee
and in total we have
\be
Q=\langle(2,1)\rangle\in\Z_4\times\Z_2\,.
\ee
Again we have
\be
Q_{\text{genuine}}=Q\cap \wh Z(F)=0\,,
\ee
implying that
\be
\cZ=Z(F)=\Z_2\,,
\ee
and the 0-form flavor symmetry group is
\be
\G0=F/\cZ=\text{Sp}(M)/\Z_2=\text{PSp}(M)\,.
\ee
\end{example}


\subsubsection{Bockstein 2-Group Symmetries}
Let us now provide a definition of Bockstein 2-group symmetry:

\bigskip

\begin{note}[Bockstein 2-Group Symmetry]{}
A Bockstein 2-group has Postnikov class of the form
\be
[\Theta]=\Bock\left(\left[w_2^f\right]\right)\,,
\ee
where $\left[w_2^f\right]$ is the $\cZ$-valued obstruction class for lifting the background $\G0$ bundle to an $F$ bundle and $\Bock$ is the Bockstein map associated to a short exact sequence
\be\label{ses3}
0\to\G1\to\cE\to\cZ\to0\,,
\ee
where $\cE$ is some extension of $\cZ$ by $\G1$. See the discussion around (\ref{Bock}) for more details on the Bockstein map.
\end{note}

\paragraph{Screening Interpretation.}
Just like 1-form symmetry group $\G1$ can be given an interpretation in terms of screening of line operators, the same holds true for the extension group $\cE$.

From the discussion in section \ref{screen}, let us recall that for $\G1$ we compute a group $\bD_1$ of equivalence classes of line operators with equivalence relation
\be
L\sim L'\,,
\ee
if there exists a local operator between the lines $L$ and $L'$. 
Then, we have
\be
\whG1=\bD_1\,.
\ee
Similarly, for $\cE$ we compute a group $\wt\bD_1$ of equivalence classes of line operators with equivalence relation
\be
L\sim L'\,,
\ee
only if there exists a local operator between $L$ and $L'$ of charge
\be
q^z=0\in\wh\cZ\,.
\ee
The charge $q^z$ for a local operator is computed from its center charge
\be
q^f\in\wh Z(F)\,,
\ee
by applying the natural surjective map
\be
\wh Z(F)\to\wh\cZ\,.
\ee
Note that even though genuine local operators all have $q^z=0$, this may not be true for non-genuine local operators. Then we have
\be
\wh\cE=\wt\bD_1\,.
\ee
Clearly the latter equivalence relation is finer than the former one, and so we have a surjective homomorphism
\be
\wt\bD_1\to\bD_1\,,
\ee
or equivalently
\be
\wh\cE\to\whG1\,,
\ee
which can be completed into the short exact sequence
\be
0\to\wh\cZ\to\wh\cE\to\whG1\to0\,,
\ee
Pontryagin dual to (\ref{ses3}). This recovers (\ref{ses3}) using this screening interpretation.

\paragraph{A 2-Group Background Field.}
By now we are used to the fact that a relationship of the form
\be
\delta B_2+\Bock\left(w_2^f\right)=0\,,
\ee
means that we should combine the 1-form background $B_2$ and the obstruction $w_2^f$ into a single $\cE$-valued closed background field
\be
B_2^w=i(B_2)+\wt w_2^f\,,
\ee
where $i$ is the injective homomorphism in (\ref{ses3}) and $\wt w_2^f$ is an $\cE$-valued lift of $w_2^f$ using the surjective homomorphism in (\ref{ses3}).

\paragraph{Bockstein 2-Groups in Gauge Theories.}
Bockstein 2-group symmetries are ubiquitous in gauge theories. We begin with a gauge theory of the form considered in section \ref{GF} where we computed the 0-form symmetry group, and include the electric 1-form symmetry into the mix. 

We would like to compute $\wt\bD_1$ which is easily recognized as
\be
\wh\cE=\wt\bD_1=\frac{\wh Z(\cG)\times\wh Z(F)}{Q}\,,
\ee
where $Q$ is the charge matrix described around (\ref{cM}). Thus, we have
\be
\cE=\text{Subgroup of $Z(\cG)\times Z(F)$ that leaves $Q$ invariant}\,.
\ee
It is easy to see that, almost by definition, $\G1$ is simply the subgroup of $\cE$ leaving $Q$ invariant comprised of elements whose projection to $Z_F$ is trivial
\be
\G1=\{(*,0)\in\cE\subseteq Z(\cG)\times Z(F)\}\,.
\ee
This can now be completed into the short exact sequence (\ref{ses3}) determining the Bockstein 2-group symmetry.

\begin{example}[Abelian Gauge Theory]{}
Let us revisit the example \ref{AGT} and compute the Bockstein 2-group symmetry present in the abelian gauge theory. We have
\be
\wh\cE=\frac{\Z\times\Z_2}{\langle(2,1)\rangle}=\Z_4\,,
\ee
which is easy to see as
\be
(0,1)\sim (-2,0),\qquad (4,0)\sim (0,0)\,.
\ee
The short exact sequence (\ref{ses3}) then has to be of the form
\be\label{ses4}
0\to\Z_2\to\Z_4\to\Z_2\to0\,.
\ee
There is a unique such short exact sequence, so we don't have to compute anything else. The 2-group relation takes the form
\be
\delta B_2=\Bock\left(w_2^f\right)=w_3^f\,,
\ee
where $w_2^f$ and $w_3^f$ are second and third Stiefel-Whitney classes of
\be
\G0=SO(3)\,,
\ee
bundles, where we have used the property that Bockstein map associated to short exact sequence (\ref{ses4}) takes second Stiefel-Whitney class to third Stiefel-Whitney class.
\end{example}

\begin{example}[Non-Abelian Gauge Theory]{}
Let us revisit the example \ref{NAGT} and compute the Bockstein 2-group symmetry present in the non-abelian gauge theory. We have
\be
\wh\cE=\frac{\Z_4\times\Z_2}{\langle(2,1)\rangle}=\Z_4\,,
\ee
and again the relevant short exact sequence is (\ref{ses4}), which implies the presence of a 2-group symmetry of the form
\be
\delta B_2=\Bock\left(w_2^f\right)\,,
\ee
where $w_2^f$ is the obstruction for lifting $\G0=\text{PSp}(M)$ bundles to $F=\text{Sp}(M)$ bundles.
\end{example}

\paragraph{Mixed Gauge-Flavor Bundles.}
Recall that the background field $B_2$ for an electric 1-form symmetry of a gauge theory describes the obstruction of lifting $\cG/\G1$ bundles to $\cG$ bundles that the gauge theory sums over. Its non-closure now means that the gauge theory does not quite sum over $\cG/\G1$ bundles anymore. The correct Lie group to consider is what is known as the \textbf{structure group}
\be
\cS=\frac{\cG\times F}{\cE}\,.
\ee
A 0-form background fixes a bundle $B_1$ for the 0-form group $\G0$. The gauge theory sums over bundles for $\cS$ that project to $B_1$ and have an obstruction for lifting to a $\cG\times F$ bundle given by the 2-group background field $B_2^w$.

\section{Higher-Form Symmetries and String Theory}\label{adv}
In this section, we will briefly explain some key ideas involved in understanding higher-form symmetries in string-theoretic constructions of quantum field theories. We will consider two kinds of constructions. The first kind involve compactifying string/M-theory and taking limits decoupling dynamical gravity. The second kind are holographic setups in string/M-theory.

\subsection{Higher-Form Symmetries from Geometric Engineering}\label{sec:HFSgeomEng}
Many interesting quantum field theories are strongly coupled and do not admit a conventional Lagrangian description. Notably, this includes superconformal field theories (SCFTs) in 5d and 6d, and Argyres-Douglas theories in 4d. Such theories can be constructed by compactifying superstring theory or M-theory on an internal compactification manifold. Such a construction is known as \textbf{geometric engineering} of the QFT under study. Here we describe how $p$-form symmetry groups are encoded in geometric engineering setups, following mainly the references  \cite{Morrison:2020ool,Albertini:2020mdx,Bhardwaj:2020phs}.

\paragraph{Setup.}
Consider a $\cD$-dimensional string/M-theory with
\be
\cD\in\{10,11\} \,,
\ee
on a manifold of the form
\begin{equation}
 M_{\mathcal{D}} = M_{d} \times M_{D} \,,
\end{equation}
where $M_d$ is the space-time where the quantum field theory $\mathfrak{T}$ that is being geometrically engineered lives, and $M_D$ is the internal compactification manifold. 
Crucially, we assume that the internal manifold is non-compact, i.e.\ it has a boundary at infinity.
As is well-known in geometric engineering, Newton's constant $G_N$ of the compactified theory is set by inverse of the volume $\text{vol}(M_D)$ of the compactification manifold $M_D$, and in order to construct a quantum field theory without dynamical gravity we have to take the limit
\be
G_N\to 0\iff\text{vol}(M_D)\to\infty \,.
\ee

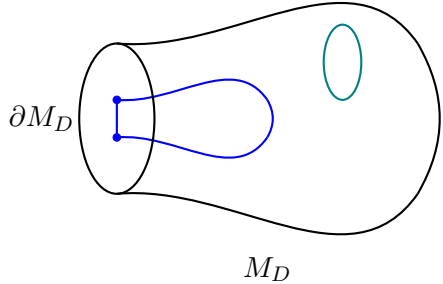
\begin{figure}
\centering
\begin{tikzpicture}
\draw[thick,blue] (0,0.25) -- (0,-0.25);
\draw[blue,fill=blue] (0,0.25) ellipse (0.05 and 0.05);
\draw[blue,fill=blue] (0,-0.25) ellipse (0.05 and 0.05);
\draw[thick,blue] (0,0.25) to [out=-5, in=125] (2,0.25);
\draw[thick,blue] (2,0.25) to [out=300, in=60](2,-0.25);
\draw[thick,blue] (0,-0.25) to [out=5, in=235] (2,-0.25);
\draw[thick] (0,0) ellipse (0.5 and 1);
\draw[thick] (0,1) to [out=-5, in=125] (4,1);
\draw[thick] (4,1) to [out=300, in=60](4,-1);
\draw[thick] (0,-1) to [out=5, in=235] (4,-1);
\draw[thick,teal] (3,0.75) ellipse (0.25 and 0.5);
\node at (-1,0) {$\partial M_D$}; 
\node at (2,-2) {$M_D$}; 
\end{tikzpicture}
\caption{Wrapping a brane on a non-compact cycle (shown in blue) gives rise to a non-dynamical defect. On the other hand, wrapping a brane on a compact cycle (shown in cyan) gives rise to a dynamical excitation.
}
\label{fig:relative_cycle}
\end{figure}

\paragraph{Defects and Dynamical Excitations.}
The question we want to investigate is how to construct higher-form symmetries in this geometric engineering set-up. For this we construct the groups of unscreened operators of various dimensions\footnote{Until recently, research in this direction within the realm of string theory realizations of QFTs focused on identifying the string theory origin of  the (not necessarily topological) charged defects under generalized symmetries. With regards to the origin of the \textit{topological} symmetry generators, \cite{GarciaEtxebarria:2019caf,Morrison:2020ool} contained some early discussions in terms of flux operators. More recent works \cite{Apruzzi:2022rei, GarciaEtxebarria:2022vzq, Heckman:2022muc, Heckman:2022xgu} explain how, in an appropriate topological limit, wrapped branes in both geometric engineering and holographic constructions describe the topological symmetry generators.}. This has two contributions:
\begin{itemize}
\item $p$-branes wrapping \textit{non-compact} $k$-cycles in the internal geometry $M_D$ give rise to $(p-k+1)$-dimensional defects in the QFT $\mathfrak{T}$ living on $M_d$. These are classified by relative homology $H_k(M_D, \partial M_D)$ of $M_D$ with respect to its boundary $\partial M_D$. See the box below for more details.
\item $p$-branes wrapping \textit{compact} $k$-cycles  give rise to $(p-k+1)$-dimensional dynamical excitations in the field theory $\fT$. These are classified by the homology group $H_k(M_D)$.
\end{itemize}
Compactification of branes on non-compact cycles gives rise to defects because the mass of the compactified object in the effective QFT $\fT$ is determined by the volume of the internal cycle, and infinitely massive objects can be regarded as defects. This is illustrated in figure \ref{fig:relative_cycle}.

\begin{tech}[Relative Homology]{}
By definition $H_k(M_D, \partial M_D)$ is given by \textit{relative} cycles modulo \textit{relative} boundaries. The definition of a relative cycle relaxes that of an ordinary cycle: a relative chain $x \in C_k(M_D)$ is a relative cycle if $\partial x \in C_{k-1}(\partial M_D)$.
In relative homology, a relative cycle $x$ is trivial if $x = \partial y + z$ with $y \in C_{k+1}(M_D)$ and $z \in C_k (\partial M_D)$.  
\end{tech}

\paragraph{Screening and $p$-Form Symmetries.}
Accounting for screening of defects by dynamical excitations, we find that $p$-branes wrapping $k$-cycles in $M_D$ give rise to the following group of unscreened $(p-k+1)$-dimensional operators
\be
\bD_{p,k} = H_k(M_D, \partial M_D;\Z) / H_k(M_D;\Z) \,,
\ee
Combining all of these we obtain what is known as the \textit{defect group} \cite{Morrison:2020ool,Albertini:2020mdx,DelZotto:2015isa}
\be
\bD = \bigoplus_{p,k} \bD_{p,k}=\bigoplus_i\bD_i\,,
\ee
where on the RHS we have reorganized $\bD$ according to defects of dimension $i$.
There is in general a non-degenerate pairing on $\bD$, which is a bi-homomorphism
\be
\langle\cdot,\cdot\rangle:~\bD\times\bD\to U(1)\,,
\ee
which captures mutual non-locality of the defects in $\bD$. The mutual non-locality captures the fact that not all of these defects can be simultaneously in the untwisted sector of the higher-form symmetries. Some of the defects have to be attached to topological operators generating the higher-form symmetries and non-locality arises as the defects pass these topological operators and acquire phases according to their charges.

There are various choices for consistently picking which defects will be twisted and which ones will be untwisted. Each such choice corresponds to what is known as a \textbf{polarization} or a \textbf{Lagrangian}, which is
\be
\L=\bigoplus_i\L_i\,,
\ee
where each $\L_i$ is a subgroup
\be
\L_i\subseteq\bD_i\,,
\ee
such that $\L$ is a maximal such subgroup on which the pairing trivializes
\be
\langle\cdot,\cdot\rangle|_{\L}=1\,.
\ee
It should be noted though that not all such choices of $\L$ may be allowed due to the presence of 't Hooft anomalies. Each allowed choice of $\L$ provides a `global form' for the $d$-dimensional QFT being constructed. One can cycle between different global forms by performing discrete gaugings (possibly with discrete torsions) of the higher-form symmetries described below.

Given an allowed $\L$, the resulting $d$-dimensional QFT has higher-form symmetries
\begin{equation}\label{ge_hfs}
	\cS_\L=\prod_{i}\G i\,,
\end{equation}
such that the $i$-form symmetry group is
\be
\G i=\wh\L_i\,.
\ee

\begin{tech}[Consistency with Swampland Conjecture]{}
According to one of the swampland conjectures, a consistent theory of quantum gravity cannot have any global symmetries, including higher-form symmetries. In the context of geometric engineering, quantum gravity theories arise if the internal compactification manifold is compact, as then compactified theory carries a non-zero value of $G_N$.
All the higher-form symmetries that we discussed are associated to non-compact cycles in the internal manifold, and hence the swampland conjecture is satisfied here. 

More generally, see \cite{McNamara:2019rup,McNamara:2020uza,Debray:2021vob,Heidenreich:2020pkc,Kaya:2022edp,Rudelius:2020orz,McNamara:2022lrw} for a sample of work at the interface of swampland and higher-form symmetries.
\end{tech}

\paragraph{Description in Terms of the Boundary.}
We can describe $\bD_{p-k+1}$ in terms of the boundary $\partial M_D$ of the internal manifold by using the long exact sequence in relative homology
\begin{equation}\label{1fs_ses}
	\dots \rightarrow H_i (\partial M_D) \xrightarrow[]{h_i} H_i (M_D) \xrightarrow[]{f_i} H_i (M_D, \partial M_D) \xrightarrow[]{g_i} H_{i-1} (\partial M_D) \xrightarrow[]{h_{i-1}} H_{i-1} (M_D) \rightarrow \dots \,.
\end{equation}
which implies
\be
H_k(M_D, \partial M_D) / H_k(M_D)=\text{im}(g_k)=\text{ker}(h_{k-1})\,,
\ee
and hence
\be
\bD_{p,k}=\text{ker}(h_{k-1})\,.
\ee
Thus $\bD_{p,k}$ has a natural interpretation as those $(k-1)$-cycles living on the boundary $\partial M_D$ that become trivial when pushed into the bulk $M_D$.

\begin{example}[$7d$ $\mathcal{N}=1$ SYM]{5dexample} 
Consider M-theory on
\begin{equation}\label{CY_5d}
	M_{D=4} = \mathbb{C}^{2} / \Gamma_{ADE} \,,
\end{equation}
where $\Gamma_{ADE}$ is a finite subgroup of $SU(2)$ using the ADE classification based on the McKay correspondence.
With a judicial choice of asymptotic fluxes \cite{Morrison:2020ool,Albertini:2020mdx}, this configuration engineers pure 7d $\mathcal{N}=1$ supersymmetric Yang-Mills theories with gauge algebra $\fg_{ADE}$. Let the corresponding simply connected group be denoted as $G_{ADE}$. 
From the gauge theory analysis discussed earlier, these theories have a defect group
\be\label{1ADE}
\bD=\bD_1\oplus\bD_{4}\,,
\ee
such that
\be
\bD_1\cong \bD_{4}\cong Z(G_{ADE})\,,
\ee
where $Z(G_{ADE})$ is the center of $G_{ADE}$. Here $\bD_1$ is the group of unscreened Wilson line operators and $\bD_{4}$ is the group of unscreened 't Hooft operators.

Now let us compute the defect group using the geometric engineering analysis. The boundary of internal manifold is
\be
\partial(\mathbb{C}^{2} / \Gamma_{ADE})=S^3/\Gamma_{ADE}\,,
\ee
whose first homology group is
\be
H_1(S^3/\Gamma_{ADE})=\text{Ab} [\Gamma_{ADE}]=Z(G_{ADE})\,,
\ee
where $\text{Ab} [\Gamma_{ADE}]$ is the abelianization of the (generally non-abelian) group $\Gamma_{ADE}$, which is isomorphic to $Z(G_{ADE})$. All of these 1-cycles are contractible when pushed into the bulk. Thus wrapping M2 and M5 branes on them, we obtain the contributions $\bD_1$ and $\bD_4$ to (\ref{1ADE}).
\end{example}

\paragraph{Description in Terms of Intersection Numbers.}
By Poincaré duality, we have\footnote{Here we are assuming that there is no torsion in homology of $M_D$.}
\be
H_k(M_D, \partial M_D)= H_{D-k}(M_D)\,,
\ee
where on the RHS we have homology group of compact $(D-k)$-cycles on the internal manifold $M_D$. The relevant quotient can now be expressed as
\be
\bD_{p,k}=H_k(M_D, \partial M_D) / H_k(M_D)=H_{D-k}(M_D) / H_k(M_D)\,,
\ee
which requires an injective homomorphism
\be
i:~H_k(M_D)\to H_{D-k}(M_D)\,.
\ee
The map $i$ maps a $k$-cycle $C_k$ to a $(D-k)$-cycle
\be
i(C_k)=\sum_{j} \left(C_k\cdot C_{D-k}^j\right) C_{D-k}^j\,,
\ee
where $C_{D-k}^j$ is a basis of $(D-k)$-cycles in $H_{D-k}(M_D)$ and $C_k\cdot C_{D-k}^j$ is the intersection number between $C_k$ and $C^j_{D-k}$ in $M_D$.

\begin{example}[5d $\cN=1$ SCFT: $E_1$ Theory]{}
An interesting and well-known 5d $\cN=1$ SCFT is the $E_1$ theory whose characteristic property is that it flows to 5d $\cN=1$ pure $SU(2)$ gauge theory (with trivial discrete theta angle) after a mass deformation.

This theory admits a geometric engineering construction in which the internal manifold is a non-compact Calabi-Yau threefold $M_6$ containing a single compact 4-cycle which is topologically
\be
C_4=\P^1\times\P^1\,.
\ee
The only compact 2-cycles are the two $\P^1$s inside $C_4$, both of which have intersection number
\be
\P^1\cdot C_4=-2\,.
\ee
Thus, we have for this geometry
\be
H_2(M_6,\partial M_6)/H_2(M_6)=\Z_2\,.
\ee
Wrapping M2 branes on these non-compact 2-cycles, we find a group
\be
\bD_1=\Z_2\,,
\ee
of unscreened line operators. These line operators are in the untwisted sector in the $E_1$ theory and thus we deduce a 1-form symmetry group
\be
\G1=\wh\bD_1=\Z_2\,.
\ee
for the $E_1$ theory.
\end{example}

\subsection{Generalised Symmetries in Holography}
In this subsection we will describe how generalised symmetries appear in holographic settings\footnote{For early works see \cite{Bergman:2020ifi,Hofman:2017vwr}, and for more recent applications see for example \cite{Apruzzi:2021phx,vanBeest:2022fss,Antinucci:2022vyk,Apruzzi:2022rei,Damia:2022bcd,Das:2022auy, Iqbal:2020lrt,Etheredge:2023ler, GarciaEtxebarria:2022vzq, Bah:2023ymy, Apruzzi:2023uma}}. 
The content of this subsection is intimately linked to that of section \ref{sec:SymTFTsection} on \textit{Symmetry Topological Field Theory}\footnote{See e.g.\ \cite{Apruzzi:2023uma} for an in-depth discussion of the precise mapping between the SymTFT and the topological bulk terms in a holographic context.}. 

\paragraph{SymTFT from Holography.}
Consider a setup with background geometry 
\be
\text{AdS}_{d+1} \times X_{D-d-1}\,,
\ee
with $D=10,11$ and $X$ representing some internal manifold\footnote{The restriction to AdS solutions here is just for illustration purposes. For example, the Klebanov-Strassler solution \cite{Klebanov:2000hb} is not of this type, but was studied using the below technology in \cite{Apruzzi:2021phx, Apruzzi:2022rei, Bah:2023ymy}}. After dimensionally reducing 10/11d supergravity, and focusing on solely the dominant contributions at the boundary, one is typically left with a bulk action of the form
\be\label{eq:bulkholographicterm}
S_{\rm bulk} = \sum_{p,i,j}\int_{\rm{AdS}_{d+1}}  N_{ij,p} \,a_{p+1}^i \wedge db^j_{d-p-1} + \cA \left( \{ a_{p+1}^i \} \right) \,,
\ee
where $a,b$ are bulk gauge fields which are differential forms on spacetime of degrees determined by their subscript, $N_{ij,p} \in \Z$ and $\cA$ is a function of gauge fields of type $a$.

Let us assume that $N_{ij,p}$ define non-degenerate matrices for each $p$. Then $S_{\rm bulk}$ is a description of a discrete (higher-form) gauge theory of the form (\ref{SymAct}) in terms of continuous fields. And the claim is that $S_{\rm bulk}$ is the SymTFT associated to some (higher-form) symmetries of the holographically dual $d$-dimensional QFT.

\paragraph{Symmetry Boundary Condition from Holography.}
As we discussed in section \ref{sec:SymTFTsection}, various gauge frames or global forms of the $d$-dimensional QFT are obtained by choosing boundary conditions of the SymTFT gauge fields. In the holographic description, this corresponds to choosing boundary conditions for the gauge fields appearing in $S_{\rm bulk}$ along the AdS boundary.

Let us discuss two examples:

\begin{example}[AdS$_5 \times S^5$: 4d $\cN=4$ SYM]{}
Consider the famous AdS$_5\times S^5$ setup in 10d IIB supergravity, which is dual to 4d $\cN=4$ pure super Yang-Mills theories with gauge algebra $\su(N)$. Beginning with the 10d IIB supergravity solution, one expands the higher-form fields around the background solution. Plugging these fluctuations into the equations of motion and Bianchi identities of 10d supergravity produces a set of equations. Since we are interested in dominant boundary contributions, we truncate them to single derivative terms which are the leading terms at large distances. From this, one can reverse-engineer the topological action reproducing these equations of motion, and finds it to be \cite{Witten:98}
\begin{equation}
\label{eq:CSIIB}
S_{\text{top}} = N \int_{\text{AdS}_{5}}b_2 \wedge d c_2\,.
\end{equation}
See \cite{Apruzzi:2021phx} for more details on the procedure. Here $b_2, c_2$ are continuous bulk 2-form gauge fields which come from expanding the NSNS and RR 2-form gauge fields of IIB supergravity around the fixed background solution.

The action $S_{\text{top}}$ describes SymTFT for 1-form symmetries of $\cN=4$ theories with $\su(N)$ gauge algebra.
We now consider the possible choices of boundary conditions.

\paragraph{(a) Boundary conditions 1.}
A simple choice of boundary conditions is
\begin{equation}
b_2 \, \text{  Dirichlet } \,, \qquad
c_2 \text{ Neumann} \,. 
\end{equation}
The bulk equations of motion of the action \eqref{eq:CSIIB} are 
\begin{equation}
Nb_2 = 0 \,, \quad Nc_2 = 0 \,.
\end{equation}
Compatibility with these equations of motion implies that these boundary conditions force $c_2$ to vary within $\mathbb{Z}_N$ in the boundary. Correspondingly Wilson surfaces for this $\Z_N$-valued gauge field on the boundary generate a 1-form symmetry, meaning that we have chosen a global form of the resulting 4d theory which carries a 1-form symmetry
\begin{equation}\label{G1h}
\G1 = \mathbb{Z}_N \,.
\end{equation}
We claim that these boundary conditions correspond to choosing the global form
which has gauge group
\be\label{cGh}
\cG=SU(N)\,.
\ee
This can be seen holographically by considering the Bianchi identity for the 7-form flux, which sources D5-branes. In particular integrating over the compact space $S^5$ we have
\begin{equation}
\int_{S^5} dF_7 = \int_{S^5} F_5 \wedge H_3 = N H_3 \,.
\end{equation}
This implies that $N$ fundamental strings (which couple to $b_2$) can be screened by ending on a D5-brane wrapped on $S^5$, which is precisely the baryon vertex of \cite{Witten:1998xy}. The fundamental strings become Wilson lines in the 4d gauge theory, and hence we learn that (\ref{G1h}) is purely an electric 1-form symmetry, thus justifying (\ref{cGh}).

\paragraph{(b) Boundary conditions 2. }
Let us exchange the Neumann and Dirichlet properties of $b_2$ and $c_2$, now imposing the boundary conditions
\begin{equation}\label{eq:swappedbc}
c_2 \, \text{  Dirichlet } \,, \qquad
b_2 \text{ Neumann} \,.
\end{equation}
By similar logic as above we learn that the 4d theory should have
\be\label{G1h2}
\G1=\Z_N\,,
\ee
1-form symmetry. However, this time the global form is
\be\label{cGh2}
\cG=PSU(N)\,,
\ee
gauge group with trivial discrete theta angle. This can again be justified holographically from the Bianchi identity for the other 7-form 
\begin{equation}
\int_{S^5} dH_7 =  -\int_{S^5} F_5 \wedge F_5 =  -N F_3 \,,
\end{equation}
which implies that $N$ D1 strings (which couple to $c_2)$ can be screened by ending on an NS5-brane wrapped on $S^5$. Since the D1 strings become 't Hooft lines in the 4d gauge theory, we learn that (\ref{G1h2}) is purely a magnetic 1-form symmetry, thus justifying (\ref{cGh2}).

\paragraph{(c) More general boundary conditions. }
More general boundary conditions give rise to other possible global forms for the gauge group and possible discrete theta angles. See \cite{Apruzzi:2021phx} for a discussion of some of these.
\end{example}

\begin{example}[AdS$_4 \times S^7/\mathbb{Z}_k$: 3d ABJM Theories]{} 
The low-energy description of a stack of $N$ M2-branes probing a $\mathbb{C}^4/\mathbb{Z}_k$ singularity is a three-dimensional Chern-Simons matter gauge theory with gauge group $U(N)_k \times U(N)_{-k}$ \cite{Aharony:2008ug}. The large $N$ holographic dual is 11d supergravity on AdS$_4 \times S^7/\mathbb{Z}_k$. There are many possible global variants \cite{Schnabl:2008wj,Tachikawa:2019dvq}
\be\ba
&U(N)_k \times U(N)_{-k} \,, \quad SU(N)_{k} \times SU(N)_{-k} \,, \quad \left( SU(N)_{k} \times SU(N)_{-k} \right) / \mathbb{Z}_N  \\
&\left( U(N)_k \times U(N)_{-k} \right) / \mathbb{Z}_{m'} \,, \quad \left( SU(N)_k \times SU(N)_{-k} \right) / \mathbb{Z}_{n'} \,,
\ea\ee
where $m'$ is a divisor of $k$ and $n'$ a divisor of $N$.

In \cite{Bergman:2020ifi} the authors give a holographic origin for \textit{all} these variants using the type of analysis discussed above, where they used 10d IIA supergravity. They argue that the 4d SymTFT contains a BF term
\begin{equation}\label{eq:bergmantopterm}
S_{4d} = \int_{\mathcal{M}_4} B_2 \wedge \left( NdA_{D0}+kdA_{D4} \right) \,.
\end{equation}
Here $B_2$ is the NS-NS 2-form gauge field and the $A$ fields are 1-form gauge fields sourced by D$0$ and D$4$ branes respectively. We explain how to derive this term using IIA supergravity below.
Applying different boundary conditions to the BF term gives a suite of 3d theories with different global symmetries and global gauge groups.

Let us now derive the topological action \eqref{eq:bergmantopterm} from the equations of motion of 10-dimensional type IIA supergravity. The AdS$_4 \times S^7/\mathbb{Z}_k$ M-theory has a description in Type IIA string theory on AdS$_4 \times \mathbb{C}P^3$. The background fluxes are as follows
\begin{equation}
\int_{\mathbb{C}P^3} \frac{F_6}{2\pi} \sim N \,, \quad \int_{\mathbb{CP}^1 \subset \mathbb{C}P^3} \frac{F_2}{2\pi} \sim k \,.
\end{equation}
We will follow the general philosophy outlined in \cite{Apruzzi:2021phx} to derive an effective topological action in four dimensions. We first expand the form fields of Type IIA on the cohomology of the internal space, making a choice of dualization frame (i.e.\ expanding $F_2$ or $F_8$ etc.). The internal space $\mathbb{C}P^3$ has homology
\begin{equation}\begin{aligned}
H^0(\mathbb{C}P^3;\mathbb{Z}) &= H^2(\mathbb{C}P^3;\mathbb{Z}) = H^4(\mathbb{C}P^3;\mathbb{Z}) = H^6(\mathbb{C}P^3;\mathbb{Z}) = \mathbb{Z}\,, \\
H^1(\mathbb{C}P^3;\mathbb{Z}) &= H^3(\mathbb{C}P^3;\mathbb{Z}) = H^5(\mathbb{C}P^3;\mathbb{Z}) = 0 \,.
\end{aligned}\end{equation}
We choose to represent $H^2, H^4$ by forms $\omega_2,\omega_4$, and normalise such that
\begin{equation}
\omega_2 \wedge \omega_4 = \text{vol}_{\mathbb{C}P^3} \,.
\end{equation}
Furthermore, we choose to expand in $F_2$ and $F_6$, for which we have explicit flux quantisation conditions without needing to resort to the Hodge dual operation
\begin{equation}\begin{aligned}
F_2 &= k \omega_2 + g_2 \,, \\
F_6 &= N \omega_2 \wedge \omega_4 + f_2 \wedge \omega_4 + f_4 \wedge w_2 \,, \\
H_3 &= h_3 + h_1 \wedge w_2 \,.
\end{aligned}\end{equation}
$F_4, F_8$ are obtained via the relations $F_4 = \ast F_6$, $F_8 = \ast F_2$:
\begin{equation}\begin{aligned}
F_4 &= N \text{vol}_{\text{AdS}_4} + \ast_4 f_2 \wedge \omega_2 + \ast_4 f_4 \wedge \omega_4\,,\\
F_8 &= k \text{vol}_{\text{AdS}_4} \wedge \omega_4 + \ast_4 g_2 \wedge \omega_2 \wedge \omega_4 \,,
\end{aligned}\end{equation}
where we have used the identity $\ast_6 \omega_2 = \omega_4$ and $\ast_{10} \text{vol}_{\text{AdS}_4} = \omega_2 \wedge \omega_4$. We now plug these expansions into the Type IIA equations of motion and Bianchi identities. We wish to identify \textit{leading order topological} contributions: we immediately drop terms that are not topological or have more than one derivative. Cycling through the various equations of motion we obtain
\begin{equation}\begin{aligned}
d \ast_{10} F_2 &= H_3 \wedge F_6 \\
0&= Nh_3 \wedge \omega_2 \wedge \omega_4 \,,
\end{aligned}\end{equation}
and
\begin{equation}\begin{aligned}
d \ast_{10} H_3 &= F_2 \wedge F_6 \\
0&= \left( kf_2 + Ng_2 \right) \wedge \omega_2 \wedge \omega_4 \,,
\end{aligned}\end{equation}
and 
\begin{equation}\begin{aligned}
d\ast_{10}F_6 &= H_3 \wedge F_2 \\
0 &= kh_3 \wedge \omega_2 \,.
\end{aligned}\end{equation}
From the Bianchi identities, $dF_6 = H_3 \wedge F_4, dF_2 = H_3 \wedge F_0, dH_3=0$, we isolate the following relevant contributions (setting $F_0=0$)
\begin{equation}
df_2 = 0 \,, \quad dh_3 = 0 \,, \quad dg_2 = 0\,,
\end{equation}
such that we can define gauge fields $a_1,b_2,c_1$
\begin{equation}
f_2 = da_1 \,, \quad h_3 = db_2 \,, \quad g_2 = dc_1 \,.
\end{equation}
We are therefore left with three topological, single derivative contributions
\begin{equation}\begin{aligned}
Ndb_2 &=0\,, \\
kdb_2 &= 0\,, \\
kda_1 + Ndc_1 &= 0 \,,
\end{aligned}\end{equation}
which are exactly the equations of motion of \eqref{eq:bergmantopterm} with the following identifications
\begin{equation}
b_2 \leftrightarrow B_2 \,, \quad c_1 \leftrightarrow A_{D0} \,, \quad a_1 \leftrightarrow A_{D4} \,.
\end{equation}
\end{example}

\section*{Acknowledgements}
We are grateful to Fabio Apruzzi, Marieke van Beest, Federico Bonetti, Mathew Bullimore, Andrea Ferrari, Davide Gaiotto, Simone Giacomelli, Max Hubner, Anton Kapustin, Ling Lin, Yasunori Lee, Jihwan Oh, Evyatar Sabag, Sakura Schafer-Nameki, Yuji Tachikawa, Apoorv Tiwari, Yinan Wang and Jingxiang Wu for collaboration on our research works related to the topic of generalized global symmetries.
LB thanks the audience at University of Oxford, ICTS Bangalore and Peking University for insightful questions, comments and discussions, where these lectures were delivered. We thank Sakura Schafer-Nameki for providing comments on a draft version of these notes. The research of LB on generalized symmetries has been supported by ERC grants 682608 and 787185 under the European Union’s Horizon 2020 programme, NSF grant PHY-1719924, and the Perimeter Institute for Theoretical Physics. Research at Perimeter Institute is supported by the Government of Canada
through Industry Canada and by the Province of Ontario through the Ministry of Economic Development and Innovation. AP is supported by the STFC under Studentship No. 2397217 through the University of Oxford and Prins Bernhard Cultuurfondsbeurs No. 40038041. 
LFT is supported by a Dalitz Scholarship from the University of Oxford and Wadham College. LG is supported by an STFC Studentship through the University of Oxford.

\bibliographystyle{ytphys}
\small 
\baselineskip=.94\baselineskip
\let\bbb\bibitem\def\bibitem{\itemsep4pt\bbb}
\bibliography{ref}

\end{document}